\documentclass{aa}  

\usepackage[colorlinks=true, linkcolor=blue, citecolor=blue]{hyperref}
\usepackage{graphicx}
\usepackage{txfonts}
\usepackage{xcolor}
\usepackage{halloweenmath}

\graphicspath{{figures/}} 

\usepackage{amsmath}	    
\usepackage{amssymb}	    
\usepackage{siunitx}        
\usepackage{nicefrac}       
\usepackage{mathtools}      
\usepackage{xcolor}         
\usepackage{textcomp}       

\usepackage{graphicx}	    
\usepackage{stfloats}  		

\newcommand{\orcid}[1]{\protect\href{https://orcid.org/#1}{\protect\includegraphics[width=8pt]{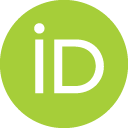}}}

\DeclareSIUnit\dex{dex}
\DeclareSIUnit\year{yr}
\DeclareSIUnit\fwhm{FWHM}
\DeclareSIUnit\ppm{ppm}
\DeclareSIUnit\ppmh{ppm\,h^{-\nicefrac{1}{2}}}
\DeclareSIUnit\arcsec{arcsec}
\DeclareSIUnit\pixel{pixel}
\DeclareSIUnit\electron{{e^-}}
\DeclareSIUnit\proton{{p^+}}
\DeclareSIUnit\electronvolt{{eV}}
\DeclareSIUnit\adu{ADU}
\DeclareSIUnit\dn{DN}
\DeclareSIUnit\hit{hits}
\DeclareSIUnit\events{events}
\DeclareSIUnit\photon{photons}
\DeclareSIUnit\volt{V}
\DeclareSIUnit\magnitude{mag}
\DeclareSIUnit\au{AU}
\DeclareSIUnit\pc{pc}
\DeclareSIUnit\exposure{exposure}

\newcommand{\comment}[1]{\textcolor{red}{#1}}

\newcommand{\paren}[1]{\left(#1\right)} 
\newcommand{\parenf}[1]{\left[#1\right]}

\newcommand{\tx}[2]{{#1}_{\text{#2}}} 			   	
\newcommand{\txxx}[2]{\text{#1}_\text{#2}} 			

\newcommand{\platosim}{\texttt{PlatoSim}}

\newcommand{\Pb}[0]{{\mathcal{P}}}

\newcommand{\focal}{{\rm \scriptscriptstyle FP}}
\newcommand{\plm}{{\rm \scriptscriptstyle PLM}}
\newcommand{\cam}{{\rm \scriptscriptstyle CAM}}
\newcommand{\equa}{{\rm \scriptscriptstyle EQ}}
\newcommand{\ccd}{{\rm \scriptscriptstyle CCD}}
\newcommand{\A}{{\rm \scriptscriptstyle A}}
\newcommand{\B}{{\rm \scriptscriptstyle B}}

\makeatletter
\newcommand{\Hline}{%
    \noalign {\ifnum 0=`}\fi \hrule height 1pt
    \futurelet \reserved@a \@xhline
}
\newcolumntype{\Vline}{@{\hskip\tabcolsep\vrule width 1pt\hskip\tabcolsep}}
\makeatother

\begin{document} 

\title{\platosim{}: An end-to-end PLATO camera simulator for modelling high-precision space-based photometry}

\author{
N.~Jannsen\inst{1}\orcid{0000-0003-4670-9616}\and
J.~De~Ridder\inst{1}\orcid{0000-0001-6726-2863}\and
D.~Seynaeve\inst{1}\orcid{0000-0002-0731-8893}\and
S.~Regibo\inst{1}\orcid{0000-0001-7227-9563}\and
R.~Huygen\inst{1}\orcid{0000-0003-2390-6051}\and
P.~Royer\inst{1}\orcid{0000-0001-9341-2546}\and
C.~Paproth\inst{2}\orcid{0000-0002-0346-2169}\and
D.~Grie{\ss}bach\inst{2}\orcid{0000-0001-7166-921X}\and
R.~Samadi\inst{3}\orcid{0000-0003-1446-8934}\and
D.~R.~Reese\inst{3}\orcid{0000-0003-4854-7550}\and
M.~Pertenais\inst{2}\orcid{0000-0002-2174-117X}\and 
E.~Grolleau\inst{3}\orcid{0000-0002-6453-8737}\and
R.~Heller\inst{4}\orcid{0000-0002-9831-0984}\and
S.~M.~Niemi\inst{5}\and 
J.~Cabrera\inst{6}\orcid{0000-0001-6653-5487}\and
A.~Börner\inst{2} \and
S.~Aigrain\inst{7}\orcid{0000-0003-1453-0574}\and
J.~McCormac\inst{8}\orcid{0000-0003-1631-4170}\and
P.~Verhoeve\inst{5}\orcid{0009-0007-7260-3846}\and
P.~Astier\inst{3}\and
N.~Kutrowski\inst{9}\and
B.~Vandenbussche\inst{1}\orcid{0000-0002-1368-3109}\and
A.~Tkachenko\inst{1}\orcid{0000-0003-0842-2374}\and
C.~Aerts\inst{1,10,11}\orcid{0000-0003-1822-7126}
}

\institute{
Institute for Astronomy, KU Leuven, Celestijnenlaan 200D bus 2401, 3001 Leuven, Belgium \\
\email{nicholasj@ing.iac.es}
\and
Institute of Optical Sensor Systems, German Aerospace Center, Rutherfordstra{\ss}e 2, 12489 Berlin-Adlershof, Germany
\and
LESIA, Observatoire de Paris, Université PSL, Sorbonne Université, Université Paris Cité, CNRS, 5 place Jules Janssen, 92195 Meudon, France
\and 
Max-Planck-Institut für Sonnensystemforschung, Justus-von-Liebig-Weg 3, 37077 Göttingen, Germany
\and
European Space Agency/ESTEC, Keplerlaan 1, 2201 AZ Noordwijk, The Netherlands
\and 
Institute of Planetary Research, German Aerospace Center, Rutherfordstr. 2, 12489 Berlin, Germany
\and
Sub-department of Astrophysics, Department of Physics, University of
Oxford, Oxford OX1 3RH, UK  
\and
Department of Physics, University of Warwick, Gibbet Hill Road, Coventry, CV4 7AL, UK
\and
Thales Alenia Space, 5 All. des Gabians, 06150 Cannes, France
\and
Department of Astrophysics, IMAPP, Radboud University Nijmegen, PO Box 9010, 6500 GL Nijmegen, The Netherlands
\and
Max Planck Institute for Astronomy, Koenigstuhl 17, 69117 Heidelberg, Germany
}

\date{\today}
 
\abstract
{PLAnetary Transits and Oscillations of stars (PLATO) is the ESA M3 space mission dedicated to detect and characterise transiting exoplanets including information from the asteroseismic properties of their stellar hosts. The uninterrupted and high-precision photometry provided by space-borne instruments such as PLATO require long preparatory phases. An exhaustive list of tests are paramount to design a mission that meets the performance requirements and, as such, simulations are an indispensable tool in the mission preparation.}
{To accommodate PLATO's need of versatile simulations prior to mission launch that at the same time describe innovative yet complex multi-telescope design accurately, in this work we present the end-to-end PLATO simulator specifically developed for that purpose, namely \platosim{}. We show, step-by-step, the algorithms embedded into the software architecture of \platosim{} that allow the user to simulate photometric time series of charge-coupled device (CCD) images and light curves in accordance to the expected observations of PLATO.}
{In the context of the PLATO payload, a general formalism of modelling, end-to-end, incoming photons from the sky to the final measurement in digital units is discussed. According to the light path through the instrument, we present an overview of the stellar field and sky background, the short- and long-term barycentric pixel displacement of the stellar sources, the cameras and their optics, the modelling of the CCDs and their electronics, and all main random and systematic noise sources.}
{We show the strong predictive power of \platosim{} through its diverse applicability and contribution to numerous working groups within the PLATO mission consortium. This involves the ongoing mechanical integration and alignment, performance studies of the payload, the pipeline development, and assessments of the scientific goals.}
{\platosim{} is a state-of-the-art simulator that is able to produce the expected photometric observations of PLATO to a high level of accuracy. We demonstrate that \platosim{} is a key software tool for the PLATO mission in the preparatory phases until mission launch and prospectively beyond.}


\maketitle

\section*{A public announcement}

\textbf{\platosim{} is now an open source software:} With less than a year before the launch of the PLATO mission \citep{rauer2025plato}, we here provide an update to the already published version of this work \citep[see][]{jannsen2024platosim}: our simulator is now fully public to the scientific community. With the public access to our GitHub\footnote{\url{https://github.com/IvS-KULeuven/PlatoSim3}} and Documentation\footnote{\url{https://ivs-kuleuven.github.io/PlatoSim3/}}, this public release is announced in the occasion of the first PLATO Guest Observer call (AO-1)\footnote{\url{https://www.cosmos.esa.int/web/plato/ao-1}}. With this release, we also provide full access to the final frozen version of the PLATO input catalogue \citep[PIC 2.2.0.1;][]{nascimbeni2025plato}, for the first long-term PLATO pointing in the South (LOPS2). With the open access to the PIC, automatic access is granted to the following sub-PICs: the Target input catalogue \citep[tPIC;][]{montalto2026tPIC}, the Target Prime Sample input catalogue \citep[tPIC prime sample;][]{nascimbeni2026tPICprime}, the Target M-dwarf input catalogue \citep[tPIC P4 sample;][]{prisinzano2026tPICm4}, the Science Calibration and Validation input catalogue \citep[scvPIC;][]{zwintz2026scvPIC}, and the technical Calibration and Fine Guidance input catalogues \citep[cPIC and fgPIC;][]{heller2026fgPIC}.


\section{Introduction}\label{sec:introduction}


Thanks to the parts-per-million (ppm) precision photometry delivered by space telescopes in the past two decades, the astrophysical frontier has undergone a revolution. As a consequence of this technological progress, answers to profound questions are now within reach, such as the habitability of other planets beyond our Solar System. With the Convection, Rotation, and planetary Transits \citep[CoRoT;][]{auvergne2009corot} mission marking the start of this endeavour, the quest for habitable planets was followed by NASA's \textit{Kepler} space mission \citep{borucki2010kepler} aimed to discover the first Earth-like planet orbiting a Sun-like star. Together with \textit{Kepler}'s extended operation, the so-called \textit{K2} mission \citep{howell2014k2}, and successor, the Transiting Exoplanet Survey Satellite \citep[TESS;][]{ricker2015tess} mission, a wealth of scientific discoveries opened gateways to research areas and synergies thought impossible at the beginning of the millennium. Beyond the continuous planet discoveries by TESS, the ongoing CHaracterising ExOPlanet Satellite \citep[CHEOPS;][]{benz2021cheops} mission likewise plays a key role in the precise characterisation of known exoplanet systems.

Also complementary science cases have flourished from the long and uninterrupted observations by CoRoT and \textit{Kepler}, and shorter baseline but all-sky coverage observations by TESS. Being a principle driver for many of these missions, the ability to finally detect acoustic oscillations within the interior of solar-like stars \citep[e.g.][]{kjeldsen1994amplitudes} was not only a success story for asteroseismology \citep[e.g.][]{michel2008first, gilliland2010kepler, chaplin2010asteroseismic, chaplin2011esemble, hekker2013corot, huber2013fundamental}, but vital for studies of exoplanets \citep[e.g.][]{christensen2010asteroseismic, silva2015ages, campante2016asteroseismic}, stellar activity \citep[e.g.][]{garcia2010corot, chaplin2014inferences, brun2015solar, kiefer2022helio}, and galactic archaeology \citep[e.g.][]{stokholm2019subgiant, aguirre2020detection, hon2021quick, stello2022tess}. The last two decades of space data (initiated with the MOST mission; \citealt{walker2003most}) have also been a treasure for pulsation studies across the entire Hertzsprung-Russell diagram -- from the pre-main sequence to the last stages of stellar evolution, and from the lowest to the highest stellar masses \citep[e.g. see the reviews of][]{brown1994asteroseismology, cunha2007asteroseismology, aerts2010asteroseismology, garcia2019asteroseismology, bowman2020asteroseismology, aerts2021probing}. 


Despite the remarkable achievements from planet hunting space missions, no Earth-Sun analogue has been discovered to date \citep[e.g.][]{hill2023catalog} and the parameter landscape of low-mass and long-orbital period planets is vastly unexplored \citep[e.g.][]{bryson2020occurrence}. Due to the limited sky coverage of CoRoT and \textit{Kepler}, both targeting stars too faint to efficiently follow up with ground-based radial velocity (RV) surveys, the hunt for small terrestrial exoplanets by similar long-baseline missions is soon to be continued with the PLAnetary Transits and Oscillation of stars (PLATO; \citealt{rauer2014plato}, Rauer et al. in prep.) mission. PLATO is the third medium (M3) mission in ESA's Cosmic Vision 2015-2025 programme with a current launch date set for the end of 2026. Compared to its predecessors, PLATO aims to obtain high precision and continuous photometric time series of more than $245\,000$ bright stars ($V$-band magnitude $<15$) over its nominal mission duration of 4 years. Using the tools of asteroseismology and ground-based spectroscopy as an integrated part of the mission strategy, the goal of PLATO is not only to detect but also to characterise exoplanets around stars of magnitude $V<10$ with a precision of 5\% in radius and 10\% in mass. Moreover, PLATO is the first mission dedicated to derive the age of planets to a 10\% precision from asteroseismic modelling of the host stars. To meet its scientific objectives, PLATO has been designed to provide a photometric precision of $\leq\SI{50}{\ppmh}$ for more than $15\,000$ solar-like stars with $V\leq11$ (\citealt{rauer2014plato}, Rauer et al. in prep.).

 
PLATO's requirements are challenging due to numerous noise sources. These involve complex interactions between the components of the instrument which can best be modelled at pixel level. Thus, prior to the in-flight operations, end-to-end simulations of the instrument have proven to be a very efficient way to scrutinise performance bottlenecks for the success of the mission. Consequently, instrument simulators have been developed for missions covering a wide range of applications, in the X-ray (e.g. \texttt{SIXTE}; \citealt{dauser2019sixte}), optical (\texttt{CHEOPSim}; \citealt{futyan2020expected}, \textsc{SimCADO}; \citealt{leschinski2016simcado}), (near)infrared (\texttt{MIRISim}; \citealt{klaassen2021mirisim}, \texttt{Specsim}; \citealt{lorente2006specsim}), and all the way to the radio (\texttt{pyuvsim}; \citealt{lanman2019pyuvsim}). Also multi-purpose software packages exist, such as \texttt{MAISIE} \citep{o2016maisie} and \texttt{SOPHISM} \citep{rodriguez2018sophism}, and simulation frameworks, such as \texttt{Pyxel} \citep{arko2022pyxel}, which are specifically designed for detectors. 


To confirm that each performance requirement is within scope, several simulation tools have been developed for PLATO. The PLATO Light Curve Simulator \citep[\texttt{PSLS}\footnote{\url{https://sites.lesia.obspm.fr/psls/}};][]{samadi2019plato} and the PLATO Instrument Noise Estimator \citep[\texttt{PINE};][]{borner2022plato} are pragmatic yet simplified tools. None of these tools alone provide the ability to simulate all of the expected observations of the PLATO space mission, including image time series, meta data, housekeeping data, and light curves. In this work, we present \platosim{}%
\footnote{\url{http://ivs-kuleuven.github.io/PlatoSim3/}}, %
a dedicated end-to-end PLATO camera simulator with all of these features. 


\platosim{} builds on the heritage of the Eddington CCD Data Simulator (\citealt{arentoft2004eddington}, \citealt{de2006modelling}) that was developed for the decommissioned Eddington (ESA) and MONS (Danish Space Agency) space missions. The original code was later expanded to meet the demands of generalising simulations for space-borne observatories (such as the \textit{ASTRIOD Simulator}; \citealt{marcos2013astroid}). Aiming at realistic applications to the PLATO mission, \cite{zima2010plato} adapted the software into a so-called first generation end-to-end \textit{PLATO Simulator}. That included a change of software language from IDL to C++ to overcome pre-existing performance bottlenecks. Shortly after PLATO's selection as ESA's M3 candidate, \cite{marcos2014plato} revisited the software to give it a more modern modular software architecture and expanded its use cases for both the PLATO mission and other (future) photometric missions operating in the optical. 


With already existing multi-instrument software packages (such as \texttt{MAISIE} and \texttt{SOPHISM}), and the increasing demand for dedicated yet diverse use cases for the PLATO mission, the development of \platosim{} over recent years has somewhat replaced the mission adaptability aspect with an in-depth applicability for the PLATO instrument. This in turn has resulted in huge advancements of the algorithms implemented and allowed the software to stay up to date with changes ranging from the observational strategy at mission level to the description of the smallest hardware components of the payload. Furthermore, a complete Python wrapper around the generic C++ code has made it easy to configure, set up, and run simulations, which has especially proven valuable for the PLATO mission consortium. \platosim{} has so far been used by multiple teams to estimate the impact of technical or programmatic tradeoffs on the final mission performance, including end of life (EOL) ageing effects, preparation of the data-processing pipelines, preparation of the engineering and scientific calibrations, development and real-time testing of the fine guiding sensor algorithms, among others. 


In this paper we describe the implementation and algorithmic design of \platosim{}, but before doing so a small overview of the payload is given in Sect.~\ref{sec:instrument}. Next we present the image acquisition model in Sect.~\ref{sec:image_acquisition_model}, the image generation model in Sect.~\ref{sec:image_generation}, and each effect implemented will be detailed in Sect.~\ref{sec:incident_light_sources}~to~\ref{sec:electron_redistribution}. \platosim{}'s photometry module will be presented in Sect.~\ref{sec:photometry}, the software architecture in Sect.~\ref{sec:platosim}, applications in Sect.~\ref{sec:applications}, and concluding remarks in Sect.~\ref{sec:conclusion}.

\section{The PLATO instrument}\label{sec:instrument}

\begin{figure*}[t!]
\center
\includegraphics[width=1.04\columnwidth]{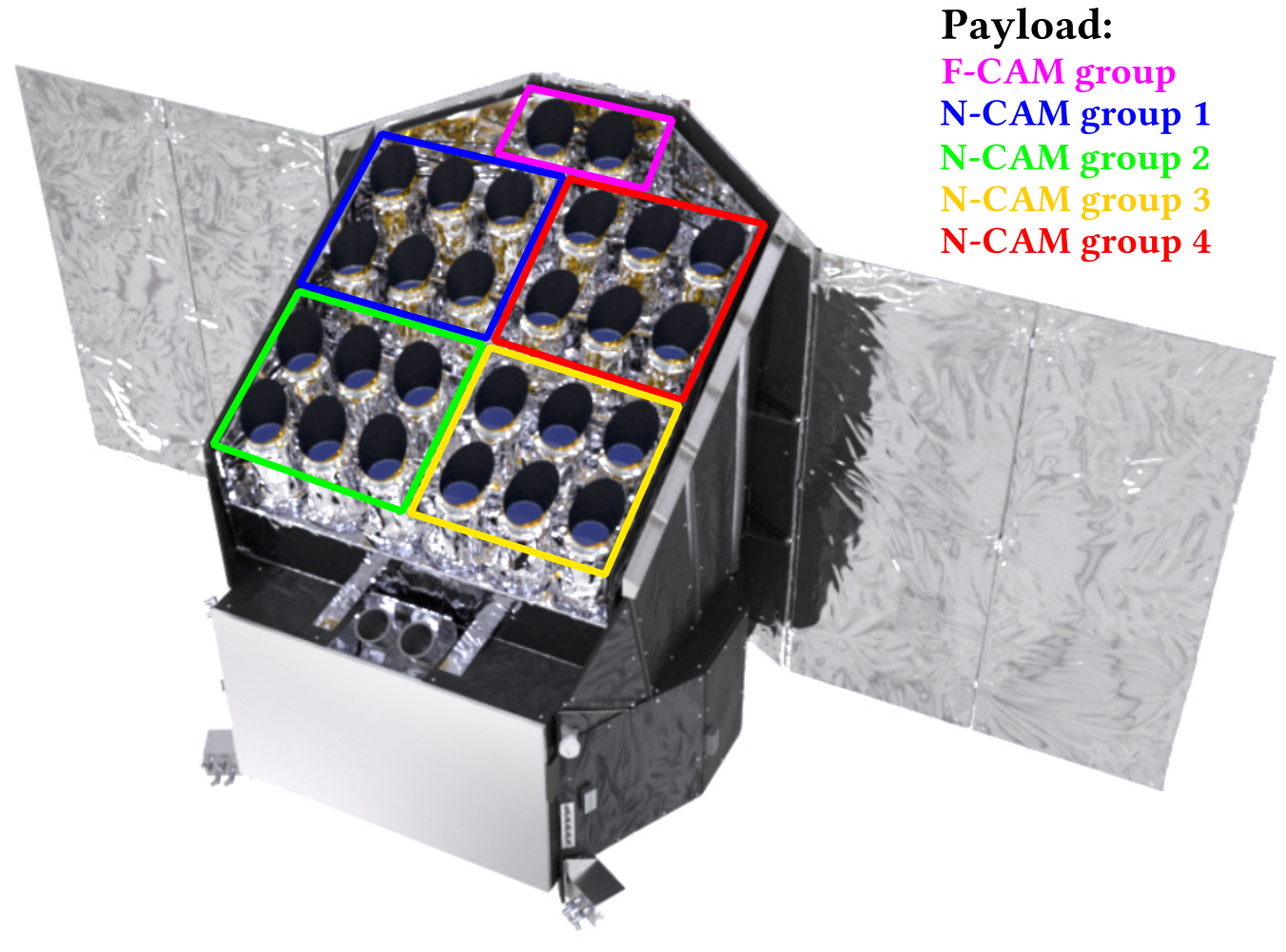}
\hspace{2mm}
\includegraphics[width=0.892\columnwidth]{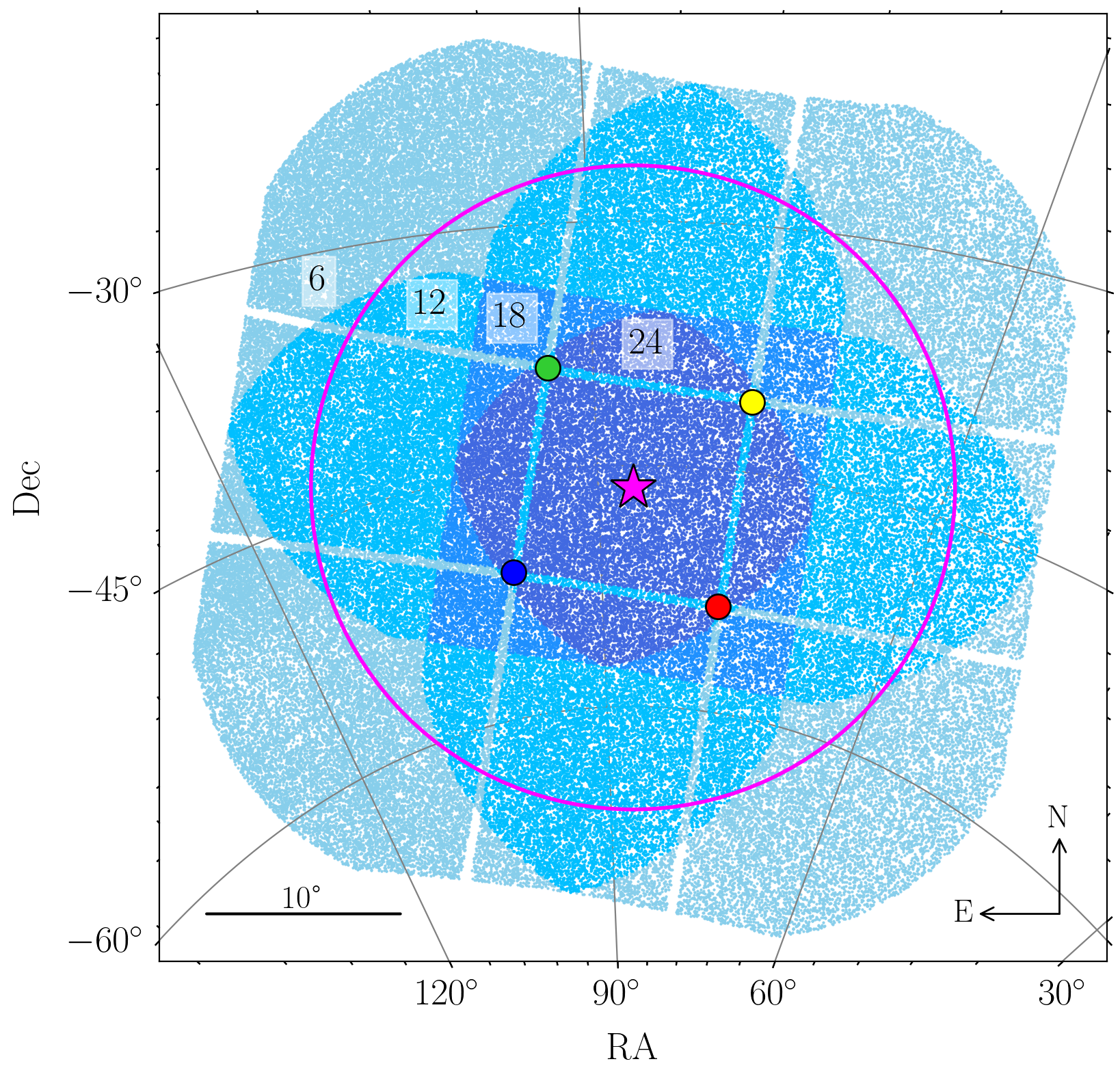}
\caption[]
{Overview of the PLATO multi-camera design. \textbf{Left:} Schematics of the PLATO spacecraft consisting of the payload module (with colour indication of the telescope groups) and the service module (bus). Credit: ESA/ATG medialab. \textbf{Right:} On-sky FOV of PLATO shown for a pointing towards the Long-duration Observation Phase (LOP) south in equatorial coordinates. The increasing darker shade of blue illustrates the increasing N-CAM overlap of $\tx{n}{CAM}\in\{6, 12, 18, 24\}$ (also indicated in the white boxes). The coloured dots show the pointing of each N-CAM group cf. the left-hand plot. The magenta star indicates the pointing of the two F-CAMs, which is parallel to the pointing of the platform, while the magenta circle shows the (camera-only) F-CAM FOV. \textit{Data courtesy: \cite{montalto2021all} and \cite{pertenais2021}}.} 
\label{fig:spacecraft}
\end{figure*}

As illustrated in the left panel of Fig.~\ref{fig:spacecraft}, the PLATO payload utilises an innovative multi-telescope concept consisting of 26 small but wide-field refractive telescopes \citep[$\sim$\SI{1\,037}{\deg\squared} each;][]{pertenais2021} mounted on a single optical bench. For historical reasons the PLATO mission consortium describes the combined unit of baffle, optical elements, and detectors as a \textit{camera} and hence for consistency we also adopt this nomenclature.

\begin{figure}[t!]
\centering
\includegraphics[width=\columnwidth]{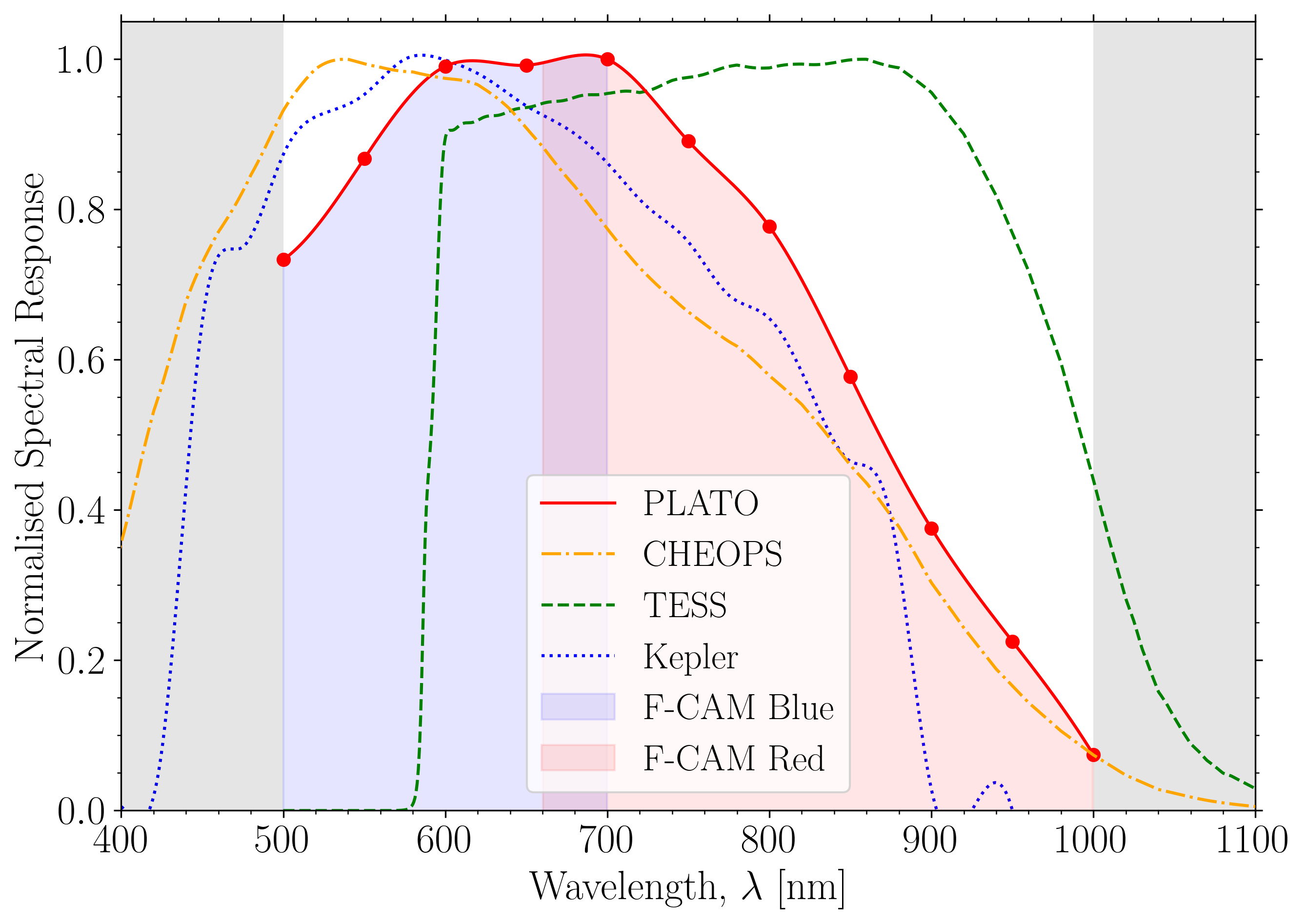}
\caption[]
{Preliminary normalised N-CAM spectral response curve at beginning of life (BOL) (with the red dots representing the mission requirements) compared to those similar planet hunting missions such as CHEOPS (orange dotted-dashed line), TESS (green dashed line) and \textit{Kepler} (blue dotted line). Each response curve is computed with cubic spline interpolation for illustrative purposes. The grey shaded areas are cut-off wavelengths which are dominated by optical transmission in the blue (left) and by the CCD anti-reflection coating in the red (right). The blue and red shaded regions illustrate the blue and red transmission regions of the two F-CAMs, respectively.\textit{ Data courtesy: ESA and NASA.}} 
\label{fig:passbands}
\end{figure}

Each camera consists of a telescope optical unit (TOU) with a \SI{12}{\centi\meter} entrance pupil diameter and a focal plane array (FPA) containing four CCDs connected to electronic controllers called front-end electronics (FEEs) -- in total comprising 104 CCDs and 26 FEEs. As visualised in Fig.~\ref{fig:spacecraft} (left) the cameras are organised in four groups each with six `normal' cameras (or N-CAMs) and one group of two `fast' cameras (or F-CAMs). A preliminary normalised spectral response curve for the N-CAM is shown in Fig.~\ref{fig:passbands}, which illustrates a strong similarity in response to the photometer of CHEOPS and \textit{Kepler} as the mission strategy of PLATO targets only slightly cooler (solar) spectral type stars (see e.g. the mission handbooks of CHEOPS%
\footnote{\fontsize{7.5}{5}\selectfont\url{https://sci.esa.int/web/cosmic-vision/-/53541-cheops-definition-study-report-red-book}}, %
TESS\footnote{\fontsize{7.5}{5}\selectfont\url{https://heasarc.gsfc.nasa.gov/docs/tess/documentation.html}}, %
and \textit{Kepler}%
\footnote{\fontsize{7.5}{5}\selectfont\url{https://archive.stsci.edu/missions-and-data/kepler/documents}})

All cameras of a given group share the same line of sight (LOS) and field of view (FOV). An opening angle of \SI{9.2}{\degree} of each N-CAM group relative to the F-CAMs, designed to increase the global sky coverage, entails a rather complex overlapping FOV arrangement as shown in the right-hand plot of Fig.~\ref{fig:spacecraft}. This plot shows the provisional long-duration observation phase (LOP) south field which is a subset of the all-sky PLATO Input Catalogue \citep[asPIC;][]{montalto2021all}. Only the FOV arrangement of the N-CAMs is shown here but the pointing of the F-CAMs is aligned with the pointing of the platform (magenta star). We note that the right-hand plot of Fig.~\ref{fig:spacecraft} shows the FOV of the spacecraft's orientation at the first \textit{mission quarter}. As PLATO will orbit the Sun from the second lagrange point (L2) in exactly one year, the spacecraft is required to realign its solar panels towards the Sun every $\sim91$ days in order to provide power to function. As we subsequently show in Sect.~\ref{sec:applications}, this is an important constraint for generating realistic simulations.

Depending on the exact location in the FOV a star may be observed with a number of overlapping N-CAMs being \linebreak $\tx{n}{CAM}\in\{6,12,18,24\}$ as indicated by the increasing colour gradient from light blue to dark blue. Considering only the effective FOV (i.e. the corrected optical FOV, taking into account the optical and mechanical vignetting and the gaps between the CCDs in the focal plane), the total estimated effective N-CAM FOV of \SI{2\,132}{\deg\squared} covers almost 19 times that of \textit{Kepler}. 

\begin{table}[t!]
\centering
\caption[]
{Characteristics of the PLATO payload.}
\begin{tabular}{ll}
\Hline
Parameter & Description/Value \\
\Hline
\multicolumn{1}{c}{TOU} & 26 units \\
\hline
Optics				& 6 refractive lenses \\
Design				& Axis-symmetric dioptics \\
Aperture diameter 	& \SI{12}{\centi\meter} \\
Full FOV 			& \SI{2132}{\deg\squared} \\
Spectral range 		& 500 -- \SI{1000}{\nano\meter} \\
\hline
\multicolumn{1}{c}{CCDs (FEEs)} & 104 units (26 units) \\
\hline
Model 				& Teledyne-e2v CCD270 \\
Design 				& Back illuminated \\
Pixel size 			& \SI{18}{\micro\meter} \\
Plate scale 		& \SI{15}{\arcsec} (on-axis) \\
\hline
\multicolumn{1}{c}{N-CAMs} & 24 units \\
\hline
Camera overlap		& 6, 12, 18, 24 (partial) \\
Effective FOV		& \SI{1037}{\deg\squared} \\
Detector FPA		& 4 full-frame CCDs \\
Detector size 		& $4510\times\SI{4510}{\pixel}$ \\
Exposure time		& \SI{21}{\second} \\
Readout time		& \SI{4}{\second} \\
Cadence 			& \SI{25}{\second} \\
\hline
\multicolumn{1}{c}{F-CAMs} & 2 units (red and blue filter) \\
\hline
Camera overlap		& 2 (full) \\
Effective FOV		& \SI{619}{\deg\squared} \\
Detector FPA	 	& 4 frame-transfer CCDs \\
Detector size 		& $4510\times\SI{2255}{\pixel}$ \\
Exposure time		& \SI{2.3}{\second} \\
Readout time		& \SI{0.2}{\second} \\
Cadence 			& \SI{2.5}{\second} \\
\hline
\end{tabular}
\label{tab:payload}
\end{table}

\begin{table}[t!]
\caption[]{List of PLATO data products downlinked to ground for pre-selected stars at different sampling rates. Some simple processing is applied to the pixel data before the extraction of for example light curves and flux centroids. We note that only imagettes are downlinked for the F-CAMs.}
\begin{center}
\begin{tabular}{lr}
\Hline
Data product 					& Sampling [s] \\
\Hline
Imagette for N-CAM (F-CAM)		& 25 (2.5) \\
Light curve: stellar aperture	& 50, 600 \\
Light curve: inverse aperture	& 50, 600 \\
Flux centroid: stellar aperture	& 50, 600 \\
Flux centroid: inverse aperture	& 50, 600 \\
Flux standard deviation 		& 600 \\
Calibration product				& 600 \\
Calibration windows				& 600 \\
\hline
\end{tabular}
\label{tab:data}
\end{center}
\end{table}

\begin{table}[h!]
\center
\caption{List of acronyms \textit{heavily} used throughout this paper.}
\begin{tabular}{ll}
\Hline
Acronym & Description  \\
\Hline
ADU 	& Analogue-to-Digital Unit \\
AIV     & Assembly, Integration, and Verification \\
AOCS 	& Attitude and Orbital Control System \\
asPIC	& All-sky PLATO Input Catalogue \\
		& \citep[See][]{montalto2021all} \\
BFE 	& Brighter-Fatter Effect \\
BOL		& Beginning Of Life \\
CCD 	& Charge-Coupled Device \\
CR 		& Cosmic Ray \\
CTI 	& Charge Transfer Inefficiency \\
CAM 	& PLATO camera \\
DPU 	& Data Processing Unit \\
DS		& Dark Signal \\
DSNU	& Dark Signal Non-Uniformity \\
EOL 	& End Of Life \\
FEE 	& Front-End Electronics \\
FP		& Focal Plane \\
FPA		& Focal Plane Array \\
FOV		& Field Of View \\
IPRNU   & Intra-Pixel Response Non-Uniformity \\
KDA 	& Kinematic Differential Aberration \\	
LOP 	& Long-duration Observation Phase \\
		& \citep[See][]{nascimbeni2021plato} \\
LOS 	& Line Of Sight \\
NSR 	& Noise-to-Signal Ratio \\
PSF 	& Point Spread Function \\
PRNU 	& Pixel Response Non-Uniformity \\
PLATO	& PLAnetary Transits and Oscillations of stars \\
		& (See \citealt{rauer2014plato}, Rauer et al. in prep.) \\
PLM 	& PayLoad Module \\
RS		& Readout Smearing \\
RTS		& Random Telegraph Signal \\
TED 	& Thermo-Elastic Drift \\
TOU 	& Telescope Optical Unit \\
\hline
\end{tabular}
\label{tab:abbreviations}
\end{table}

Sharing identical optical designs (as shown later in Fig.~\ref{fig:tou}) the main difference between the F-CAM and N-CAM is the operational mode, the readout cadence, and the wavelength transmission. With a readout cadence of \SI{2.5}{\second} secured by a CCD frame-transfer mode, the F-CAMs are foremost fine guidance sensors for the attitude and orbit control system (AOCS). Featuring frame-transfer CCDs implies that their FOV is about half the FOV of a single N-CAM \citep{pertenais2021}. Being equipped with respectively a blue and red colour filter makes the F-CAMs ideal science instruments for asteroseismology of bright stars. The N-CAMs operate without an optical filter in a full-frame readout mode at a cadence of \SI{25}{\second} and are the primary photometers used to meet the core science goals. 

\begin{figure*}[h!]
\centering
\includegraphics[width=2\columnwidth]{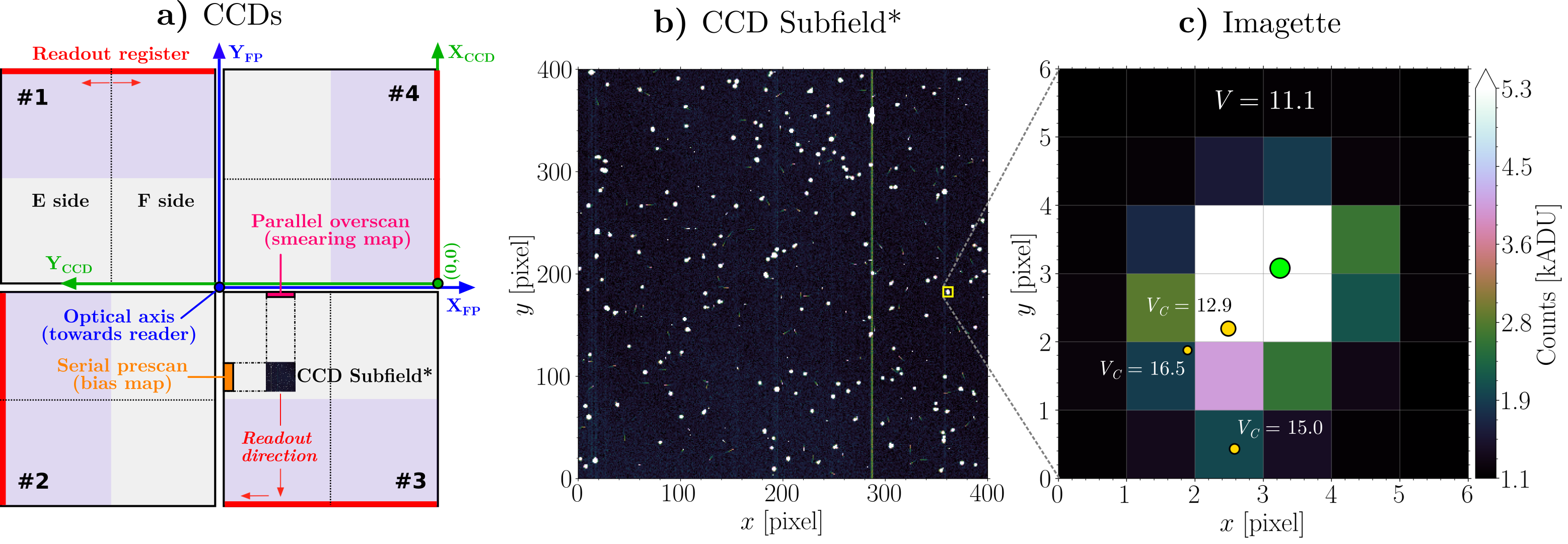}
\caption[]
{Schematic overview of how \platosim{} generates a CCD image in the FPA. \textbf{a)} Illustrative overview of the N-CAM FPA with the 4 CCDs. The blue axes indicate the focal plane with the central blue dot (in the middle of the 4 CCDs) represents the optical axis $\tx{Z}{FP}$ pointing in the positive direction towards the reader. The green axes illustrate the origin for $\tx{n}{CCD}=4$ as a reference and the readout register of each CCD is highlighted with a red bar. Each CCD is divided into an F and E side (with independent CCD and FEE gain appliances) with anti-parallel serial readout directions. For the F-CAM the location of the metallic shields of frame-transfer CCDs are highlighted as purple shaded rectangles. \textbf{b)} Simulated $400\times\SI{400}{\pixel}$ subfield of the LOP south. The subfield location on $\tx{n}{CCD}=3$ is displayed in panel a) together with the corresponding parallel overscan and serial prescan region which is used by \platosim{} to reconstruct a proper smearing and bias map, respectively. \textbf{c)} A so-called imagette showing a zoom-in on a target star from panel b). Indicated by the subpixel barycentres, the PIC target (green dot) has three significant fainter stellar contaminants (yellow dots scaled in size to the target star magnitude).} 
\label{fig:ccdFocalPlane}
\end{figure*}

\platosim{} is designed to model the Teledyne-e2v CCD270 detector that assemble the FPA of PLATO as shown in Fig~\ref{fig:ccdFocalPlane}a. A main characteristic of this detector is the large photosensitive area (see Table~\ref{tab:payload}) together with a division into two CCD halves (a F and E side; see dashed lines) each with an independent readout register to speed up the readout process. On top of this design, the frame-transfer CCDs of the F-CAM are divided into two CCD halves parallel to the readout register: a photosensitive area and a charge storage area. The charge storage area is covered by a metallic shield which is illustrated by the purple shaded regions in Fig.~\ref{fig:ccdFocalPlane}a. With the exception of an initial CCD frame-transfer for the F-CAM compared to the N-CAM, each CCD half is read-out following a standard \textit{rolling shutter} technique first in parallel direction towards the readout register and then in serial direction towards the corner of the register (i.e. left for the F side and right for the E side as seen in the CCD reference frame).

We elaborate more on the details of FPA in the following, however, before doing so, an overview of the PLATO data flow (from acquisition to downlink) is here placed in context to \platosim{}. While images are collected by the CCDs, the FEEs are responsible for extracting windows around preselected targets, sky background regions, and CCD regions used for calibration. By analogy with the FEE windowing strategy, the schematic overview of Fig.~\ref{fig:ccdFocalPlane} highlights an important feature which is a part of \platosim{}'s image acquisition model, namely the concept of a \textit{CCD subfield}. Being a CCD area under consideration of size ($\tx{n}{row}\times\tx{n}{col}$), the subfield is introduced to make long-duration simulation studies of modest sized subfields (such as Fig.~\ref{fig:ccdFocalPlane}b) feasible. It is however often more computationally efficient to only simulate an \textit{imagette} which is a $6^2$ pixel subfield (or $9^2$ pixel subfield for the F-CAMs) around a target star (illustrated in Fig.~\ref{fig:ccdFocalPlane}c). The imagette is a key data product of PLATO as all targets will have their photometry extracted from an imagette (see Sect.~\ref{sec:photometry}). Unlike the designated strategy to only extract imagettes in flight (except for saturated stars), the \platosim{} subfield can take any rectangular shape (smaller than the CCD dimensions) thus allowing more versatile simulations.

Following the data flow post the FEE, all data are sent to the data processing unit (DPU) that extracts and prepares each product for compression, archiving, and lastly transmission to ground. The PLATO data products that will be downlinked to ground are shown in Table~\ref{tab:data} and can be categorised into time series of: imagettes, fluxes, flux standard deviations, flux centroids, and data for calibration. To observe as many stars as possible, only imagettes will be sampled at the rate of the nominal N-CAM cadence, while the remaining data products will be sampled every \SI{50}{\second} or \SI{600}{\second} (i.e. averages over two or 24 exposures, respectively). Accompanying these measurements, time series of the inverted aperture mask (called an \textit{extended aperture}) will be computed as well. The calibration data consist of imagettes (or windows), for example, to compute the local sky background flux, axillary data, and housekeeping data. Except for the flux centroids, all data products of Table~\ref{tab:data} can be simulated by \platosim{} as will be addressed in this paper. 

A high level instrument summary of this section is provided in Table~\ref{tab:payload}. Moreover, since space mission terminology is known to be notoriously heavy, a glossary of acronyms that are used throughout this paper is presented in Table \ref{tab:abbreviations}.

\section{Image acquisition model}\label{sec:image_acquisition_model}

Every PLATO measurement starts with reading out the CCD subfield after an exposure to obtain the pixel values $S_{ij}$ expressed in analogue-to-digital units (ADU) where we use $i$ and $j$ as the row and column pixel coordinates. These pixel values can be modelled with
\begin{equation}\label{S}
S_{ij}(t) = \parenf{\left(I_{ij}(t) \ g_{{\rm CCD}, ij}(t) + B_{ij}(t)\right) \ g_{{\rm FEE}, ij}(t) + \epsilon_{\text{RN},ij}(t) }_{\tx{n}{bit}} \,.
\end{equation}
Here, $I_{ij}(t)$ is the number of photo- and thermal electrons accumulated in pixel $(i,j)$ during the exposure, which is caused by the target star, the sky background, open shutter smearing, dark signal, among others, and which we describe in Sect.~\ref{sec:image_generation}. The product of the CCD gain $\tx{g}{CCD}$ (expressed in \si{\micro\volt/\electron}) and the FEE gain $\tx{g}{FEE}$ (expressed in \si{\adu/\micro\volt}) make up the total gain $g$ (expressed in \si{\adu/\electron}),
\begin{equation}\label{gain}
g(t)=\tx{g}{CCD}(t) \ \tx{g}{FEE}(t) \,.
\end{equation}
We highlight the pixel dependence of the gain: since each CCD are read out in two halves, the left-hand side and the right-hand side have in practice slightly different gains.

Moreover, these gains are not constant but depend on the number of electrons in the well, leading to the well-known effect of CCD non-linearity. 
The underlying reason is that the CCD output amplifier and the FEE re-amplifier do no longer amplify linearly for a high number of electrons, leading to a sublinear increase of the output voltage (and thus ADU) with an increasing number of electrons. Measurements using the PLATO flight model CCDs revealed that the actual (bias subtracted) pixel signal deviates no more than \SI{100}{\adu} ($< 1\%$) from what a linear model would predict from the number of electrons in the well. This deviation can be reasonably well modelled using a simple polynomial, leading to
\begin{equation}\label{gain_nonlinearity}
    g_{{\rm FEE},ij} = a_0 + a_1 I_{ij} + a_2 I_{ij}^2 + a_3 I_{ij}^3 \, .
\end{equation}
The model of Eq.~\eqref{gain_nonlinearity} accounts both for the CCD non-linearity occurring at low and high flux levels.
 
The CCD gain and FEE gain also have a weak temperature dependence (of the order of $10^{-4}$ in units of \si{\adu/\micro\volt/\kelvin} and \si{\micro\volt/\electron/\kelvin}, respectively), and may therefore cause a small drift, which is relevant for high-precision photometry. In addition, the PLATO CCDs operate at a temperature of $-70^\circ$C while the FEEs operate around $30^\circ$C and their temperature variations will not be synchronised in time. \platosim{} allows to take this time-dependence into account which can help to define the temperature stability requirements related to the gains.

The quantity $B_{ij}$ in Eq.~(\ref{S}) is the bias level (expressed in \si{\micro\volt}), that is, the electronic offset produced by the FEE to ensure that the analogue-to-digital (A/D) converter always receives a positive signal. This bias level can be pixel dependent, as is the case for the PLATO detector. Laboratory measurements indicated that the bias is higher at the very first column -- the so-called \textit{line start effect} -- and then decreases and levels out to the median value in the subsequent columns at a rate of about 1--\SI{2}{\adu} per 100 pixels. The data reduction pipeline therefore needs to use a more elaborate method to determine the bias than simply using the median pixel value of the \textit{prescan} region. Simulations that take into account the proper bias pixel dependency are very useful to test this method. In addition, the bias level can be temperature and thus time-dependent. In the case of \platosim{} we allow for a small temperature instability at the level of $\SI{1}{\adu\per\pixel\per\kelvin}$ and the actual bias map for each CCD half is generated for a serial prescan and a (virtual) overscan region of $25 \times 15$ pixels, respectively (with the former shown in Fig. \ref{fig:ccdFocalPlane}a).

The readout noise $\epsilon_{\text{RN},ij}$ in Eq.~\eqref{S} is assumed to be Gaussian,
\begin{equation}\label{read_noise}
\epsilon_{\text{RN},ij} \sim \mathcal{N}\paren{\mu=0, \ \sigma^2=\tx{\sigma}{RN,CCD}^2+\tx{\sigma}{RN,FEE}^2} \,,
\end{equation}
where $\tx{\sigma}{RN,CCD} = \SI{24.5}{\electron\per\pixel}$ and $\tx{\sigma}{RN,FEE} = \SI{32.8}{\electron\per\adu}$ are the readout noise levels of the CCD and the FEE at mission BOL, respectively. 

The quantity $[x]_{\tx{n}{bit}}$ in Eq.~\eqref{S} denotes the \textit{quantisation} done by the A/D converter, i.e. flooring the quantity $x$ to the nearest integer value in the range $[0, 2^{\tx{n}{bit}}-1]$. In case of PLATO's $\tx{n}{bit}=16$ bit CCDs, this leads to a theoretical upper pixel value of $65\,535$ ADU. In practice the total gain (expected to be $g\approx\SI{0.05}{\adu/\electron}$) controls the digital saturation, hence saturation is expected to happen around \SI{55\,000}{\adu} and non-linear effects for even lower digital counts cf.~Eq.~\eqref{gain_nonlinearity}.

\section{Image generation model}\label{sec:image_generation}

CCD (sub)images are generated with a measurement cadence of \SI{25}{\second} (i.e. $\Delta t_{\rm exp} \simeq \SI{21}{\second}$ exposure and $\Delta t_{\rm ro} \simeq \SI{4}{\second}$ readout) for the N-CAMs, and with a cadence of \SI{2.5}{\second} (i.e. $\Delta t_{\rm exp} \simeq \SI{2.3}{\second}$ exposure and $\Delta t_{\rm ro} \simeq \SI{0.2}{\second}$ frame transfer) for the F-CAMs. Our simulations take into account that the readout of the four detectors in each N-CAM FPA (being four CCDs connected to a single FEE) is synchronised. I.e. all CCDs with the same identification number $\tx{n}{CCD} \in \{1,2,3,4\}$ across all cameras are read out simultaneously. Depending on $\tx{n}{CCD}$, the timing of each CCD readout is delayed by an amount
\begin{equation}
\Delta \tx{t}{CCD} = \SI{6.25}{\second}\cdot(\tx{n}{CCD}-1) \,,
\end{equation}  
due to the spacecraft's limited budget of electrical power, the limited bandwidth between the FEEs and the DPU, as well as to minimise cross-talk between CCDs and FEEs. The same does not hold for the CCDs in the F-CAMs which are read out simultaneously to secure a stable fine guidance and an exact timing of the colour photometry. We note that the time stamps mentioned here are on-board time stamps, not barycentric ones.

The number of electrons $I_{ij}(t)$ accumulated in a pixel $(i,j)$ during one observation as explained in Eq.~\eqref{S} can be divided in two parts: those gathered during the exposure of the CCD and those accumulated during the CCD readout phase as there is no mechanical shutter in a PLATO camera that prevents light from hitting the detector during readout. The latter leads to so-called \textit{open shutter smearing} and will be explained in more detail in Sect.~\ref{sec:open_shutter_smearing}. Mathematically we can write 
\begin{equation}\label{I_ij}
\begin{aligned}
I_{ij}(t) & = \int\limits_t^{\mathclap{t+\Delta t_{\rm exp}}} I^{\rm (exp)}_{ij}(t')\ dt' 
         \ \ + \ \ \int\limits_{\mathclap{t+\Delta t_{\rm exp}}}^{\mathclap{t+\Delta t_{\rm exp}+ \Delta t_{\rm ro}}} I^{\rm (ro)}_{ij}(t')\ dt' \\
          & \approx \sum_n I^{\rm (exp)}_{ij}(t+n\ \delta t)\ \delta t 
            \ + \Delta t_{\rm ro} \cdot \bar{I}^{\rm (ro)}_{ij}(t) \,,
\end{aligned}
\end{equation}
where $I^{\rm (exp)}_{ij}$ and $I^{\rm (ro)}_{ij}$ are the electron accumulation rates during the exposure and readout phase, respectively. The integration over time during the exposure is discretised in time steps $\delta t$ which we take one-tenth of the time scale of the most rapidly varying time-dependent phenomenon that is affecting the electron accumulation rate (which is usually the spacecraft jitter). Since the readout time $\Delta t_{\rm ro}$ is considerable shorter than the exposure time our simulations do not track any time-dependence during the readout phase, but instead we compute an averaged electron accumulation rate and simply multiply it with the readout time, as outlined in more detail in Sect.~\ref{sec:open_shutter_smearing}.

The number of electrons per second $I^{\rm (exp)}_{ij}$ accumulated during a CCD exposure can be broken down further into photo-electrons coming from stars $F_{\!\star}$, the sky background $F_{\!\rm sky}$, stray light $F_{\!\rm stray}$, together with electrons induced by cosmic particles $F_{\!\rm cosmics}$, and thermal electrons that are generated even in the absence of incident light, causing the so-called dark current $F_{\!\rm dark}$. We note that each of these quantities are stochastic, for example $F_{\!\star}$ is modelled as a Poisson distributed random variable to include photon noise. Mathematically, $I^{\rm (exp)}$ can be written as
\begin{equation}\label{I_ij_exp}
\begin{aligned}
I^{\rm (exp)}_{ij}(t) = 
& \iiint\limits_{i \ j \ \lambda} 
\parenf{ F_{\!\star}(t,x,y,\lambda) + F_{\!{\rm sky}}(x,y,\lambda) + F_{\!{\rm stray}}(t,x,y, \lambda) }  \\
& \qquad\,\, \cdot T(t,x,y,\lambda) \cdot E(t,x,y,\lambda) \cdot Q(t,x,y,\lambda) \ d\lambda \ dx \ dy \\[2mm]
& + F_{\!{\rm cosmic}}(t,i,j) + F_{\!{\rm dark}}(t,i,j) \,,
\end{aligned}
\end{equation}
where the integration happens in two spatial dimensions in the focal plane covering pixel $(i,j)$ as well as over the wavelength $\lambda$ in the relevant spectral range. In this expression, $T(t,x,y,\lambda)$ denotes the optical throughput of the PLATO camera, $E(t,x,y,\lambda)$ is the detector efficiency (including e.g. the pixel response non-uniformities and bad pixels), and $Q(t,x,y,\lambda)$ is the quantum efficiency of the CCD. Each of the quantities mentioned above will be described in more detail in this or subsequent sections.

In addition to the \textit{electron accumulation} described in Eq.~\eqref{I_ij_exp}, there are also several \textit{electron redistribution} effects that are not as easily described by the expression in Eq. \eqref{I_ij_exp}. The most relevant ones are charge diffusion in the silicate, charge-transfer inefficiency during readout, and full-well saturation causing blooming. These effects cause a redistribution of electrons in surrounding pixels, and are modelled in \platosim{} as an additional process after the CCD exposure. We refer to Sect.~\ref{sec:electron_redistribution} for more details.

Equations \eqref{I_ij} and \eqref{I_ij_exp} highlight the computational challenge that simulating space-based images poses. PLATO's primary science goal, exoplanets, requires knowledge about instrumental noise in the low-frequency regime (i.e. a time scale similar to both the transit duration and the orbital period of a planet) as well as the noise in the higher-frequency regime (i.e. a time scale similar to several phenomenae of stellar variability). It is therefore sometimes needed to simulate an observational run of 90 days (after which the observations are interrupted to turn the solar panels back towards the Sun). With a measurement cadence of \SI{25}{\second} for the N-CAMs this implies more than $311\,000$ measurements for a time series, while keeping track of the low-frequency drifts as well as the high-frequency jitter of the spacecraft using a sufficiently small time step $\delta t$. The computational burden of the spatial integration over each pixel is driven by the resolution needed to simulate the intra-pixel sensitivity variations, which requires to discretise each pixel in $64^2$ subpixel elements \citep[cf.][]{zima2010plato}. The integration over the wavelength range is needed to take into account the variation of the point spread function (PSF) as well as the optical throughput with wavelength. 

In practice some simplifications are needed to make the computations feasible. A first approximation is to eliminate the integration over the wavelengths by using a wavelength-averaged PSF weighted with a stellar spectrum of a Sun-like star, as well as using wavelength-averaged values of the throughput $T$, the detector efficiency $E$, and the quantum efficiency $Q$. This reduces Eq. \eqref{I_ij_exp} to the simplified expression
\begin{equation}\label{I_ij_exp_nowave}
\begin{aligned}
I^{\rm (exp)}_{ij}(t) = 
& \iint\limits_{i\  j} \parenf{ \bar{F}_{\!\star}(t,x,y) + \bar{F}_{\!{\rm sky}}(x,y) + \bar{F}_{\!{\rm stray}}(t,x,y) }  \\
& \quad\,\,\,\, \cdot \bar{T}(t, x, y) \cdot \bar{E}(t, x, y) \cdot \bar{Q}(t,x,y) \ dx\ dy \\[2mm]
& + F_{\!{\rm cosmics}}(t,i,j) + F_{\!{\rm dark}}(t,i,j) \,.
\end{aligned}
\end{equation}
A second approximation is to neglect the intra-pixel variations and only take into account the pixel variations for those use cases 
where it has a limited effect.

In the following sections, we provide more details on how we take into account the different quantities included in Eq.~\eqref{I_ij_exp_nowave}.
Section~\ref{sec:incident_light_sources} deals with the incident radiation fluxes $\bar{F}_{\!\star}(t,x,y)$, $\bar{F}_{\!{\rm sky}}(x,y)$,
and $F_{\!{\rm cosmics}}(t,i,j)$. This also involves computing the time-dependent focal plane coordinates of the stars, which is 
described in Sect.~\ref{sec:star_positions}. Section~\ref{sec:optical_throughput_detector_efficiency} deals with the 
throughput and efficiency quantities $\bar{T}(t,x,y)$, $\bar{E}(t,x,y)$, and $\bar{Q}(t,x,y)$.

\section{Light and electron sources}\label{sec:incident_light_sources}

This section focuses on part of the ingredients of the image generation described in Sect.~\ref{sec:image_generation}, 
more particularly on the polychromatic sources 
$\bar{F}_{\!\star}(t,x,y)$, $\bar{F}_{\!{\rm sky}}(x,y)$, $\bar{F}_{\!{\rm stray}}(t,x,y)$, $F_{\!{\rm cosmics}}(t,i,j)$, $F_{\!{\rm dark}}(t,i,j)$ 
in Eq.~\eqref{I_ij_exp_nowave}, representing respectively flux originating from incident stellar light, sky background light, stray light, 
and photoelectrons coming from cosmic hits and dark current.

\begin{figure*}[t!]
\centering
\includegraphics[width=\columnwidth]{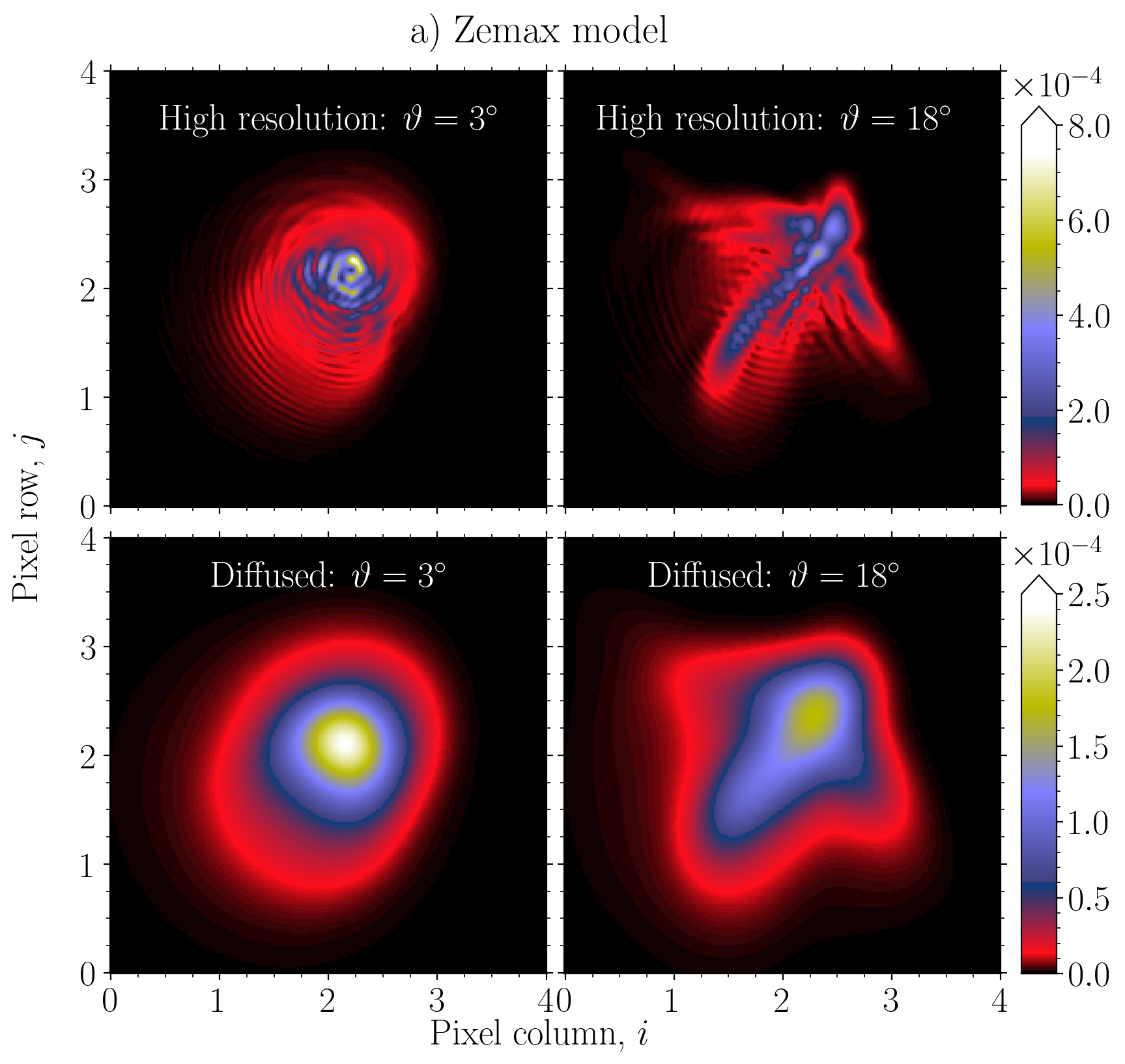}
\includegraphics[width=\columnwidth]{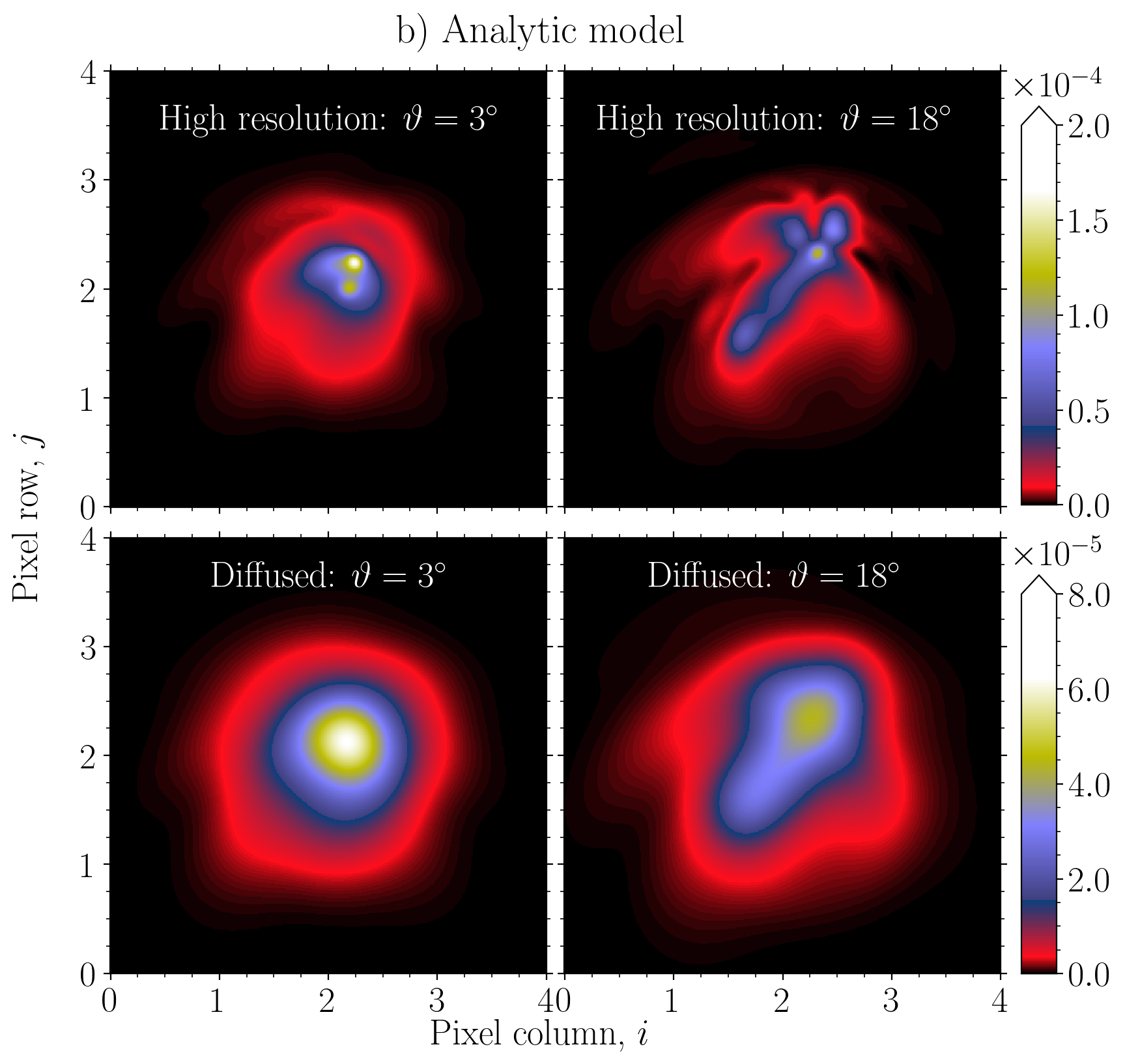}
\caption[]
{Illustration of a synthetic PLATO PSF generated at different optical axis distances $\vartheta$ with \textbf{a)} Zemax OpticStudio and \textbf{b)} an analytic model. The top panels show the high resolution PSF for $\vartheta=\SI{3}{\degree}$ (left) and $\vartheta=\SI{18}{\degree}$ (right). The lower panels show the corresponding PSF after a $\SI{0.2}{\pixel}$ Gaussian diffusion kernel has been applied. Each PSF is constructed at an azimuth angle of \SI{45}{\degree} and has a resolution of 64 subpixel elements corresponding to a $\SI{1}{\arcmin} \times \SI{1}{\arcmin}$ field on the sky. The image is normalised such that the sum over all pixels is equal to 1.} 
\label{fig:PSF}
\end{figure*}

\subsection{Incident stellar light}\label{sec:stellar_signal}

The number of monochromatic photons per second $F_{\!\star}(t,x,y,\lambda)$ in Eq. (\ref{I_ij_exp}) coming from incident light can be further broken down as
\begin{equation}\label{eq:Fstar_monochromatic}
F_{\!\star}(t,x,y,\lambda) = A \cdot f_{\!\star}(\lambda, t) \cdot g(x,y,x_0,y_0,\lambda,t) \,,
\end{equation}
where $A$ is the light collecting area of one camera ($\SI{113.1}{\cm\squared}$ in the case of a PLATO camera), $f_{\!\star}(\lambda, t)$ is the spectral photon distribution of the star (expressed in \si{\photon\per\second\per\meter\squared\per\nm}) and $g(x,y,x_0,y_0,\lambda,t)$ is the monochromatic normalised PSF at focal plane coordinates $(x,y)$ of a star centred around the focal plane coordinates $(x_0, y_0)$. 

The main time dependence of the PSF comes from a temperature dependence, which can lead to a slight change of the focus. For PLATO, focus changes are dominated by thermal variations of the TOU structure (changing the distance between the lenses), the optical lenses (changing the diffractive index), and temperature differences between the bipods that interface the FPA to the optical bench \citep{borsa2022}. \platosim{} uses a grid of monochromatic (Huygens) PSFs computed with Zemax OpticStudio%
\footnote{\url{https://www.zemax.com/pages/opticstudio}} %
with a spatial sampling of $8^2$ pixels times $64^2$ subpixels per pixel. In-flight, the TOU will be temperature controlled around the pupil
of the camera to the optimal focus temperature, and hence from the above discussion, a fixed and homogeneous temperature of $-70^{\circ}\text{C}$
throughout the camera is assumed in the Zemax model. We note that this model realistically reflects the expected optical performance as it includes
image distortion together with typical manufacturing and integration tolerances (e.g. refractive index, irregularities, lens and lens surface tilt and/or decentre, and inter-lens distances).

In practice the point spread function of a star is also affected by so-called \textit{charge diffusion} \citep[see e.g.][]{Rodney2006, Fairfield2007, widenhorn2010charge, Lawrence2011}. When a photon enters the CCD silicate it frees one or more electrons, which then wander away, including laterally, for a short distance before they are collected by a gate electrode. The net result is that electrons can also end up in neighbouring pixels which slightly diffuses (blurs) the PSF. \platosim{} models this effect by convolving the PSF with a spherical Gaussian having a half-width of 0.2 pixels. In the remainder of this section, when we refer to the PSF we always designate the PSF that has been convolved with a diffusion kernel. Figure~\ref{fig:PSF} illustrates some PSF examples that are used by \platosim{}, both the Zemax as well as the analytical model, with and without charge diffusion taken into account.

As mentioned in the discussion leading to Eq. \eqref{I_ij_exp_nowave}, the integration over the wavelength is computationally cumbersome and in practice \platosim\ therefore uses a polychromatic normalised PSF $\bar{g}(x,y,x_0,y_0)$ 
derived as a weighted average of monochromatic PSFs, weighted with the spectral energy distribution (SED) of a G0 dwarf star in the wavelength range 
of the PLATO passband $\mathcal{P}$
\begin{equation}
    \bar{g}(x,y,x_0,y_0) = \frac{\displaystyle \int_{\mathcal{P}} g(x,y,x_0,y_0,\lambda) \ f_{\rm G0V}(\lambda) \
    d\lambda}{\displaystyle\int_{\mathcal{P}} f_{\rm G0V}(\lambda) \ d\lambda} \,,
    \label{eq:psf_polychromatic}
\end{equation}
where the SED $f_{\rm G0V}(\lambda)$ was taken from \citet{Coelho2005}.
This allows to approximate Eq. (\ref{eq:Fstar_monochromatic}) with the polychromatic photon flux
\begin{equation}
    \bar{F}_{\!\star}(t,x,y) = A \cdot \ F_{\!\star}(t) \cdot \bar{g}(x,y,x_0,y_0) \,,
    \label{eq:Fstar_polychromatic}
\end{equation}
used in Eq.~(\ref{I_ij_exp_nowave}). In the expression above $F_{\!\star}(t)$ denotes the polychromatic stellar photon flux
integrated over $\mathcal{P}$
\begin{equation}
    F_{\!\star}(t) = \int_{\mathcal{P}} f_{\!\star}(\lambda,t) \ d\lambda \approx \Delta\lambda_{\mathcal{P}} \cdot F_0 \cdot 100^{m_V(t) / 5} \,.
    \label{eq}
\end{equation}
Here, $\Delta\lambda_{\mathcal{P}}$ is the full width at half maximum of the PLATO passband (about \SI{532}{\nm} for a normal camera), 
$V$ is the Johnson-Cousin visual magnitude of the star, and $F_0 = 1.00179 \cdot 10^8\, \si{\photon\per\second\per\meter\squared\per\nm}$ is the 
zero-point reference flux corresponding to a $V = 0$ G0-dwarf star. Alternatively, it is possible to use PLATO magnitudes
$\Pb$, which can be derived from the Johnson-Cousin $V$ magnitude using the transformation derived from synthetic stellar spectra by \cite{marchiori2019flight}
\begin{equation}\label{V-P}
V - \Pb = c_0 + c_1 \, \tx{T}{eff} + c_2 \, \tx{T}{eff}^2 + c_3 \, \tx{T}{eff}^3 \,,
\end{equation}
where $T_{\rm eff}$ is the effective temperature of the star. The best fit coefficients $\{c_0, c_1, c_2, c_3\}$ tabulated in \cite{marchiori2019flight} has recently been revisited by Fialho et al. (in prep.). The flux can then be derived (as was done in the same article) using
\begin{equation}\label{F_P}
F_\Pb = 100^{-(\Pb-\Pb_{\rm zp})/5} \,,
\end{equation}
with a mission BOL zero point of $\Pb_{\rm zp} = 20.77$ for an A0 dwarf star of $\Pb = 0$ being the current best fit estimate for the N-CAM (and correspondingly F-CAM zero-points of $\Pb_{\rm zp, blue} = 20.18$ and $\Pb_{\rm zp, red} = 19.81$ for the blue and red filter, respectively). 

The practical implementation of Eqs. (\ref{I_ij_exp_nowave}) and (\ref{eq:Fstar_polychromatic}) requires the focal plane coordinates $(x_0, y_0)$
of the star around which the PSF is centred, as well as a numerical integration of the PSF over the relevant pixels. The former are derived from the equatorial sky coordinates $(\alpha, \delta)$ of the star using the transformations outlined in Appendix \ref{app:reference_frames}. However, an accurate derivation also requires to take into account optical distortion, kinematic aberration, the (imperfect) attitude control of the spacecraft, and the thermal drift of the camera, all of which displace the PSF in the focal plane. More details of this description are given in Sect.~\ref{sec:star_positions}. The integration of the PSF is computationally non-trivial to implement because the PSF varies over the focal plane. Fast convolution of the PSF using the fast Fourier transform (FFT) is therefore only justified when the simulated region of the CCD is sufficiently small. Alternatively, Appendix~\ref{app:plato_psf} shows how an analytical approximation can be constructed that allows to efficiently integrate over the pixels, and which is used by \platosim{} in case of larger CCD regions. Such approximation is also beneficial to realistically characterise the PSF's change in shape and size over time induced typically by a change in the thermal environment, also known as \textit{PSF breathing} \citep[e.g. see][for orbital focus variations of HST]{bely1993orbital}.

\begin{figure}[t!]
\centering
\includegraphics[width=\columnwidth]{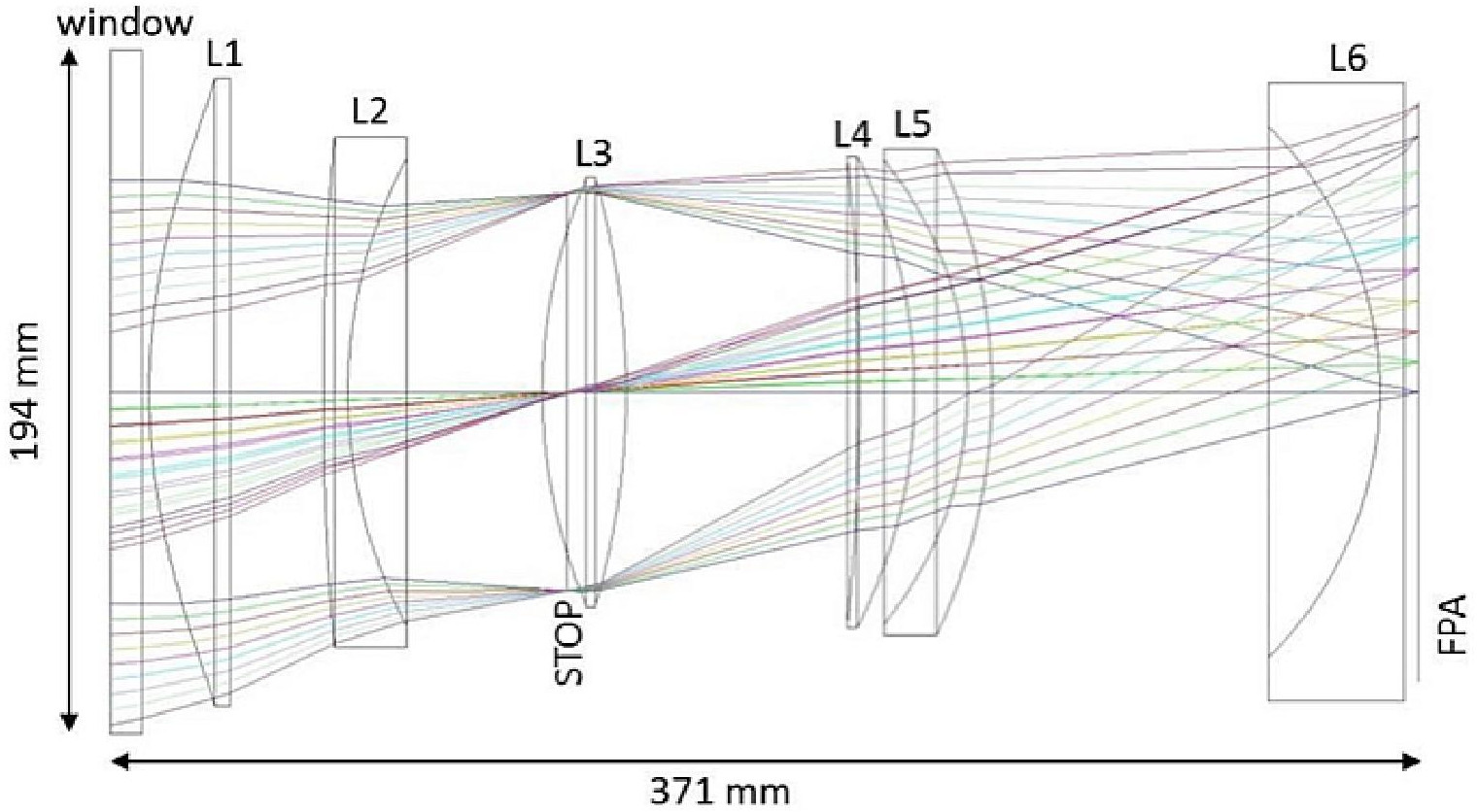}
\caption[]
{Layout of the Telescope Optical Unit (TOU) together with the detectors on the right forming the Focal Plane Array (FPA). Light passes the entrance window to the left (which for the F-CAMs is a dedicated optical filter) and propagates through the refractive optical lenses (L1--L6) unto the FPA on the right. \textit{Credit: ESA.}} 
\label{fig:tou}
\end{figure}

On top of all of this, \platosim\ takes into account that the same star is projected multiple times on the focal plane at different locations. Apart from the nominal PSF that carries more than 99.9\% of the light, there are two so-called \textit{ghost} ($\mathghost$) images of the star. The latter are caused by the fact that the optical elements of each camera (as shown in Fig.~\ref{fig:tou}) not only refract the light but also cause internal reflections so that part of the light is also projected elsewhere in the focal plane.

The most important ghost is a \textit{point-like} image that carries no more than 0.08\% of the light. Results from the test campaign of the engineering camera model revealed that the intensity of the point-like ghost decreases exponentially from the optical axis outwards. This ghost image will be extremely weak for stars for which the nominal PSF beyond $\SI{8}{\degree}$ away from the optical axis, however, in practise heavily saturated stars ($V \sim 0$) will show visible ghosts for nominal PSF positions $<\SI{12}{\degree}$ from the optical axis (at the level of a few tens of ADU). The point-like ghost is caused after reflection of the light on the CCD 
surface and on both surfaces of the entrance window (front and back). Its PSF is thus very similar to the nominal one, and is located diametrically 
opposite of the optical axis, that is
\begin{equation}\label{ghost_centroid_pointlike}
\begin{pmatrix}
x_\focal \\
y_\focal
\end{pmatrix}_{\mathghost \text{P}}
= -
\begin{pmatrix}
x_\focal \\
y_\focal
\end{pmatrix} \,,
\end{equation}
where we used the optical axis as the origin of the focal plane reference frame.
Point-like ghosts are therefore created by stars whose nominal PSF is on another CCD (e.g. see Fig.~\ref{fig:ccdFocalPlane}a). 
 
The second ghost image is a so-called \textit{extended} ghost, and is caused by reflection of the CCD surface and the back of lens L6 
(see again Fig.~\ref{fig:tou}). It carries only 0.003\% of the light (well below the mission requirements), and is located on the same 
CCD as the nominal PSF but radially shifted towards the edge of the FOV
\begin{equation}\label{ghost_extended_centroid}
\begin{pmatrix}
x_\focal \\
y_\focal
\end{pmatrix}_{\mathghost \text{E}}
= 1.0672
\begin{pmatrix}
x_\focal \\
y_\focal
\end{pmatrix} \,.
\end{equation}
An analysis with Zemax shows that its PSF can be well approximated with a homogeneous disk with a large diameter (hence the name extended) that depends on the distance of the optical axis, ranging from about 200 pixels close to the optical axis to more than 370 pixels at the edge of the FOV. The exact dependence was tabulated using Zemax, and then approximated with a second order polynomial interpolant that is used in \platosim\ simulations. The nominal source PSF will be inside the extended ghost for a star up to \SI{6}{\degree} away from the optical axis. As an example, a (saturated) star of $V=0$ located \SI{7}{\degree} from the optical axis will create a ghost of $\sim\SI{220}{\pixel}$ diameter distributed with $\sim\SI{800}{\electron\per\pixel}$ (as shown later in Fig. \ref{fig:alignment}). Clearly, with such a spatial dilution of a tiny fraction of the light, extended ghosts are only relevant for the brightest stars such as Canopus in LOP south and Vega in the LOP north \citep[following the updated results of][]{nascimbeni2021plato}, and will be well below the background noise for all other stars.

\subsection{Incident light from the sky background}\label{sec:sky_background}

The diffuse sky background $\bar{F}_{\!{\rm sky}}$ that affects every PLATO CCD measurement consists mainly of zodiacal light, and light from
unresolved Milky Way stars. The former is caused by sunlight being scattered by inter-planetary dust particles agglomerated across the ecliptic
plane. To model the zodiacal light for simulating space-based photometry, \cite{de2006modelling} used the monochromatic values of the zodiacal light
at $\lambda=500$ nm in the vicinity of the Earth tabulated by \cite{leinert19981997}, and assumed that the spectral distribution of the zodiacal
light is the same as the one of the Sun ($F_{\odot}(\lambda)$, tabulated in \citealt{wehrli1985extraterrestrial}) to estimate the amount of zodiacal
light flux that hits the detector. 

\cite{marchiori2019flight} adopted the same approach but improved upon it by including the reddening factor $f_{\rm red}(\lambda)$ of the solar
spectrum, the small correction factor $f_{L2}=0.975$ for a spacecraft in L2 rather than the direct vicinity of the Earth, and by including the
passband (i.e. the spectral response $S(\lambda)$) of a PLATO camera when integrating over the zodiacal spectrum. \platosim{} adopts the same approach which leads to the following expression for the zodiacal flux
\begin{equation}\label{F_ZL}
\tx{F}{ZL}(\alpha, \delta) = \frac{\tx{F}{ZL}(\alpha, \delta, \SI{500}{\nano\meter}) \cdot f_{L2}}{F_{\odot}(\SI{500}{\nano\meter})} 
\int F_{\odot}(\lambda) \ f_{\rm red}(\lambda) \ S(\lambda) \ \text{d}\lambda \,,
\end{equation}
where $\tx{F}{ZL}(\alpha, \delta, \SI{500}{\nano\meter})$ is the monochromatic zodiacal flux at \SI{500}{\nano\meter} derived from
\cite{leinert19981997}. To model the Galactic sky background \platosim{} adopts the same approach as in \cite{de2006modelling}, using tabulated Pioneer 10 observations from beyond \SI{2.8}{\au} (where the contribution of the zodiacal light is negligible). 

\begin{figure}[t!]
\centering
\includegraphics[width=\columnwidth]{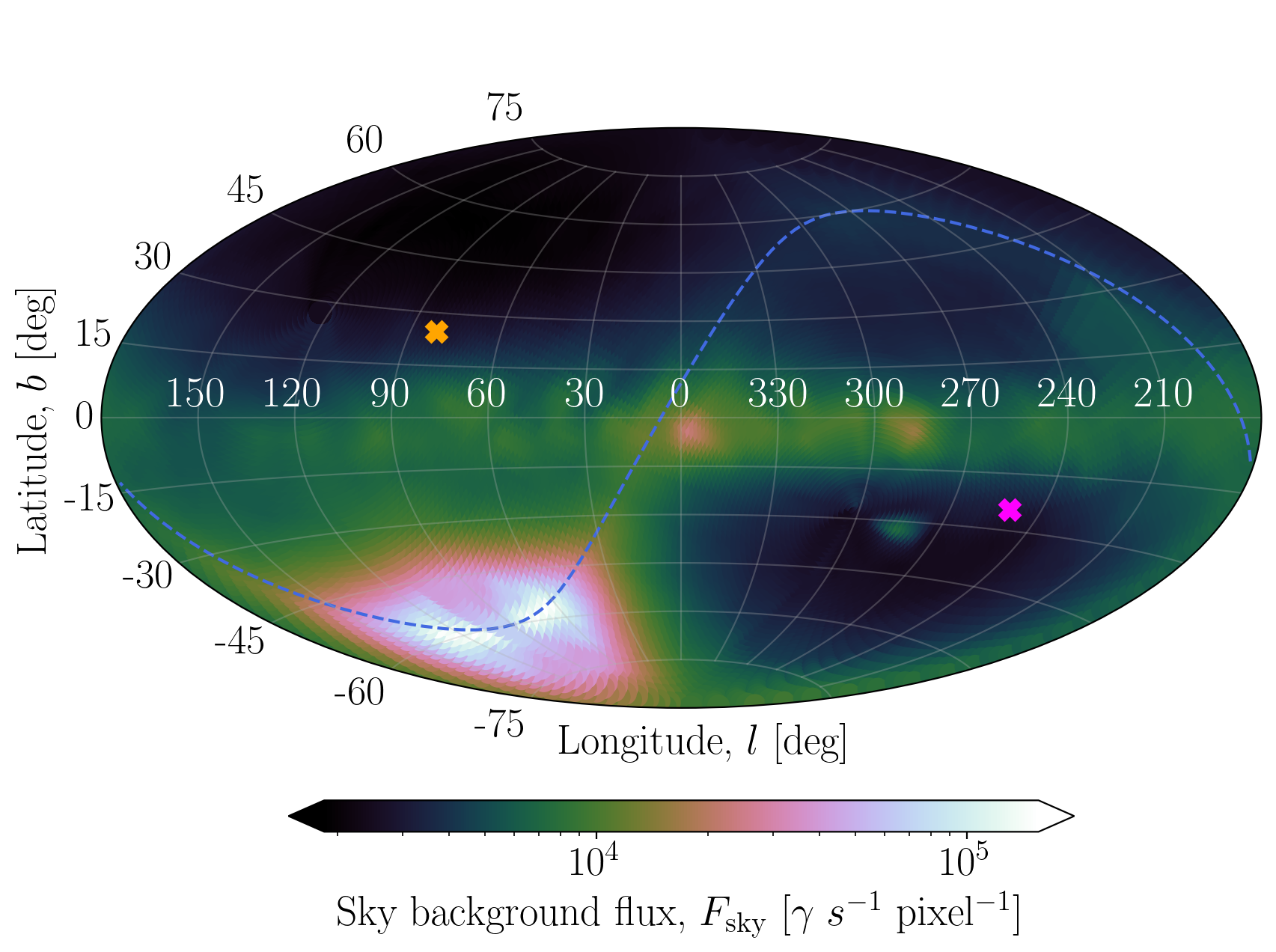}
\caption[]
{Aitoff projection in Galactic coordinates $(l,\,b)$ of the all-sky background model used by \platosim{}. The model includes zodiacal light and diffuse galactic light in units of incoming photons per second per pixel. The blue dashed line shows the ecliptic plane (with the location of the Sun clearly visible as the highest intensity area) and with the crosses illustrating respectively the LOP north (orange) and LOP south (magenta) from \cite{nascimbeni2021plato}. We note that data gaps in the zodiacal map of \cite{leinert19981997} and in the galactic Pioneer 10 map have been interpolated using a cubic spline.} 
\label{fig:skybackground}
\end{figure}

Figure~\ref{fig:skybackground} shows the combined sky background model of \platosim{} in an all-sky aitoff projection in Galactic coordinates together with the suggested LOPs of \cite{nascimbeni2021plato}. Since \platosim{} do not include the sky background of extra galactic sources, care must be taken for simulations that covers the FOV of the large Magellanic Clouds (which are partially overlapping with the LOP south). The final selected LOP(s) of PLATO will be chosen such that the combined sky background flux will be below the mission requirement of $\SI{20}{\electron\per\pixel\per\second}$.

Although the PLATO camera design has a baffle and a stray light mask, these do not perfectly block reflected light of celestial bodies in our Solar System. The main stray light contributors for PLATO are reflected light of the Earth and the Moon, for which the requirement of the combined flux is set to be less than $\SI{40}{\electron\per\pixel\per\second}$. The straylight differs from one camera group to the other, as different camera groups are pointing in a different direction. Detailed modelling of the straylight requires the exact sky position of the Earth and the Moon as well as an optical straylight model resulting from an in-depth analysis of the camera surfaces, coatings, and paintings. Due to the importance of stray light, the inclusion of such a model is currently under development for \platosim{}.

\subsection{Cosmic rays}\label{sec:cosmic_rays}

Cosmic rays (CRs) are high-energy cosmic particles that leave a high intensity trail over multiple pixels when colliding with a detector. The exact morphology of the dissipated energy on the detector depends both on the properties of detector (e.g. material and front/back illumination) and on the properties of the cosmic particle (e.g. particle type, energy, and incident angle). Furthermore, since the frequencies of cosmic rays depend on the time-varying space weather (dependent on the solar cycle, coronal mass-ejection events, galactic processes, etc.), and their impact strongly depends on spacecraft properties (such as physical orientation, material shielding/penetration, etc.), a realistic model is non-trivial. CR simulators do exist, such as \texttt{STARDUST} \citep{rolland2008stardust}, \texttt{Geant4} \citep{allison2016recent}, \texttt{GRAS} \citep{santin2005gras}, and \texttt{CosmiX}%
\footnote{\url{https://gitlab.com/david.lucsanyi/cosmix}} %
\citep{lucsanyi2020simulating}. All of these codes model particle transport using the Monte Carlo technique. \texttt{CosmiX} is the fastest code due to its semianalytical approach. \texttt{CosmiX} was initially developed for space-borne missions such a Gaia and PLATO, but is presently too time consuming to be efficiently integrated into \platosim{}. 

Instead a simplified model is implemented in \platosim{}: the number of CR hits during an exposure is drawn from a Poisson distribution around a mean value that scales with the exposure time $\tx{t}{exp}$ and subfield size ($\tx{n}{row}\times\tx{n}{col}$) of the CCD area under consideration
\begin{equation}\label{N_cos}
    \tx{N}{CR} \sim \mathcal{P}(\mu = \tx{R}{CR} \ \tx{\Delta t}{exp} \ n_{\rm row} \ n_{\rm col} \ \tx{s}{pix}^2) \,,
\end{equation}
where $\tx{R}{CR}$ is the cosmic hit rate (\si{\events\per\second\per\centi\meter\squared}), $\Delta t_{\rm exp}$ is the exposure time, and $s^2_{\rm pix}$ is the surface area of one pixel. The impact locations on the CCD are randomly chosen over the entire subfield. The trail length on the CCD of each cosmic hit is drawn from a uniform distribution, and the intensity in electron counts is drawn from a skew-normal distribution (given a location $\psi_{\rm CR}$, scale $\omega_{\rm CR}$, and shape parameter $\alpha_{\rm CR}$).

\begin{figure}[t!]
\centering
\includegraphics[width=\columnwidth]{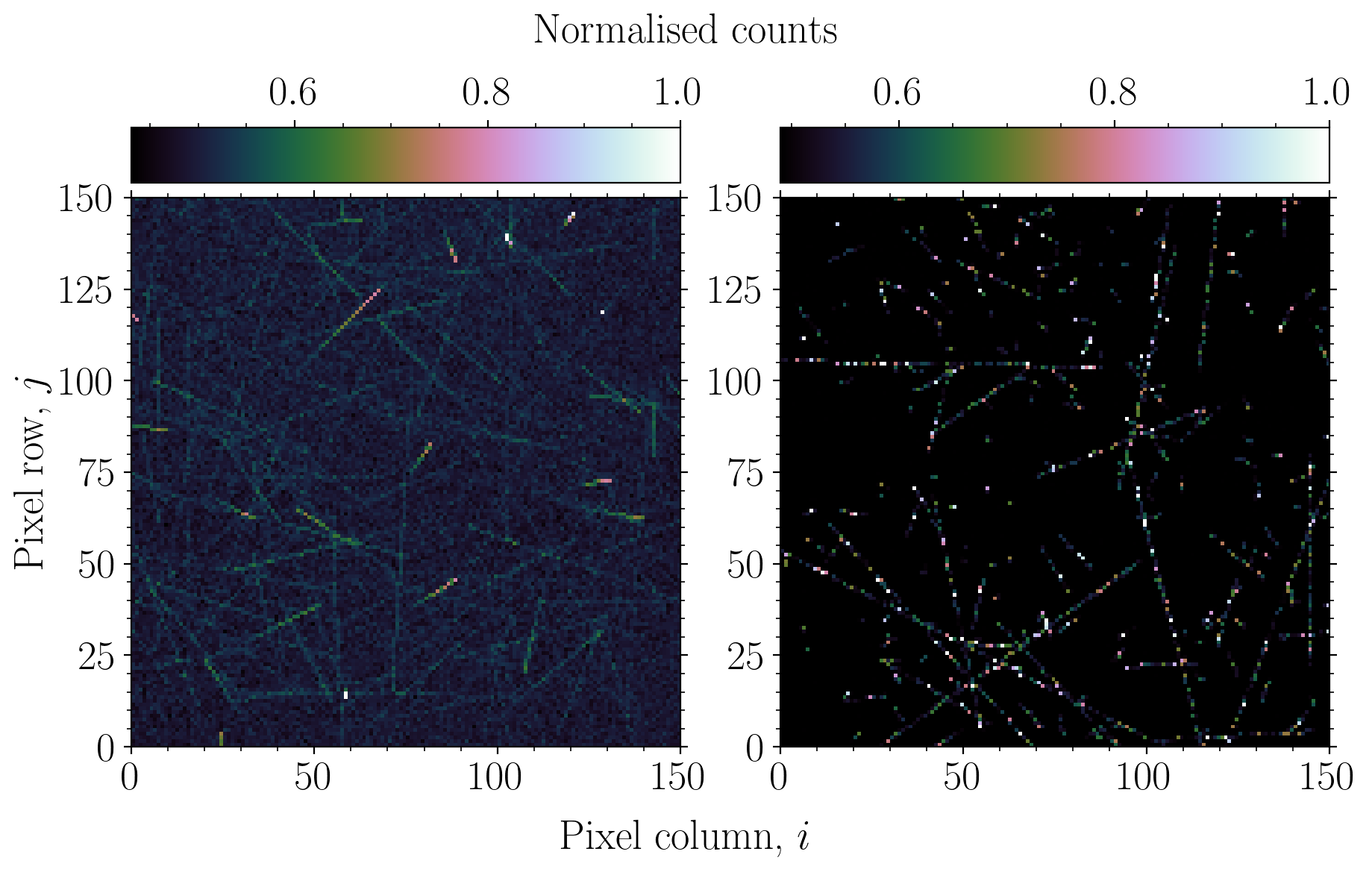}
\includegraphics[width=\columnwidth]{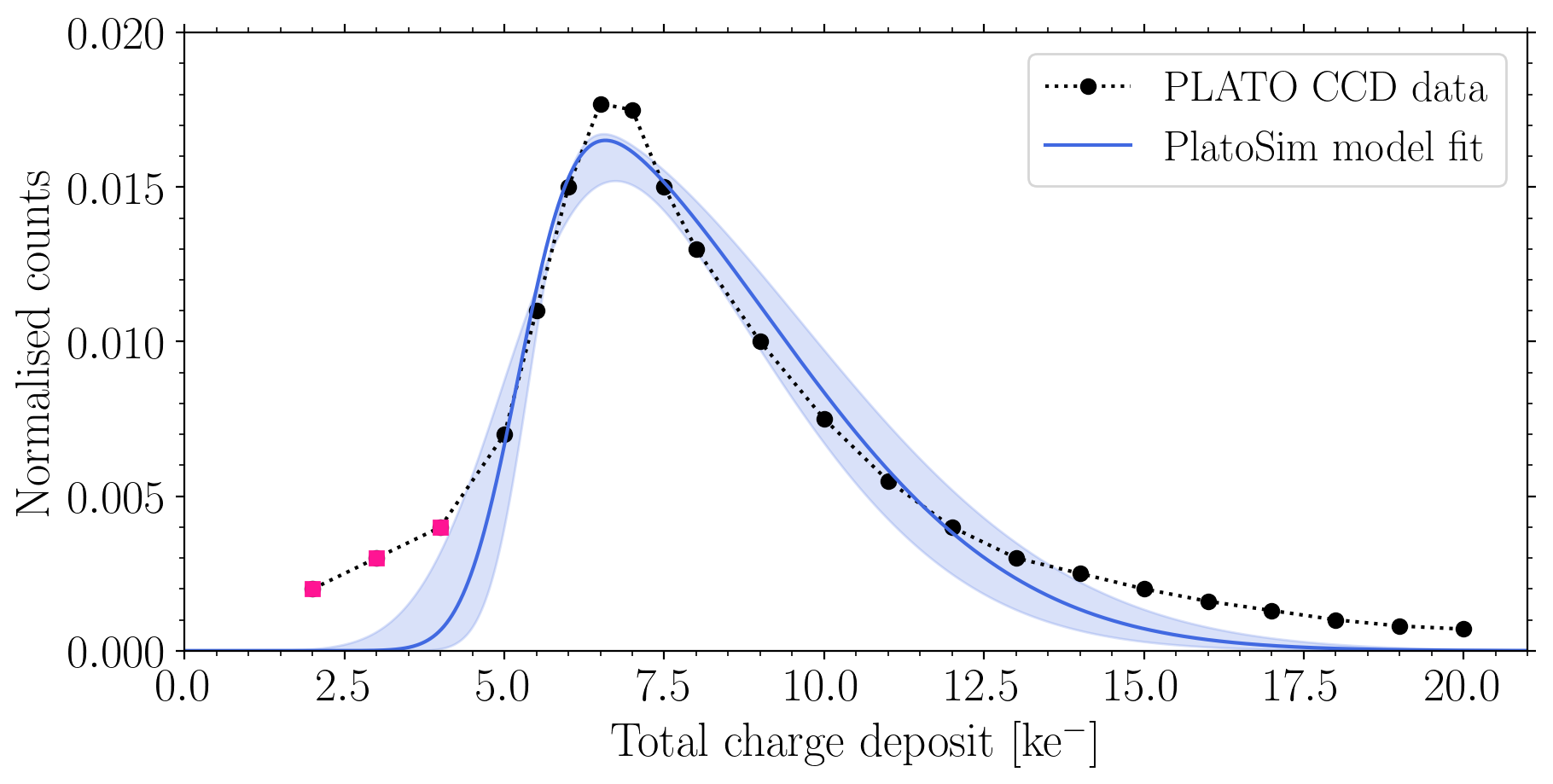}
\caption[]
{A CR model comparison. \textbf{Top panels:} A visual comparison between \platosim{} (left) and \texttt{CosmiX} (right), shown for a small $150^2$ pixel subfield. The images are generated using a cycle time of \SI{25}{\second}, a CR hit rate of \SI{100}{\events\per\second\per\centi\meter\squared} (corresponding to solar maximum), and a CR trail length for \platosim{} of up to \SI{300}{\pixel}. As input for \texttt{CosmiX}, we used a unidirectional \SI{55}{\mega\electronvolt} proton beam and a \SI{40}{\micro\meter} Si pixel volume. \textbf{Bottom panel:} Number distribution of total charge deposit per event. The black dots show the measurements from the proton irradiation test campaign on a PLATO flight model CCD at \SI{203}{\kelvin} \citep{prod2018investigating}. The blue solid line is a best fit model of \platosim{}'s skew-normal distribution to the test data and the blue shaded region is the $2\sigma$ confidence interval. The first three data points (pink squares) were excluded in the fit as they originates from secondary $\delta$-electrons from the setup. The best fit parameters $(\psi_{\rm CR}, \omega_{\rm CR}, \alpha_{\rm CR}) = (5232\pm76, 3842\pm197, 6\pm1)$ were used to produce the \platosim{} simulation above.} 
\label{fig:cosmics}
\end{figure}
  
We have validated \platosim{}'s CR model to the \texttt{CosmiX} simulator, since this software is open source and in excellent agreement with more complex CR codes such as \texttt{Geant4} and \texttt{GRAS}. Since \texttt{CosmiX} is a dedicated module in the open source detector framework \texttt{Pyxel}%
\footnote{\url{https://esa.gitlab.io/pyxel/}} %
\citep{arko2022pyxel}, we used this to generate representative PLATO CCD images by loading in a dark frame generated by \platosim{} and then applied \texttt{CosmiX}. A configuration file made it easy to select settings representative of the PLATO CCD needed for \texttt{CosmiX}.

Figure~\ref{fig:cosmics} shows a visual model comparison between \platosim{} (top left) and \texttt{CosmiX} (top right). The bottom panel shows a proton irradiation test performed at cold temperature on a PLATO flight model CCD \citep[black dots from][]{prod2018investigating} and a best model fit of the \platosim{} skew-normal distribution to the test data (blue solid line). The corresponding \texttt{CosmiX} model \citep[see][]{lucsanyi2020simulating} shows a 93\% model agreement to the test data, where the remaining 7\% mainly corresponds to so-called secondary $\delta$-electrons emerging from the setup itself (i.e. when the proton beam outside the cryostat collides with the aluminium flange). These electrons leave their imprint as an increased fraction of events below \SI{4}{\kilo\electron}, hence, we choose to exclude the first three data points (pink squares) from the \platosim{} model fit.

It is clear that the simplified CR model of PlatoSim naturally shows discrepancies with \texttt{CosmiX} and the PLATO CCD test data (especially at higher deposits of charge). Most noticeable from the pixel data is the difference in CR morphology. In particular \texttt{CosmiX} shows a more discrete nature of the charge deposits along the tracks. The underlying reason is because \texttt{CosmiX} assumes that, while it tracks groups of electron clusters around vertices created by interactions with the incoming ionising particle, the loss of energy of the primary particle through the Si depletion zone of the detector is negligible. On the other hand, \platosim{} models the total charge deposit more continuous and with a gradient that is effectively determined by the distributions from which the energy, incident angle/location, and trail length are drawn. Despite these discrepancies, overall the two models agree sufficiently well in order for \platosim{} to fulfill its original purpose, namely to train the on-board CR rejection algorithm of the PLATO reduction pipeline.

\subsection{The dark signal}\label{sec:dark_signal}

General for all CCDs, the PLATO detectors show a \textit{dark signal}, that is thermal electrons that are generated even in the absence of incident light. This dark signal contributes to the total noise budget with a temporal and spatial component which we model. The dark signal accumulated during an exposure of duration $\Delta t_{\rm exp}$ and a readout of duration $\Delta t_{\rm ro}$, as occurring in Eqs.~(\ref{I_ij_exp}) and (\ref{I_ij_exp_nowave}), is modelled in \platosim{} using a Poisson distribution
\begin{equation}\label{eq:darksignal_poisson}
    F_{\rm dark}(i,j) \sim \mathcal{P}\paren{\mu = n_{{\rm DS},ij} \cdot (\Delta t_{\rm exp} + \Delta t_{\rm ro})} \,,
\end{equation}
where $n_{\rm DS}$ is the dark signal. In practice the latter is not fixed for a particular CCD, but shows a fixed-pattern spatial variation over the CCD which is usually characterised by the dark signal non-uniformity (DSNU; $\sigma_{\rm DSNU}$). \platosim{} models the DSNU by drawing $n_{\rm DS}$ from a normal distribution
\begin{equation}\label{eq:darksignal_DSNU} 
\tx{n}{DS} \sim \mathcal{N}\paren{\mu=\tx{\bar{n}}{DS}, \ \sigma=\tx{\sigma}{DSNU}} \,.
\end{equation}
The nominal values of $\bar{n}_{\rm DS}$ and $\sigma_{\rm DSNU}$ for a PLATO CCD, as tabulated by the manufacturer e2v, are respectively $\tx{\bar{n}}{DS}=\SI{1.2}{\electron\per\second}$ and $\tx{\sigma}{DSNU}/\bar{n}_{\rm DSNU}=15\%$ root-mean-square (rms) at mission BOL. These values are slightly conservative compared to on-ground calibration estimates \citep{verhoeve2016optical} of $\tx{\bar{n}}{DS}\approx\SI{1.03}{\electron\per\second}$ and $\tx{\sigma}{DSNU}/\bar{n}_{\rm DSNU}\approx12\%$ rms. Apart from the exposure time, the dark signal also depends on the detector temperature, which \platosim{} models as a linear function of the CCD temperature using a slope of \SI{5}{\electron\per\second\per\kelvin}, being the mission requirement value.

\subsection{Readout smearing}\label{sec:open_shutter_smearing}

Similar to many space-borne instruments the PLATO cameras do not use a mechanical shutter to block light from reaching the CCD during readout. This implies that during the readout the CCD continues to gather photons. During the row transfers at readout, pixels will therefore be shifting `under' the PSFs of stars in the same column and accumulate photons of these stars during the short time that it takes to shift one row of pixels to the next one. In practice this will lead to a uniform bright vertical trail that is imprinted in each column, but is most visible for those columns that contain bright stars. This effect, called \textit{readout smearing}, is described with the second term of Eq.~(\ref{I_ij}), and is illustrated in Fig. \ref{fig:ccdFocalPlane}b for pixel column $\sim290$, mainly caused by the saturated star in the top of the column. 

If, with a slight abuse of notation, we denote with $I^{\rm (exp)}_{ij}$ the number of electrons that were accumulated in pixel $(i,j)$ during an exposure of duration $\Delta t_{\rm exp}$, the total number of electrons per second collected in the entire column $j$ is
\begin{equation}\label{eq:open_shutter_smearing}
\bar{I}_j = \frac{1}{\Delta t_{\rm exp}} \sum_{i=1}^{n_{\rm row}} I^{\rm (exp)}_{ij} \,.
\end{equation}
Here $n_{\rm row}$ is the number of rows that are illuminated. We note again that this is different for the F-CAMs than for the N-CAMs as the former feature readout using frame-transfer and the bottom half of their CCDs are covered to prevent illumination. During readout, each pixel $(i,j)$ in column $j$ will collect $\delta t_{\rm trans} \cdot \bar{I}_j$ electrons, where $\delta t_{\rm trans}$ is the time it takes to transfer one row of photoelectrons to the next one. This happens partly during the readout phase of the previous exposure when the row was transferred from the top of the CCD to location $i$, and during the readout phase of the current exposure when the row is transferred from location $i$ to the readout register at the bottom of the CCD.

Since also photons collected during readout are subjected to photon noise, the final value for the accumulated flux during readout is taken from a Poisson distribution
\begin{equation}\label{eq:smearing_flux}
    I_{ij}^{\rm (ros)} = \mathcal{P}\paren{\mu=\delta\tx{t}{trans} \ \bar{I}_{j}} \,.
\end{equation}
The readout smearing is measured using a parallel overscan region whose size spans 30 virtual pixel rows for the N-CAMs (see pink box of Fig.~\ref{fig:ccdFocalPlane}a).

\section{Focal plane positions of the stars}\label{sec:star_positions}

This section describes how \platosim\ computes the positions $(x_0, y_0)$ of a star in the focal plane mentioned in Eqs. (\ref{eq:Fstar_monochromatic}) and (\ref{eq:Fstar_polychromatic}) and the corresponding pixel positions $(i_0, j_0)$ on the CCD. A realistic model of the time dependence of these positions is crucial to understand the non-white noise it induces.

\subsection{Differential kinematic aberration}

Given the true equatorial sky coordinates $(\alpha, \delta)$ of a star, the \textit{apparent} sky coordinates are slightly 
different because of stellar kinematic aberration: the overall motion of the spacecraft relative to 
the star induces a shift of the apparent stellar position due to the aberration of the light rays as they enter the
camera. The amplitude and direction of the shift of the apparent stellar position depends on the angle between the position
vector $\vec{s}$ of the star and the velocity vector $\vec{v}$ of the spacecraft with respect to the star. 
More specifically, if $\theta$ is the unaberrated angle between 
these two vectors (i.e. $\cos\theta = \vec{v}\cdot\vec{s}$), then the aberrated angle $\theta_{\rm ab}$ between them is given by
\begin{equation}
    \theta_{\rm ab} = \tan^{-1}\left( \frac{\sqrt{1-\beta^2} \sin\theta}{\beta + \cos\theta}\right) \,,
\end{equation}
where $\beta = v/c$ with $v$ the velocity of the spacecraft with respect to the star and $c$ the speed of light.
The corresponding aberrated position vector of the star $\vec{s}_{\rm ab}$ is given by 
\begin{equation}
    \vec{s}_{\rm ab} = \vec{v} \cos\theta_{\rm ab} + 
    \frac{\vec{s} - \vec{v} \cos\theta}{|\vec{s} - \vec{v} \cos\theta|} \ \sin\theta_{\rm ab} \,.
    \label{eq:aberrated_pos_vector}
\end{equation}
The corresponding pixel displacement of a star therefore depends on its position in the focal plane and varies in time as PLATO's velocity vector changes over its orbit. However, the AOCS of PLATO uses bright fine guidance stars in the same FOV observed by the F-CAMs to continuously stabilise its pointing. These fine guidance stars experience roughly the same aberration, so that the aberration is therefore largely and continuously corrected by the AOCS. The correction is only approximate since a given pixel's LOS depends on its exact location in the focal plane and the small pixel-to-pixel variation of the aberration that still occurs. This \textit{differential aberration} is not radially symmetric around the optical axis of a camera but is offsets from the pointing axis of the platform determined by the F-CAMs. Hence the maximum amplitude of the differential aberration of an N-CAM is at the FOV edge furthest from the platform pointing, resulting in a shift of up to $\sim\SI{0.8}{\pixel}$ in 3 months. 

The differential aberration is taken into account using a realistic orbit read by \platosim{}. Generally PLATO orbits around the L2 following a libration point orbit (a so-called Lissajous orbit), but the dominant velocity component for the kinematic aberration is the one following the orbit of L2 around the Sun. In the remainder of the paper, when we use sky coordinates, we always refer to the apparent sky coordinates subjected to kinematic aberration.

\subsection{Projection of the star on the focal plane}

The exact position of the star on a CCD depends on where in the sky the spacecraft platform is pointing, how the camera is mounted on the platform, how the focal plane reference frame is defined in the camera, and finally how the CCDs are orientated inside the focal plane. In practice, \platosim{} models the CCD coordinates $(x_\ccd, y_\ccd)$ through a set of reference frame transformations 
\begin{equation}\label{rf_eq2ccd}
\begin{pmatrix}
x_\ccd \\ 
y_\ccd \\ 
1
\end{pmatrix}
= \mathbf{R}_\focal^\ccd \cdot \mathbf{R}_\cam^\focal \cdot \mathbf{R}_\plm^\cam \cdot \mathbf{R}_\equa^\plm \,.
\begin{pmatrix}
    x_{\rm eq} \\
    y_{\rm eq} \\
    z_{\rm eq} 
\end{pmatrix} \,.
\end{equation}
Here, $(x_{\rm eq}, y_{\rm eq}, z_{\rm eq}) = (\cos\delta \cos\alpha, \cos\delta\sin\alpha, \sin\delta)$ are the components of the unit vector
pointing towards a star with apparent sky coordinates $(\alpha, \delta)$ in the equatorial reference frame. The rotation matrices in
Eq.~(\ref{rf_eq2ccd}) are used to transform from the equatorial reference frame (EQ), to the payload module (PLM) reference frame, to the camera
(CAM) boresight reference frame, to the (undistorted) focal plane (FP) reference frame, and finally to the CCD reference frame. The rotation matrix
$\mathbf{R}_\equa^\plm$ depends on three angles $(\alpha_\plm, \delta_\plm, \kappa_\plm)$ defining the orientation of the spacecraft, and
$\mathbf{R}_\plm^\cam$ depends on two angles $(\eta_\cam, \rho_\cam)$ defining the orientation of a camera on the payload module. All five of these
angles are time dependent because of spacecraft pointing jitter and slow thermo-elastic drifts as explained in Sections \ref{sec:jitter} and
\ref{sec:TED}. The matrix $\mathbf{R}_\cam^\focal$ involves a pinhole projection as explained in Appendix \ref{app:reference_frames}. The resulting 
focal plane coordinates are moreover subjected to optical distortion which is explained in Section \ref{sec:distortion}.

\subsection{Payload module pointing jitter}\label{sec:jitter}

The AOCS controls the stability of the spacecraft pointing and is affected mainly by the reaction wheels and
the fine guidance star system. As the AOCS is not perfect, the payload module jitters around a mean pointing, causing the stars to slightly move over the CCD (with a typical distance smaller than a pixel). The high-frequency components of the jitter cause PSF blurring, while the low-frequency components can displace the barycentre of the PSF from one pixel to the next. Due to the non-uniformity of the pixel-to-pixel response, the pointing jitter leads to increased photometric noise, and is thus a key driver for the photometric performance. Several correction algorithms have therefore been published so far in the literature \citep[e.g.][]{drummond2006jitter}.

\begin{figure}[t!]
\centering
\includegraphics[width=\columnwidth]{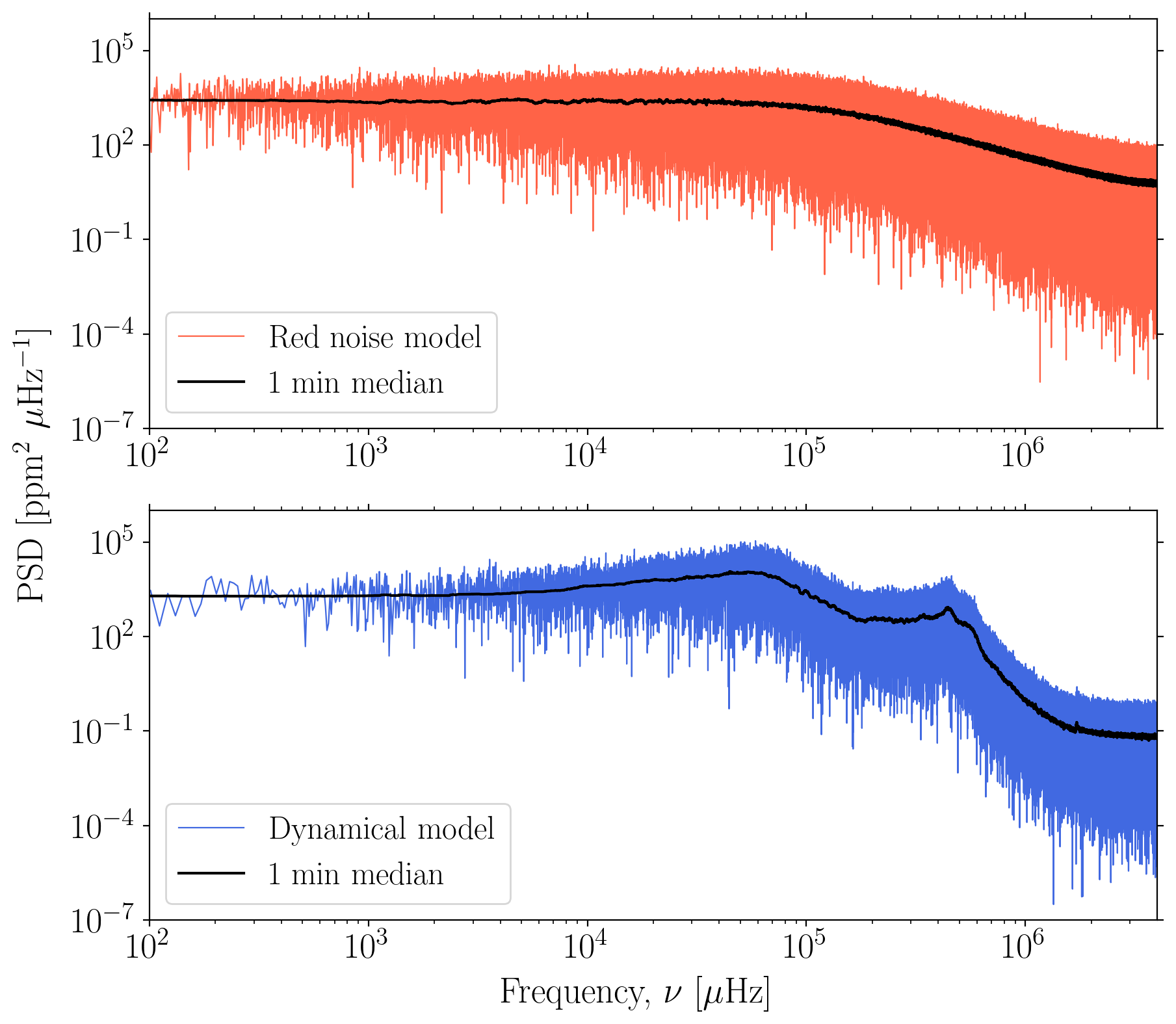}
\caption[]
{Power spectral distribution (PSD) for a high frequency jitter time series of the yaw angle produced with \platosim{}'s red noise model (top) and a dynamical OHB/TAS model (bottom). The two simulated models are sampled at \SI{8}{\hertz} and have a time duration of \SI{27}{\hour}. The solid black line corresponds to a \SI{1}{\min} moving median filter.} 
\label{fig:jitter}
\end{figure}

To model the pointing jitter we define the platform yaw $\phi$, the pitch $\theta$, and the roll $\psi$ as rotation angles around respectively the $X_\plm$, $Y_\plm$, and $Z_\plm$ axes, such that the angles increase with a clockwise rotation, when looking along the positive axes. At any given time a perturbation to the pointing direction and the roll angle of the spacecraft is calculated by first performing a roll rotation around the $Z_\plm$ axis, then a pitch rotation around the rotated $Y_\plm$ axis, and finally a yaw rotation around the twice-rotated $X_\plm$ axis. The combined rotation matrix (in the reference frame of the platform) is thus given by
\begin{equation}\label{rotMatrixJitter}
\mathbf{R}(\theta, \varphi, \psi) = \mathbf{R}(\theta) \, \mathbf{R}(\phi) \, \mathbf{R}(\psi) \,.
\end{equation}
In \platosim{} the update of the platform pointing using the rotation matrix above is done for every time step $\delta t$ (cf.~Eq.~\eqref{I_ij}).

The yaw, pitch, and roll time series used in the simulations are either taken from a detailed perturbation dynamical model of the spacecraft 
(not included in \platosim{}), or are simulated using red noise. In the latter case the jitter angles are modelled as in \cite{de2006modelling}, 
using a first-order auto-regressive model
\begin{equation}
    \label{redNois}
    \theta_{n+1} = e^{-\delta t / \tau}\ \theta_{n} + \varepsilon_{n+1} \,.
\end{equation}
Here $\theta_{n}$ is the yaw angle a time $t_n$, $\tau$ is the jitter time scale, $\delta t \ll \tau$ is the discretised time step, and $\varepsilon$ is a Gaussian distributed noise fluctuation with zero mean and a variance equal to
\begin{equation}
    {\rm Var}[\varepsilon_n] = \sigma^2 \frac{\delta t}{\tau} \,,
\end{equation}
where $\sigma$ is the amplitude scale of the jitter. 

Figure~\ref{fig:jitter} shows for the yaw angle a power spectral density (PSD) function of the red noise model from \platosim{} (top panel using a rms amplitude scale of \SI{0.04}{\arcsec}) and dynamical model (bottom) produced by PLATO's prime contractor Otto Hydraulic Bremen (OHB)/Thales Alenia Space (TAS). These simulations have a duration of 27 hours and a (fast) jitter time scale of \SI{8}{\hertz}. Compared to the dynamical model description of OHB/TAS, which assumes the instrument feedback from the F-CAMs to correct the pointing, the PSD proportionality $1/f^2$ for the red noise model is lacking the correlated systematics between yaw, pitch, and roll (as seen by the additional kink of the OHB/TAS model). Nevertheless it has been shown that at a cadence of \SI{25}{\second} (i.e. \SI{e4}{\micro\hertz}) the residual jitter noise of the dynamical model behaves Gaussian, which is, such as red noise, a stochastic process.

\subsection{Thermo-elastic drift}\label{sec:TED}

To maintain the solar panels in the direction of the Sun, PLATO will perform a \SI{90}{\degree} rotation every three months after completing
a quarter of the orbit around the Sun. During a run of three months, the spacecraft is not rotated to maintain a fixed field of view,
which implies that the part of the spacecraft that is directly pointing towards the Sun gradually changes. In turn this means that the
thermal profile of the spacecraft also slowly changes in time. In particular the thermal flexure of the optical bench will 
introduce slight changes to the pointing direction of each camera. This effect, also known as thermo-elastic drift (TED), will 
lead to a slow drift of the stars over the focal plane, up to 80\% of a pixel as a worst case estimate for PLATO. The camera drift in 
\platosim{} is modelled using the same formalism as done for the AOCS jitter (see Sect.~\ref{sec:jitter}) taking the Euler 
angles (yaw, pitch, roll) as input either from a thermo-dynamical model or from a red noise model.

\subsection{Optical field distortion}\label{sec:distortion}

A last physical effect that impacts the positions of the star in the focal plane is the optical field distortion. In every real-life application a camera is subjected to image distortions due to slight manufacturing errors of the optical lens and relative optical alignment errors. Building from the heritage of the Brown-Conrady model \citep{brown1971close}, a unified distortion model was formulated by \cite{wang2008new} (referred to as the \textit{Wang model}), which classifies the lens distortion into a radial, a tangential, and a thin prism distortion. Radial distortion is caused by an imperfect radial curvature of a lens, whereas the tangential (or decentring) distortion is caused by misalignments between different lens elements, and the thin prism distortion arises from a slight tilt of a lens with respect to the detector. 
        
\platosim{} uses the self-contained distortion model in the case of the Zemax PSF, and implements the Wang distortion model for the analytic PSF. The Wang model is applied to all cameras and transforms the undistorted to the distorted (D) focal plane coordinates using
\begin{equation}\label{distortionLens}
\begin{pmatrix}
x_\focal \\
y_\focal
\end{pmatrix}_{D}
= 
\begin{pmatrix}
x_\focal \\
y_\focal
\end{pmatrix}
+
\begin{pmatrix}
D_x \\
D_y
\end{pmatrix} \,,
\end{equation}
where 
\begin{align}
D_x &= x_\focal(k_1 r^2 + k_2 r^4 + k_3 r^6) + x_\focal(p_1 x_\focal + p_2 y_\focal) + q_1 r^2 \,, \\
D_y &= y_\focal(k_1 r^2 + k_2 r^4 + k_3 r^6) + y_\focal(p_1 x_\focal + p_2 y_\focal) + q_2 r^2 \nonumber \,.
\end{align}
Here $r=\sqrt{x_\focal^2 + y_\focal^2} / l$ is the undistorted radial distance from the optical axis normalised by the focal length $l$. The set of coefficients $(k_1, k_2, k_3, q_1, q_2, p_1, p_2)$ belongs to the model description of respectively the radial $(k)$, the tangential $(p)$, and thin prism $(q)$ component. The coefficients used by \platosim{} to model the direct and inverse distortion have been derived using a Zemax model as part of the mission preparation. In addition, \platosim{} allows the coefficients to be provided as a time series, to allow modelling the effect of a changing thermal environment on the optical distortion.

\begin{figure}[t!]
\centering
\includegraphics[width=\columnwidth]{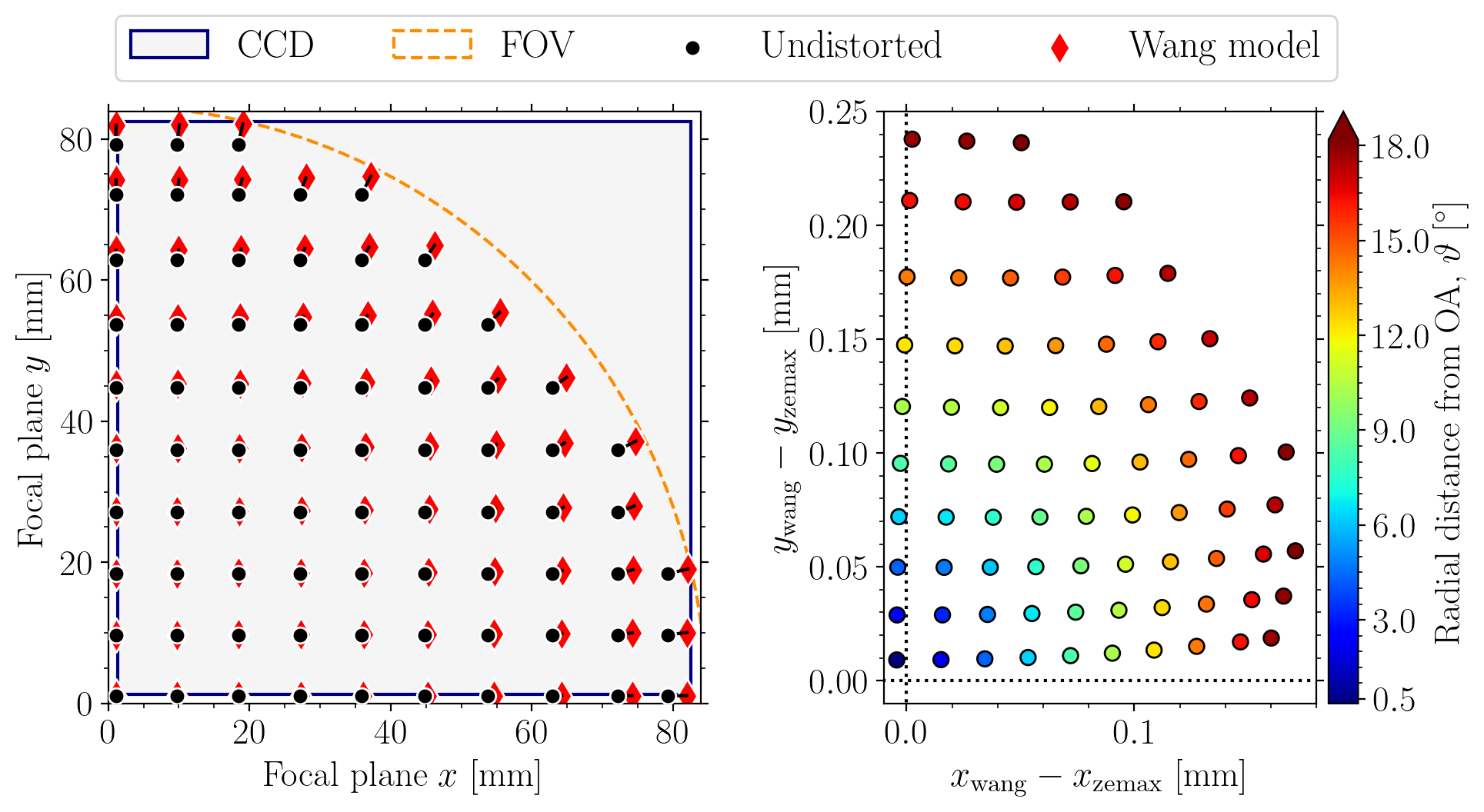}
\caption{Illustration of the field distortion models. \textbf{Left:} Distortion over one quadrant of the FPA with a grid step size of \SI{2}{\degree}. Shown are the undistorted paraxial chief ray coordinates (black dots) and the real (distorted) chief ray coordinates calculated by the \cite{wang2008new} distortion model (red diamonds), together with the CCD area (dark grey area enclosed by dark blue lines) and the effective size of the camera FOV (dashed orange line). \textbf{Right:} Residuals between the Wang and the Zemax distortion model evaluated in the FPA grid points shown in the left panel. The colour bar serves as a reference of the radial distance to the optical axis, $\vartheta$. We note that with a PLATO plate scale of \SI{18}{\micro\meter} a residual of \SI{0.1}{\milli\meter} corresponds to $\sim\SI{5.6}{\pixel}$.} 
\label{fig:distortion}
\end{figure}

An illustration of the expected field distortion model representative for the PLATO camera is shown in the left-hand panel of Fig.~\ref{fig:distortion} 
for one quadrant of the focal plane. Here the black dots represent the undistorted (i.e. unobservable distortion-free) chief ray positions and the red diamonds are the distorted chief ray positions from the Wang model. The right-hand panel of the same figure shows the residual plot between the Zemax and Wang model computed at the same FOV grid shown in the left-hand plot. Overall the residual plot shows a good agreement between the two models, with the Wang model generally resulting in a slightly stronger distortion compared to the Zemax prediction, and where the largest discrepancies are farthest from the optical axis (dashed orange line in the left-hand plot).

\section{Optical throughput and detector efficiency}\label{sec:optical_throughput_detector_efficiency}

This section describes the throughput and efficiency quantities $\bar{T}(t,x,y)$, $\bar{E}(t,x,y)$, and $\bar{Q}(t,x,y)$ that
occur in Eq.~\eqref{I_ij_exp_nowave}.

\subsection{Optical throughput}\label{sec:optical_throughput}

\begin{figure}[t!]
\centering
\includegraphics[width=\columnwidth]{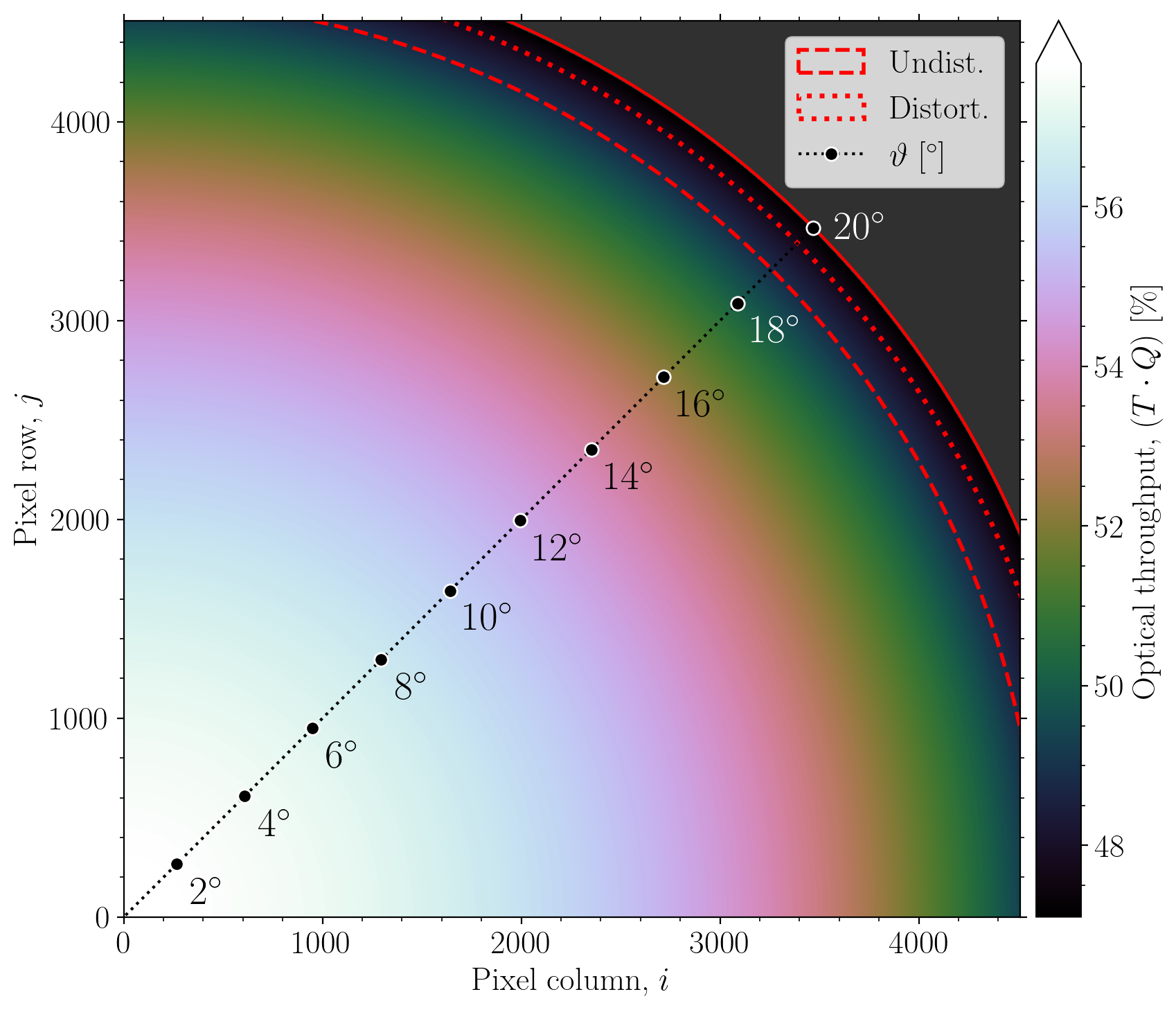}
\caption[]
{Illustration of the total throughput map for one full frame CCD. The dotted diagonal line shows the distance from the optical axis in degrees ($\vartheta$) and the red dashed lines show the angular position of the stray light mask. We note that the FOV in the focal plane physically extents beyond $\tx{\vartheta}{max}$ (to $\sim\SI{19.6}{\degree}$ indicated by the red dotted line) due to the effect of optical distortion cf. Sect.~\ref{sec:distortion} and is followed by an exponential intensity decay of vignetting (modelled out to $\SI{20}{\degree}$ shown by the red solid line).} 
\label{fig:throughput}
\end{figure}

The total \textit{optical throughput} (also known as the spectral response) of an optical instrument represents its efficiency to convert incident photons into counts of electrons at detector level. Hence it is a product of the dimensionless optical transmission $T_{ij}(t, \lambda)$ and the quantum efficiency $Q_{ij}(t, \lambda)$. The time-dependence comes from a slow degradation between the beginning and the end of the mission (BOL $\rightarrow$ EOL), which is modelled in \platosim{} using a linear relation. We discuss this model choice in the following. Figure~\ref{fig:throughput} illustrates the total optical throughput integrated over the PLATO passband, at the beginning of life over one full-frame CCD.

The monochromatic transmission $T_{ij}(t, \lambda)$ combines the effects of the transmission efficiency of a photometric passband filter $\tx{T}{fil}$, the vignetting $\tx{T}{vin}$, the particulate and molecular contamination $\tx{T}{con}$, and the polarisation transmission efficiency $\tx{T}{pol}$,
\begin{equation}\label{T_ij}
T_{ij}(t,\lambda) = \tx{T}{fil}(t,\lambda) \ T_{\text{vin},ij}(\vartheta, \lambda) \ \tx{T}{con}(t,\lambda) \ T_{\text{pol},ij}(\vartheta,\lambda) \,.
\end{equation}
The polychromatic version, as appearing in Eq.~(\ref{I_ij_exp_nowave}) and used by \platosim{}, is derived by taking the average over the PLATO passband
\begin{equation}\label{T_ij_nowave}
    \bar{T}_{ij}(t) = \bar{T}_{\rm fil}(t) \ \bar{T}_{\text{vin},ij}(\vartheta) \ \bar{T}_{\rm con}(t) \ \bar{T}_{\text{pol},ij}(\vartheta) \,.
\end{equation}

The phenomenon of vignetting, that is the brightness attenuation towards the edge of the FOV, can be divided in three components: natural, optical, and mechanical vignetting. The attenuation by natural vignetting is caused by the fact that off-axis light rays not only have a longer travel distance but also see a projected (i.e. reduced) area of the entrance pupil, leading to a decreasing light intensity at angles far away from the optical axis. Optical vignetting is induced by the optical design of a camera that features multiple optical elements. A lens earlier in the light path causes a reduction of the effective opening of the next lens because the output angles of the former are limited. Also this causes a decrease in intensity towards the edge of the FOV. Mechanical vignetting is due to the blocking of light rays by the straylight mask, and causes a semi-hard circular border of the FOV at a maximum angle $\tx{\vartheta}{max}$ from the optical axis. The vignetting for a PLATO camera was analysed using a Zemax optical model, as well as measured during the camera assembly, integration, and verification, which led to the following best fitting parametric model
\begin{equation}
    \label{eq:vignetting}
    \bar{T}_{\rm vin}(\vartheta) = 
    \begin{cases}
        1 - k_1 \vartheta^2 - k_2 \vartheta^4 - k_3 \vartheta^6  & \text{for } \vartheta \le \vartheta_{\rm max} = 18.9^{\circ} \,, \\[2mm]
        c + e^{-(\vartheta-\vartheta_{\rm max})/\sigma}  & \text{for } \vartheta > \vartheta_{\rm max} \,, \\
    \end{cases}
\end{equation}
where $\{k_1, \ k_2, \ k_3\} = \{4.18 \cdot 10^{-2}, \ -5.65 \cdot 10^{-5}, \ 2.37 \cdot 10^{-7}\}$. Here $\sigma = 0.6^{\circ}$ and $c$ is a constant so that the function is continuous in $\vartheta = \vartheta_{\rm max}$.

The particulate contamination is the (unintended) presence of particles on (mostly optical) surfaces, whereas the molecular contamination is the layer of molecules on top of a surface caused by out-gassing of materials in the first phase of the mission \citep[for an in depth discussion see e.g.][]{zhao2009effect}. A major part of the particulate contamination takes place during the fairing ejection (i.e. well before PLATO will start its journey to the L2) which, by and large, sets the level of contamination in the camera entrance window. The story of the molecular contamination is on the other hand more complex. The majority of out-gassing typically takes place during the cooldown of the spacecraft (being mostly during the first three days after launch), but will in reality never stop completely. Furthermore, the out-gassing can also take place from various materials at different rates. However, we assume that out-gassing during launch is neglectable for PLATO, due to the spacecraft's limited duration within Earth’s atmosphere, together with a slow rate of change for out-gassing at the time PLATO starts operating in the L2. Hence, as mentioned before, a linearly decreasing model of the transmission efficiency is a good approximation if radiation damage is the dominating factor for the degradation. \platosim{} uses a throughput value of 0.972 due to particulate contamination and a value of 0.9573 due to molecular contamination, which are the BOL requirement values.

The polarisation transmission efficiency is modelled using the following (fairly arbitrary monotonically decreasing) parametric model
\begin{equation}\label{T_pol}
\bar{T}_{\text{pol},ij}(\vartheta) = \tx{\bar{T}}{pol,max} \cos\paren{\frac{\vartheta}{\tx{\vartheta}{ref}}\cos^{-1}\parenf{\tx{\bar{T}}{pol}(\tx{\vartheta}{ref})}} \,,
\end{equation}
where $\tx{\bar{T}}{pol,max}$ is the $\tx{\bar{T}}{pol}$ maximal value and $\tx{\bar{T}}{pol}(\tx{\vartheta}{ref})$ is the value at a certain reference angular 
distance $\tx{\vartheta}{ref}$ away from the optical axis.

Lastly, the quantum efficiency (QE) first presented in Eq.~\eqref{I_ij} is generally defined as
\begin{equation}\label{Q_ij}
Q_{ij}(t,\lambda) = \frac{Q_{\text{ext},ij}(t, \lambda)}{1-T_{\text{ref},ij}(t, \lambda)} \,,
\end{equation}
where $\tx{Q}{ext}$ is the external quantum efficiency, that is the ratio of the number of electrons over the number of incident photons, and $\tx{T}{ref}$ is the 
reflectivity, that is the fraction of photons reflected at the surface despite the anti-reflection coating. This expression implies that the quantum efficiency also depends to second order on the angle of the incident light. Since the PLATO payload's optical design leads to CCD illumination over a wide range of incident angles (up to \SI{40}{\degree}; \citealt{rauer2014plato}, Rauer et al. in prep.), the variation of the QE with incidence angle is taken into account in \platosim{}. As before we avoid dealing with the wavelength dependence of the QE by averaging it over the passband
$\bar{Q}_{ij} \approx \langle Q_{ij}(\lambda) \rangle_{\lambda}$, and we use a similar parametric model as above
\begin{equation}\label{QE_theta}
    \bar{Q}(\vartheta) = \bar{Q}_{\rm max} \cos\paren{\frac{\vartheta}{\tx{\vartheta}{ref}}\cos^{-1}\parenf{\bar{Q}(\vartheta_{\rm ref})}} \,.
\end{equation}

\subsection{Detector efficiency}\label{sec:prnu}    

The detector efficiency $\bar{E}_{ij}(t)$ in Eq.~(\ref{I_ij_exp_nowave}) encompasses both spatial pixel sensitivity variation as well as defective pixels. The former is caused by the fact that the electric field structure within a pixel is affected by small variations in pixel size, the structure of the gate electrodes, the thickness of the $\text{SiO}_2$ insulation layer, and the doping uniformity in the epitaxial (crystalline) Si layers \citep[see e.g.][]{jorden1994nonuniformity}. The combination of pixel sensitivity variations and spacecraft pointing jitter introduces tiny flux variations that increase the noise level. \platosim{} therefore includes the pixel-response non-uniformity (PRNU; often referred to as the \textit{flat-field} for space-borne instruments) in its simulations.

\begin{figure}[t!]
\centering
\includegraphics[width=\columnwidth]{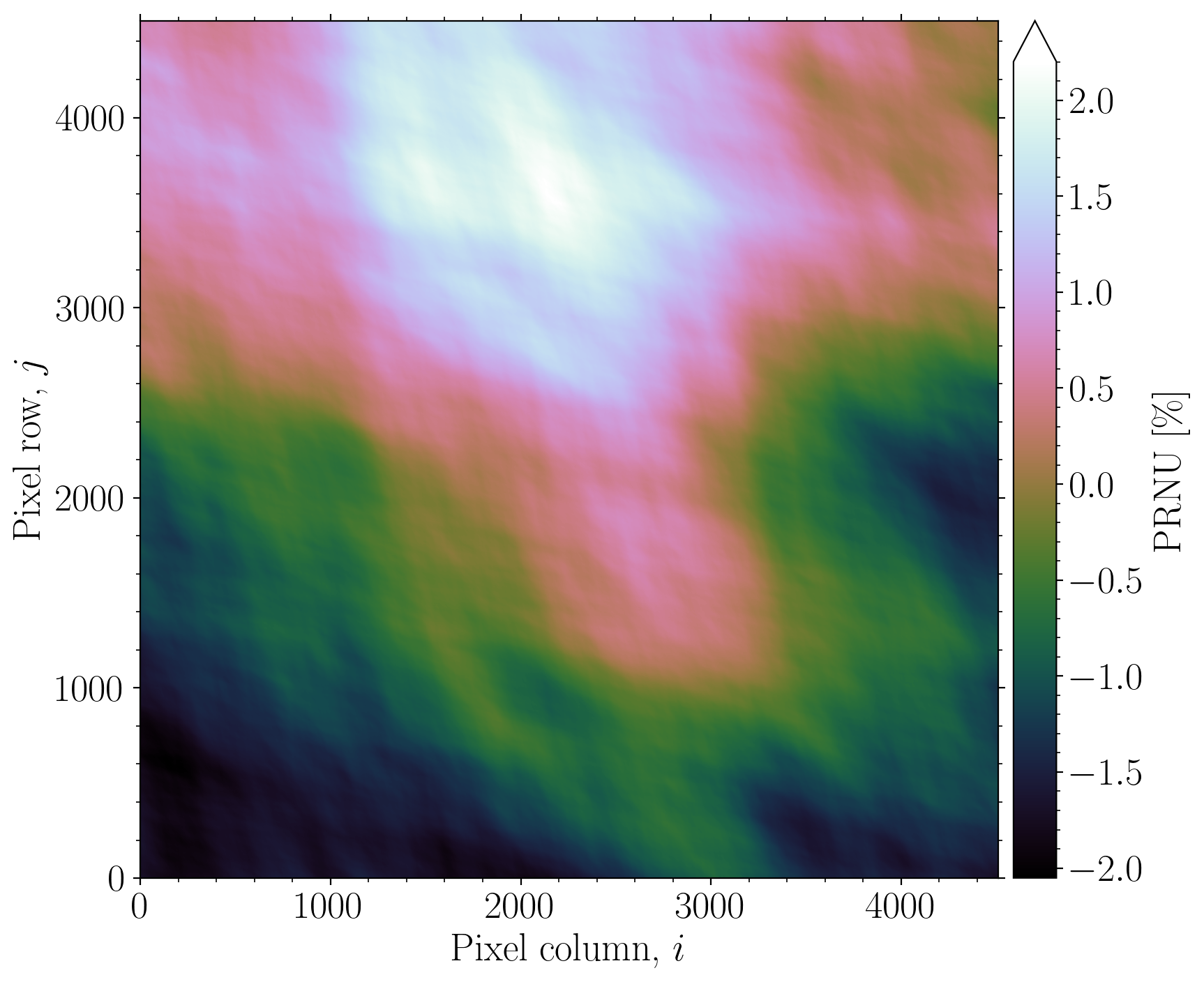}
\caption[]
{Illustration of the automatically generated flat-field (PRNU) for a full-frame CCD image. This image represent the flat-field used to construct the subfield in Fig. \ref{fig:cti} and has a peak-to-peak pixel sensitivity variation $\sim4\%$ and a local rms noise level of $\sim1\%$.} 
\label{fig:prnu}
\end{figure}

The most reliable flat-field is an empirical one obtained from measuring the PRNU at different wavelengths using a flight model of the camera. These measurements are then combined in a weighted average over the PLATO passband. Early in the design of the space mission such measurements are not available, in which case \platosim{} resorts to a simulated flat-field. The important feature here is that the spatial variation in pixel sensitivities usually do not follow a white noise pattern, but shows a spatial correlation. To simulate this, \platosim{} first models the 2D Fourier transform $\bar{E}_{\rm FT}(m, n)$ of the PRNU map as
\begin{equation}
    \bar{E}_{\rm FT}(m,n) = \frac{\varepsilon_{m,n}}{1 + m^{\beta} + n^{\beta}} \,,
\end{equation}
where $(m,n)$ are the spatial wavelengths, $\varepsilon_{m,n} \sim \mathcal{N}(0,1)$ is Gaussian distributed noise fluctuation, and the exponent $\beta$ determines the strength of the correlation at larger spatial distances on the CCD. Subsequently, the inverse Fourier transform is taken and scaled to have a given mean and rms, to obtain the actual PRNU map $\bar{E}(i,j)$. Figure~\ref{fig:prnu} shows an example of a relative flat-field for $\beta=2$, which shows a strong spatial correlation over the CCD. 

Defective pixels also impact the detector efficiency and fall under one of the following three categories: \textit{dead}, \textit{hot}, or \textit{telegraphic}. Compared to a normal pixel a dead pixel has an anomalously low sensitivity, whereas a hot pixel has an anomalously high dark current. A telegraphic (or RTS; random telegraph signal) pixel, on the other hand, is a pixel which periodically switches between an active state with high dark current and an inactive state with normal dark current, and vice versa. While the identification of defective pixels has been determined as part of the on-ground calibration on a flight model CCD \citep{verhoeve2016optical}, the relative fraction of defective pixels is less than $0.1\%$ and are thus not expected to pose a problem for PLATO. Therefore, defective pixels are currently not implemented in \platosim{}, however, we do acknowledge the potential importance of implementing these (in particular RTS pixels) later in the mission.

\section{Electron redistribution models}\label{sec:electron_redistribution}
    
\subsection{Brighter-fatter effect}\label{sec:implementation_ccd_bfe}

The brighter-fatter effect (BFE) is an electron redistributing effect caused by electrostatic interaction between neighbouring pixels. Pixels that
already collected a large number of electrons during an exposure shrink in effective collecting area, so that they collect and retain less electrons
because some of them are now attracted to a neighbouring pixel. The BFE is different from charge diffusion discussed in
Sect.~\ref{sec:stellar_signal} in the sense that the former is caused by a changing electrostatic field configuration during an exposure, while the
latter is caused by electrons drifting laterally during a random walk before ending up in a pixel. The BFE phenomenon is fairly well understood, and
the model implemented in \platosim{} uses the approximative framework outlined by \citet{Antilogus2014}, \citet{Guyonnet2015} and \citet{Astier2019}. 

The electrostatic field lines caused by the pixel gate electrodes define the path of the incoming electrons and therefore determines to which pixel
the electron is attracted. In the case of a perfect regular grid of electrodes, the pixel boundaries fall at the geometrical midpoint between two
electrodes as is nicely illustrated in Fig.~4 of \citet{Antilogus2014}. When a pixel accumulates electrons the electrostatic field lines change as
the accumulated electrons have a repulsive effect on new incoming electrons. Electrons close to the geometrical midpoint are now attracted more by
the neighbouring electrode, effectively shifting the pixel boundary towards the first electrode so that its corresponding collecting area becomes
smaller. The boundary between a `central' pixel ($C$) and its neighbouring pixel `north' ($N$) can be affected by the number of electrons in a nearby
(although not necessarily neighbouring) pixel `$P$'. In a first approximation the boundary shift $\delta X^{C \leftrightarrow N}$ scales linearly
with the charge $Q_{P}$ already accumulated in pixel $P$. Since the electric field is additive, the total effect can be computed by summing up the
effects of all nearby pixels $\{P\}$
\begin{equation}
    \delta X^{C \leftrightarrow N} = \frac{1}{2} \sum_{\{P\}} a^{C\leftrightarrow N}_{P}\ Q_{P} \,.
    \label{eq:BFE_boundary}
\end{equation}
where the factor $1/2$ is added to be compatible with the definition of the coefficients $a^{C\leftrightarrow N}_{P}$ as defined in
\citet{Guyonnet2015}. 
The sum of the linear coefficients $a^{C\leftrightarrow N}_{P}$ should be zero as we do not expect any net shift to happen when all charges $Q_P$
are equal. 

\citet{Antilogus2014} argue that the change in charge $\delta Q^{C\leftrightarrow N}$ in pixel $C$ due to the boundary shift $C \leftrightarrow N$
scales in a first approximation linearly with the shift $\delta X^{C \leftrightarrow N}$ as well as with the charge density at the boundary between
the two pixels which scales to first order by the total amount of charge $Q_C+ Q_N$ 
\begin{equation}
    \delta Q^{C\leftrightarrow N} = \frac{1}{4} \sum_{\{P\}} a^{C\leftrightarrow N}_{P}\ Q_{P} \ (Q_C + Q_N) \,,
    \label{eq:BFE_deltaQN}
\end{equation}
where we added an extra scale factor $1/2$ instead of absorbing it in the coefficients $a^{C\leftrightarrow N}_{P}$ to be compatible with \citet{Guyonnet2015}. The coefficients $a^{C\leftrightarrow N}_{P}$ in Eq. (\ref{eq:BFE_deltaQN}) still sum up to zero. The change in charge in pixel $C$ is also affected by the change in boundary with the other neighbouring pixels `south' ($S$), `west' ($W$), and `east' ($E$), 
so that we can write
\begin{eqnarray}
    \delta Q_C & = & \delta Q^{C\leftrightarrow N} + \delta Q^{C\leftrightarrow S} + \delta Q^{C\leftrightarrow E} + \delta Q^{C\leftrightarrow W} 
                     \\[3mm] \nonumber
               & = & \frac{1}{4} \sum_{\{X\}} \sum_{\{P\}} a^{C\leftrightarrow X}_{P}\ Q_{P} \ (Q_C + Q_X)
\label{eq:BFE_deltaQ}
\end{eqnarray}
where $\{X\}$ stands for $\{$north, south, west, east$\}$. The coefficients $a^{C\leftrightarrow X}_{P}$ were derived in two steps. First, the
inter-pixel variance and covariance curves were computed using a set of flatfields with a large range of fluxes, using the prescription of
\cite{Astier2019}. Then, the electrostatic model of \cite{Astier2023} was fitted to obtain the coefficients $a^{C\leftrightarrow X}_{P}$.

\subsection{Charge transfer efficiency}\label{sec:implementation_ccd_cti}

The effect of charge-transfer inefficiency (CTI) happens during the readout of the CCD because of imperfections in the CCD silicon substrate lattice, which creates electron traps. Due to the stochastic capture and release of electrons into and out of these traps, part of the signal is left behind when some electrons are trapped while being transferred during readout. The delayed release of electrons is subsequently causing a smearing effect leading to the well-known CTI tails. As radiation damage is the leading cause for the creation of these traps, the CTI deteriorates fast with time for any space mission due to the increased radiation dose received compared to that of ground-based instruments \citep[e.g.][]{massey2014improved}.
 
A first simple CTI model implemented in \platosim{} follows the fraction of the total charge in a pixel $\theta_{\rm CTE}$ that is left behind, hence, $\theta_{\rm CTI} = 1-\theta_{\rm CTE}$. As charge transfer happens both during parallel transfer as well as serial transfer, the charge that is lost during each of these transfers is \citep{janesick2001scientific}
\begin{equation}\label{eq:cti}
Q_{N+n} = \frac{Q_0 \,N!}{(N-n)! \, n!} (1-\theta_{\rm CTI})^n \, \theta_{\rm CTI}^{N+n} \,,
\end{equation}
where $Q_0$ is the initial charge contained in the target pixel, $N$ is the number of pixel transfers, $n$ is the trailing pixel number following the target pixel (with $n=0$ being the target pixel itself), and $Q_{N+n}$ is the remaining charge in the $N+n$ pixel.

\begin{figure}[t!]
\centering
\includegraphics[width=\columnwidth]{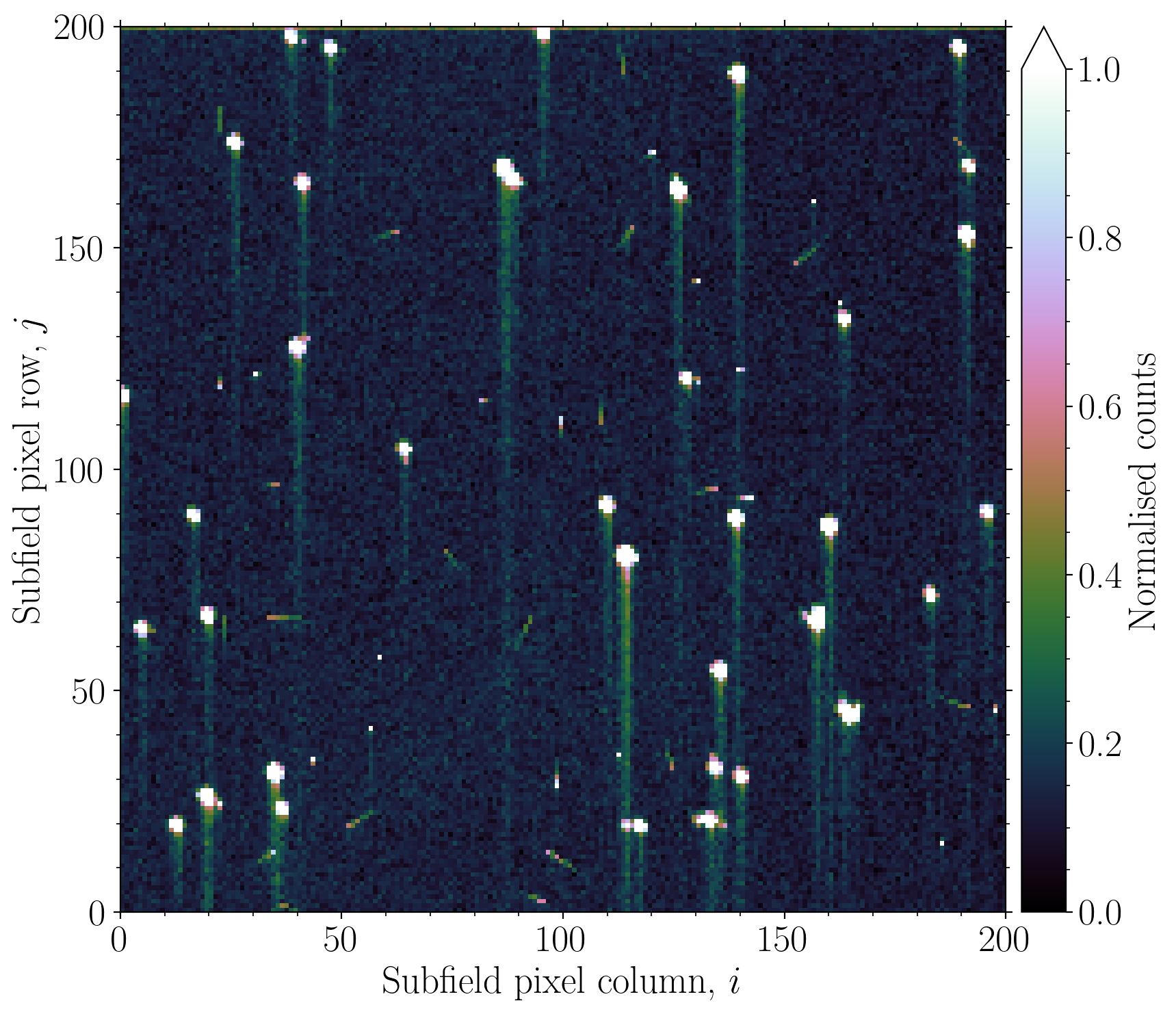}
\caption[]
{Illustration of the effect of CTI at post mission EOL (i.e. \SI{6.5}{\year} after commissioning) using the Short model. The plot shows a centrally placed $200\times\SI{200}{\pixel}$ CCD subfields of PIC stars from the LOP south including cosmic rays simulated using a hit rate of \SI{10}{\events\per\second\per\centi\meter}. The images has been clipped by a $2\sigma$ cut and then normalised for illustrative purposes. The readout register is located towards the bottom of the image.}
\label{fig:cti}
\end{figure} 

As a more physical description (naturally introducing the time dependence), the analytic model proposed by \cite{short2013analytical} (abbreviated the \textit{Short model}) for radiation-induced CTI for CCD detectors is also implemented into \platosim{}. This implementation includes 4 different CTE trap species, each with its own trap density $n_t$ (traps~\si{\per\pixel}), trap capture cross-section $\sigma_t$ (\si{\meter\squared}), and a release time scale $t_t$ (\si{\second}). In addition, the model also requires a parameter $\beta$ describing whether adding electrons to a charge package is increasing the volume of the package or its density. \cite{prod2016technology} derived BOL and EOL values for these quantities for the PLATO mission, assuming a nominal operating temperature of \SI{203}{\kelvin}. Figure~\ref{fig:cti} illustrates a worst case example of the effect of CTI at EOL (\SI{6.5}{\year} after commissioning) using the Short model. While the CTI increases with time the photometric quality decrease correspondingly, hence, the PLATO data reduction pipeline therefore contains a CTI correction step \citep[see e.g.][for good overview of such corrections]{israel2015well}.

\subsection{Blooming}\label{sec:blooming}

For bright stars ($V \le 8.5$ for N-CAM observations) the full-well capacity of some of the pixels is reached, and the electrons start to overflow to neighbouring pixels in the same column, which is usually referred to as \textit{blooming}. Although this saturation happens during the CCD exposure, \platosim{} models it pragmatically as a post-exposure effect that is independent of CCD non-linearity (which is applied later). The caveat of this implementation results in a slightly enhanced blooming pattern at the expense of computational speed, compared to an iterative approach which models their interaction every time step $\delta t$. In a first model, the electron excess is simply distributed evenly between the pixels above and below the saturated pixel. If these pixels also get saturated, the overflow goes to the next pixel in the column, etc. The result is a blooming pattern that is symmetric with respect to the central pixel of the star, that is the upward blooming trail has the same length as the downward one. 

Tests with the PLATO CCDs revealed, however, that in practice blooming can be highly asymmetric, with one trail being significantly longer than the other one. Although the exact underlying cause is still being investigated, the working hypothesis is that the electrons experience barriers in one direction, so that the excess electrons follow the path of less resistance in the opposite direction. Rather than trying to mimic the still uncertain physical causes, \platosim{} simply allows to specify the fraction of excess electrons that goes downwards (i.e. towards the readout register), which is sufficient to test the design and testing of the mask creation and photometry extraction of saturated stars.

\section{Photometry}\label{sec:photometry}

One of the main applications of \platosim{} is its ability to provide realistic light curves extracted at pixel level using its build-in photometric algorithm. Besides alleviating \platosim{}'s ability to generate light curves on demand, this feature is especially important when running large batches of simulations where the raw pixel data and house keeping data may be of the order of several gigabytes (or even terabytes). Due to the new software design of \platosim{}, generating light curves at run time significantly reduces the storage memory for output and adds a minimal time of execution per simulation.

To date a noteworthy list of literature exists on software for PSF photometry \citep{da1992basic, schechter1993dophot, anderson2006psfs, popowicz2018psf, hedges2021linearized}, aperture photometry \citep{howell1989two, stetson1987daophot, naylor1998optimal, libralato2016psf, bryson2010kepler, handberg2014automated, lund2015k2p2, aigrain2016k2sc, smith2016finding, marchiori2019flight, hoyer2020expected}, or pipelines that either combine both or provide both methodologies in a single software package \citep{kjeldsen1992high, still2012pyke, bradley2016photutils, cardoso2018lightkurve}. Generally the PLATO light curve generation of non-saturated stars follows two separate data processing chains, namely PSF photometry designed for on-ground data products (i.e. imagettes of $V<11$ stars belonging to the asPIC sample called P1) and aperture photometry designed for on-board data products (i.e. flux measurements of fainter targets ($11<V<15$) primary consisting of stars from the asPIC sample called P5). For now only the a subset of the on-board algorithms are implemented in \platosim{} whereas a functional coupling to the full processing chains, on-board and on-ground, have been established as will be explained in Sect.~\ref{sec:pipeline_validation}.

The performance reached by the photometry (both on-ground and on-board) directly depends on the knowledge of the PSF across the focal plane for each independent pointing. However, since the PSF morphology is expected to change notably after launch (due to slight changes of the optical mount during launch and to changes in the thermal environment throughout the mission) the `true' PSF cannot be measured from ground. Furthermore, acquiring accurate knowledge about the PLATO PSF in-flight is a main challenge due to the sparse pixel sampling of the PSF. Thus, we first elaborate on the procedure to overcome this challenge by reconstructing the true but unknown PSF across the CCD focal plane.

\subsection{PSF inversion using microscanning}\label{subsec:micrscanning}

PSF reconstruction builds from the idea of extracting a high resolution PSF, $\mathbf{x}$, from a series of corresponding lower resolution PSFs, $\mathbf{y}$, following \citep{park2003super}
\begin{equation}\label{PSF_inversion}
\mathbf{A} \ \mathbf{x} = \mathbf{y} \,,
\end{equation}
where $\mathbf{A}$ is a PSF projection matrix onto the low-resolution pixel grid. Mathematically the inversion is solved by discretising the PSF using a sum of basic functions $\phi_i$
\begin{equation}\label{basic_functions}
f(x,y) = \sum_i a_i \ \phi_i(x,y) \,,
\end{equation}
with $a_i$ being the unknown inversion coefficients. Solving Eq. \eqref{PSF_inversion} can be tackled using a least-squares procedure. The preferred method for PLATO is discussed in \cite{samadi2019plato}.

\begin{figure}[t!]
\centering
\includegraphics[width=\columnwidth]{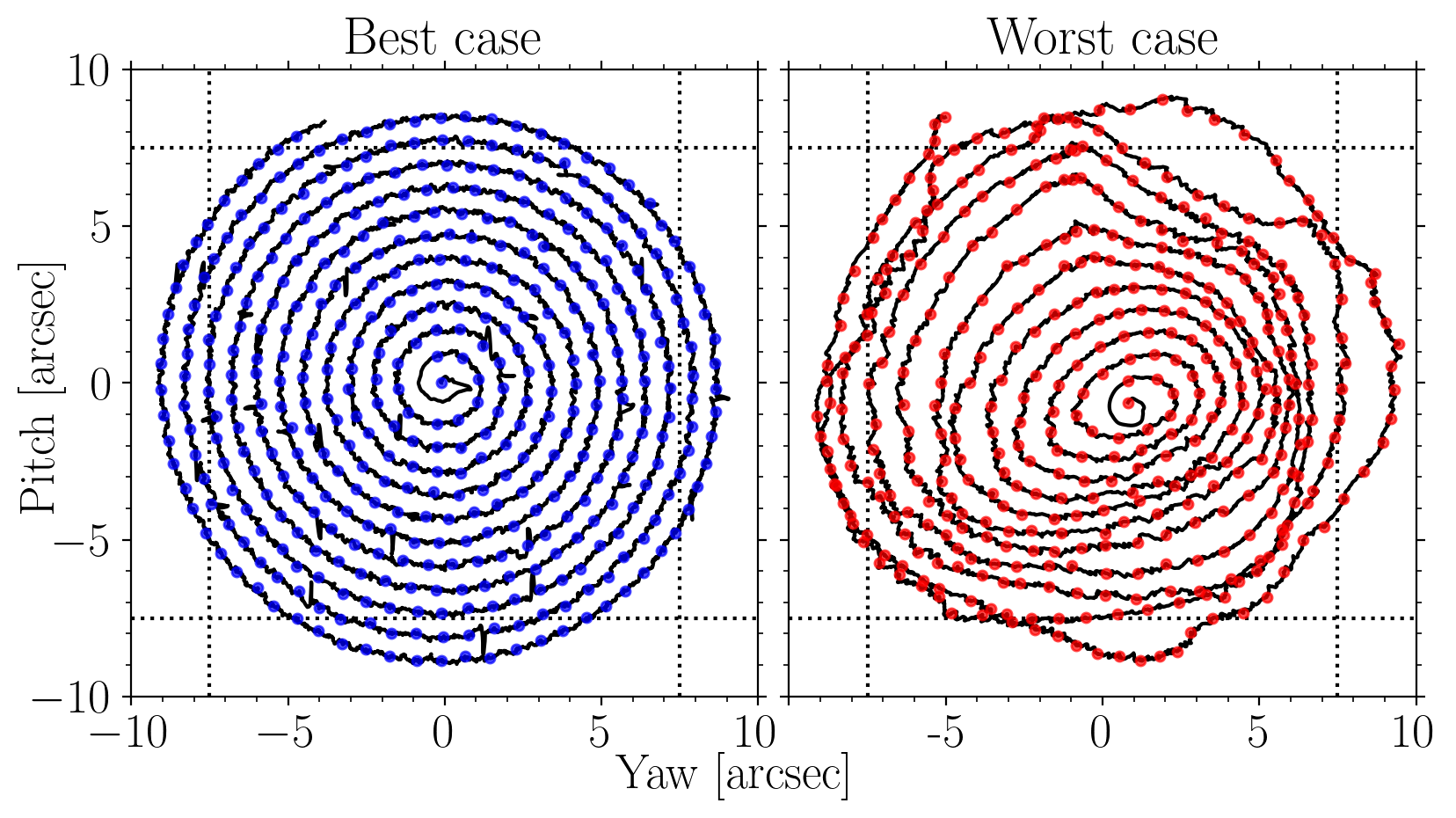}
\caption[]
{Illustration of the Archimedean spiral jitter pattern including; \textbf{Left:} minor residual AOCS jitter as a best case; \textbf{Right:} major AOCS residual jitter as a worst case. For an ideal pattern the distance $D$ between consecutive measurements is approximately constant and distance between consecutive spiral arms is $D\sqrt{3}/2$. These files contain 430 scans (shown as coloured circles marking the start of each exposure) and forms a near-equilateral triangular grid, which in turn provides a dense but near complete sampling over PLATO's pixel grid (dotted lines), needed for a successful PSF inversion. \textit{Data courtesy: OHB/TAS}.}
\label{fig:microscan}
\end{figure}

Following from the PSF inversion technique first employed by CoRoT \citep{auvergne2009corot} and later matured with Kepler \citep{bryson2010kepler}, in practice the series of low resolution pixel frames, $\mathbf{y}$, is acquired by intentionally commanding the AOCS to follow a certain pattern of small coordinate displacements. Such selective jittering is also known as a \textit{microscanning} session. Figure~\ref{fig:microscan} displays the Archimedean spiral pattern decided for PLATO, shown for a best case (left) and worst case (right) scenario in terms of pointing stability during the session.

Microscanning sessions will be performed during the quarterly interruptions necessary to realign the spacecraft's solar panels and will have a duration of around three hours each. The current in-flight strategy for microscanning is going to provide the inverted PSF for all P1 sample stars, whereas only a subset of carefully selected P5 (named R2) sample stars across the FOV will have their PSFs inverted for each pointing. Thus, the remaining P5 targets will have their high resolution PSFs determined from interpolation. The exact scheme of interpolation and for which magnitude range inverted PSFs reliably can be determined has been established by the PLATO team responsible for the data processing and algorithms. A transition to this interpolation strategy is currently being integrated into \platosim{}. As an approximate but realistic approach, \platosim{} allows to use a precomputed grid of inverted PSFs generated from both the worst and best case microscanning examples displayed in Fig.~\ref{fig:microscan}.  

Currently the grid of inverted PSFs are only fully representative for a single camera and quarter. In practice a grid of inverted PSFs for each camera is ideal, since the PSFs are strongly dependent on the alignment of optical lenses in each TOU, and for each mission quarter, since changes in PSF morphology together with ageing effects (such as the CTI) impact the accuracy of the inversion increasingly over time. Thus, particularly important for EOL conditions, more microscans are needed in the future to bring the photometry closer to the current mission strategy. However, as the state-of-the-art PSF- and aperture photometry algorithms require a high resolution PSF as input, this approach is already more realistic than the unphysical use of the `true' PSF (be it Zemax or analytical).

\subsection{Pre-processing steps}\label{sec:preprocessing}

Prior to the photometry extraction several pre-processing steps are applied to the simulated CCD subfield. First the bias offset is subtracted by computing the mean over a serial prescan region (orange box in Fig. \ref{fig:ccdFocalPlane}a) and a virtual%
\footnote{Virtual as in extra readouts of the register.} %
overscan region from either the F or E side of the detector, dependent on the subfield location. Next, readout smearing is corrected for by subtracting the bias corrected smearing map obtained from the parallel overscan region (pink box in Fig. \ref{fig:ccdFocalPlane}a). The gain is then used to convert the pixel values from ADU to counts of photoelectrons. Lastly, the bias subtracted sky background map is multiplied with the overall throughput to get counts of $\si{\electron\per\pixel\per\exposure}$ and then subtracted from the pixel map.

\subsection{Optimal aperture algorithm}\label{sec:aperture} 

From the wealth of aperture photometry pipelines mentioned above the on-board algorithm implemented in \platosim{} follows from, a study by \cite{marchiori2019flight}. By design, it optimises the photometric quality towards planet transit searches. This study found that a binary mask serves as the best compromise between noise-to-signal ratio (NSR) and the ratio of stellar contamination. 

The general idea is to build a mask starting with the pixel having the lowest NSR, then adding one pixel at a time under the condition that adding it should contribute more to the aggregated signal than to the aggregated noise. For an imagette, this procedure can be formulated mathematically by first arranging all $n$ pixels in increasing order of NSR
\begin{equation}\label{NSR_n}
\text{NSR}_n = \frac{\sqrt{ \sigma_{F_{T_n}} + \sum_{k=1}^{N_C} \ \sigma_{F_{C_{n,k}}} + \sigma_{B_n} + \sigma_{D_n} + \sigma_{Q_n} }}{F_{T_n}} \,,
\end{equation}
where $F_{T_n}$ and $\sigma_{F_{T_n}}$ are respectively the mean flux and photon noise of the target star, $\sigma_{F_{C_{n,k}}}$ is the photon noise of each $k$ stellar contaminant, $\sigma_{B_n}$ is the sky background noise, $\sigma_{D_n}$ is the combined detector noise, and $\sigma_{Q_n}$ is the quantisation noise. The second step consists in determining the aggregated NSR over the imagette's $m$ pixels conforming to the aforementioned pixel order of increasing NSR
\begin{equation}\label{NSR_agg}
\txxx{NSR}{agg}(m) = \frac{ \sqrt{ \sum_{n=1}^m \paren{ \sigma_{F_{T_n}} + \sum_{k=1}^{N_C} \ \sigma_{F_{C_{n,k}}} + \sigma_{B_n} + \sigma_{D_n} + \sigma_{Q_n} } }}{ \sum_{n=1}^m F_{T_n} } \,.
\end{equation}
The last step is simply to construct the aperture from the collection of pixels $m$ that minimises Eq. \eqref{NSR_agg}.
 
Pointing performance degradations over long time scales (due to long-term drifts) and on short time scales (due to reaction wheel momentum dumps, attitude tweaks, loss of fine guidance, pre/post safe mode events, etc.) may introduce significant pixel displacements. Compared to PSF photometry which by design is more robust against such instrumental perturbations, for aperture photometry this lead to systematic errors in the form of flux loss outside the pixel mask. To mitigate the loss of photometric precision over the course of a mission quarter the strategy for PLATO is to update the aperture of each star periodically under the condition that a lower NSR can be achieved by the updated mask \citep{marchiori2019flight}. 

\begin{figure}[t!]
\centering
\includegraphics[width=\columnwidth]{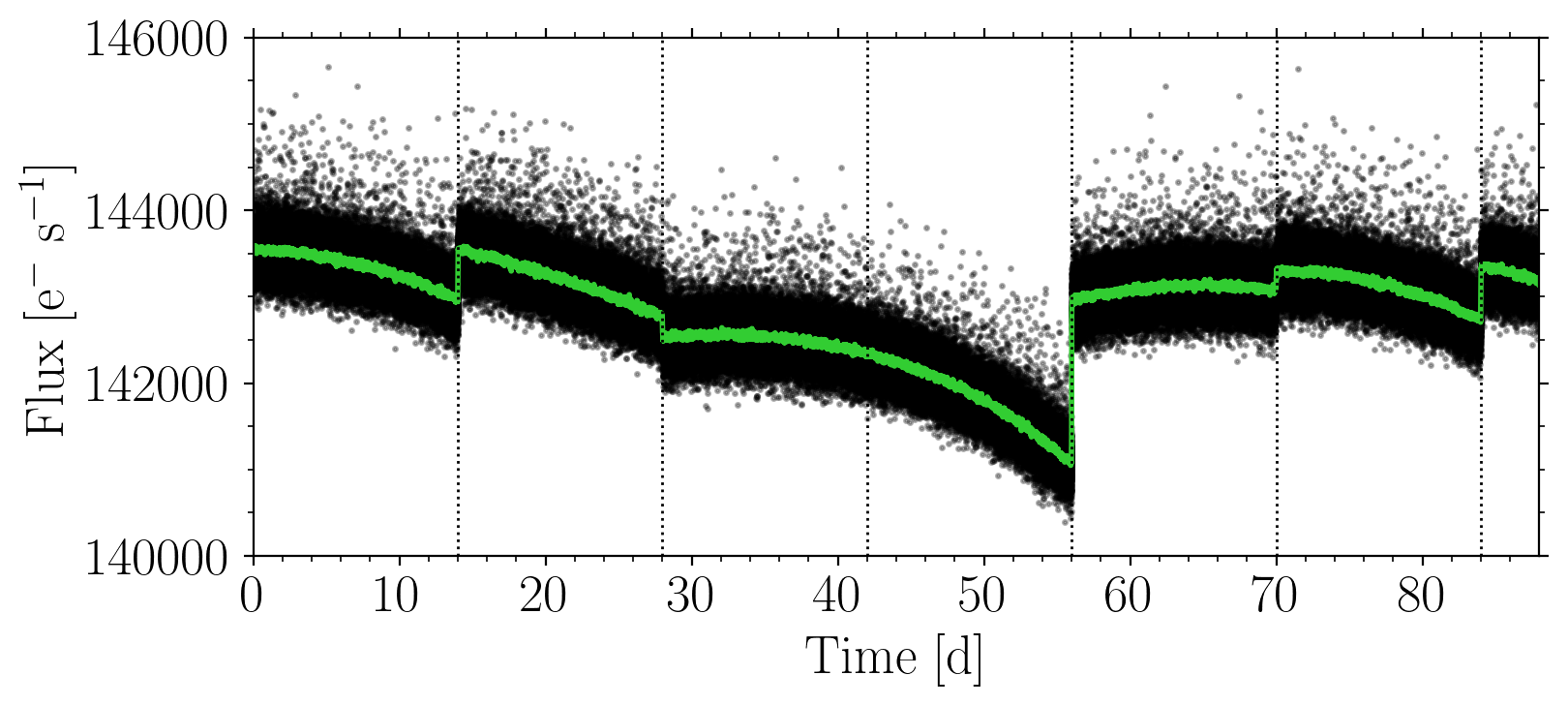}
\caption[]
{On-board photometry performed with the optimal aperture algorithm of \cite{marchiori2019flight} for a $V=10$ star. With a central barycentric pixel position and a (worst case) systematic drift of \SI{1.3}{\pixel} over the course of one mission quarter, the figure shows the algorithm in action with the automatic pixel-mask updates triggered every 14 days (grey-dotted lines) if a lower NSR can be achieved (which is not the case for the update at 42 days). Since the NSR does not scale linearly with flux, the mask-update strategy does not necessarily increase the flux level (as the case at 28 days). The positive flux outliers are due to contamination from cosmic rays (with a hit rate of \SI{10}{\events\per\second\per\centi\meter\squared}).}
\label{fig:photometry}
\end{figure}

\begin{figure*}[t!]
\centering
\includegraphics[width=1.9\columnwidth]{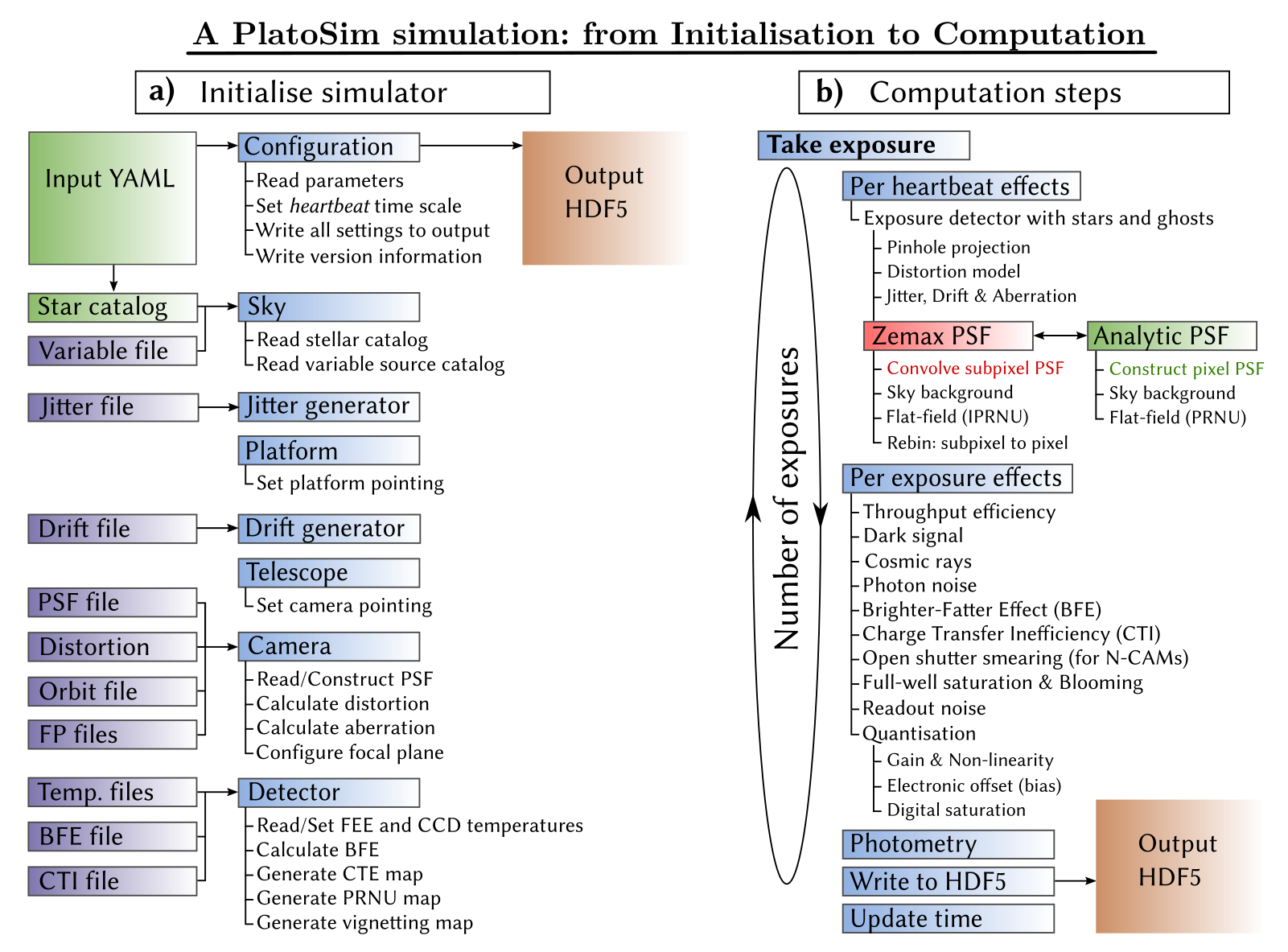}
\caption[Schematics of the \platosim{} software package.]
{Schematic of the \platosim{} software package. \textbf{a)} Overview of initialisation and configuration of \platosim{} prior to simulation execution. \textbf{b)} Overview of each simulation step as a loop over the total number of exposures. The boxes represent input files (purple), the output file (orange), software modules (blue), and the general simulation steps (green). The two flowcharts combined illustrates \platosim{}'s events of execution by a) first constructing a simulation (and all input parameters needed) followed by b) creating synthetic pixel images for a given number of exposures.}
\label{fig:platosim}
\end{figure*} 

Investigations led by the PLATO performance team have shown that the combined effect of TED and KDA can lead to a barycentric displacement up to \SI{1.3}{\pixel} over a three months duration. We illustrate this worst case scenario in Fig. \ref{fig:photometry} for a $V=10$ star positioned \SI{5}{\degree} from the optical axis at an initial central intra-pixel position. As the star drifts over the pixel array the mask update events (grey dotted lines) manifest themselves as flux jumps (which is amplified with a \SI{1}{\hour} running flux median shown in green in Fig.~\ref{fig:photometry}). This highlights a strategy trade-off between the mask update frequency vs. potential error propagation from the pipeline corrections. To minimise the mask update frequency developers of the PLATO pipeline are currently investigating if suboptimal apertures can accommodate the fully predictable KDA contribution on the stellar barycentric displacement across the FPA. For further information we refer to \cite{samadi2019plato} for a discussion on the impact of the mask updates on the final light curve and how long term and short term pixel displacements can be corrected in the post-processing procedure.


It is noteworthy that systematic noise sources acting on time scales shorter than the cadence are not included in the mask creation of Eq.~\eqref{NSR_agg}. This is the case for AOCS jitter and to lesser extent CCD effects that degrade the photometric quality (such as CTI, cosmetic defects, etc.). Including the contribution of jitter (and larger attitude tweaks) is a challenge since it depends on the final shape of the mask. Nevertheless, instrumental perturbations such as jitter has been shown to have a negligible impact on the photometry in nominal conditions \citep{marchiori2019flight}, and can partially be corrected for during the PLATO pipeline chain \citep[cf.][]{samadi2019plato}. Moreover, as soon as the aperture mask of Eq.~\eqref{NSR_agg} has been defined, the stellar pollution ratio (SPR) from nearby contaminant can be derived, which can help for rejecting false-positive planet detections from blended eclipsing binaries.

\section{\platosim{} software architecture}\label{sec:platosim}

The creation of a synthetic CCD image starts with a set of input parameters that defines the general properties of the spacecraft's hardware components, the stellar field, and the simulated observation itself. As depicted in Fig.~\ref{fig:platosim} the \platosim{} software generates a simulation in two steps: \textbf{a)} it configures all input parameters and constructs an output file and then \textbf{b)} it follows all (requested) algorithm steps in a loop over a total number of exposures defined by the user.

The simulation construction relies on a YAML configuration file that initialises all the input parameters and optionally reads further supplementary files (purple boxes). As a minimum \platosim{} needs a YAML input file and a star catalogue to successfully run (green boxes). Identical to the physical hardware components that make up the PLATO payload, \platosim{} consist of a platform, telescope%
\footnote{The `Telescope' nomenclature is introduced to separate the AOCS model and the optical effects in the software architecture.}, %
camera, and detector module, in combination with a sky module to include sky background and stellar variable signals, and two time series generators (jitter and drift) for the inclusion of pointing systematics in an automated way (blue boxes). Each module deals with a particular effect or subsystem. Combined, these are controlled by a global simulation object that is directly configurable in Python. As mentioned in Sect.~\ref{sec:image_generation}, the smallest time varying phenomenae, referred to as PlatoSim's `heartbeat', is initialised in order to partition each exposure into smaller time steps.

The construction of a simulation with respect to input and output is completely standardised in modules. This is highly beneficial for understanding subsystems or individual effects. All supplementary input files should be provided in ascii format and output files are in HDF5 format. As indicated by the purple boxes in Fig.~\ref{fig:platosim}a, whether an effect is included and/or a model selected leaves an enormous flexibility for the user to conduct highly diverse (and complex) simulations.

Upon execution, after the setup has finished in step a), \platosim{} generates a time series of synthetic CCD images in a loop over the total number of requested exposures, as illustrated in Fig.~\ref{fig:platosim}b. Over the course of a single exposure, the algorithmic steps are organised into two consecutive classes: first those that are computed per heartbeat, and next those that are computed only once per exposure. The flow of the algorithmic steps are approximately according to the light path of the incident photons, placing each effect or subsystem in a logical order after their physical occurrence. However, as mentioned earlier this is only approximate as different effects physically do not follow a deterministic entry point of occurrence which algorithmic architectures necessarily need to comply to. For example various CCD and FEE effects (such as BFE, blooming, and non-linearity) strongly depend on each other. As a final action of each acquisition, the photometry module is (optionally) applied, the output to written to disk, and the internal clock is updated. 

Lastly, \platosim{} allows for the possibility of parallelisation. The execution of a simulation on multiple cores (or CPUs) is not only possible per subfield (as was done for the simulations in Sect.~\ref{sec:applications}), but also in time. The former execution is a standard \textit{sequential} workflow (i.e. where each simulation run independently on a designated core), whereas the latter options is a \textit{partitioning} (i.e. a single simulation is chopped into smaller time series and  deliver each to a different core). The latter option is more complex as it requires the time series of any supplementary input file (e.g. the variable source file) to be computed ahead of the simulation, however, the random seeds are handled automatically by \platosim{} during the partitioning.

\section{Applications to the PLATO mission}\label{sec:applications}

Due to the end-to-end and modular design, \platosim{} is used for many different disciplines within the PLATO mission consortium as will be discussed in the following. We note that the data and results of Sects.~\ref{sec:performance_studies} and \ref{sec:hare_and_hound} are made available to everyone through the \platosim{} repository.

\subsection{Mission preparation}\label{sec:mission_preparation}

The assembly, integration, verification and test (AIV/AIT) of a spacecraft such as PLATO involves a considerable amount of preparatory work. This is done first of all to design the details of operations for each of the tests, which will impact the quality and the relevance of the results as well as the planning. Secondly, preparatory work also helps defining the necessary Ground Support Equipment (optical stimuli, telemetry acquisition systems, etc.), designing the data processing algorithm for each test, etc. \platosim{} played a key role for a series of tests in this respect.

The first delicate operation when assembling PLATO is to align the cameras. This involves assembling the FPA, bearing the detectors with the TOU within very tight accuracy budgets \citep{pertenais2021}, and ensuring the optimal performance of the camera in operations. While the PLATO cameras will operate in vacuum and around $-70^{\circ}\text{C}$, the alignment and assembly occur at room temperature, that is  $\sim100^{\circ}\text{C}$ warmer and under normal atmospheric pressure. Consequently, during the alignment, not only is the optical quality of the cameras considerably degraded, but the detectors also produce tens of thousands of \si{\electron\per\second\per\pixel} of dark current. The optical verification necessary for a proper alignment hence requires dedicated operational modes. This includes short integration times and partial readout of the CCDs to limit the dark current, CCD clearout between every frame, continuous image-dump between observations, etc. \platosim{} was used to simulate data obtained at ambient temperature to make the trade-off between observations under full pupil illumination and observations with a hartmann mask. Hence, \platosim{} was used to design the test approach and data-reduction process, and estimate the resulting accuracy in each case. The details on the simulations and on the alignment of the engineering camera are presented in \cite{royer2020optical} and \cite{royer2022alignment}, respectively.

\begin{figure}[t!]
\centering
\includegraphics[height=4.35cm]{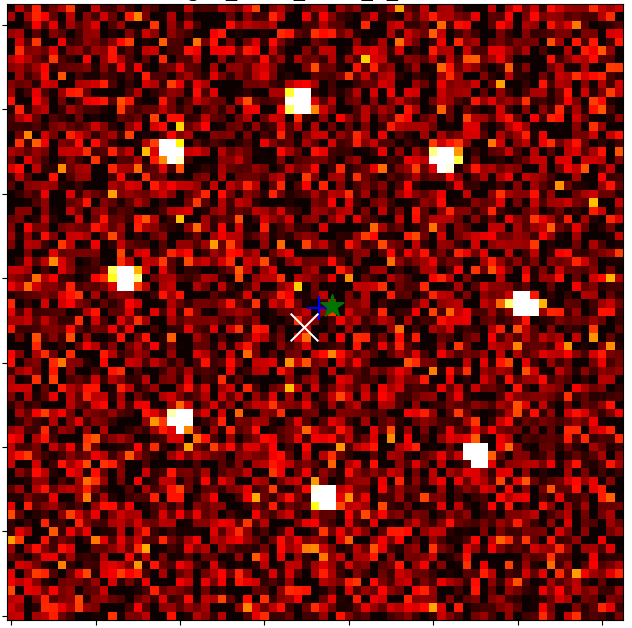}
\includegraphics[height=4.30cm]{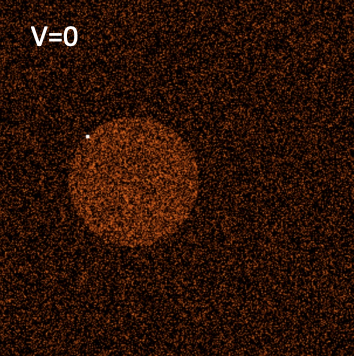}
\caption{Examples of simulated data generated in preparation of the AIT/AIV of PLATO mission. \textbf{Left:} Simulation of a hartmann pattern obtained at ambient temperature and far out of focus, in preparation of the camera alignment. \textbf{Right:} Simulation of a $V=0$ star close to the optical axis of the camera (white dot) and the extended ghost it creates on the detector via parasitic reflections, ran in preparation of the thermal-vacuum tests.}
\label{fig:alignment}
\end{figure}

\platosim{} is also instrumental in the preparation of the environmental tests performed at thermal vacuum (TVAC). We here cite four examples where \platosim{} was used in the testing. First the long term stability test, aimed at testing the measurement stability at camera-level. Secondly, the characterisation of the image-ghosts due to multiple reflections in between various optical surfaces and on the detector~\citep{pertenais2022}. Third, the characterisation of the camera image geometry, that is of the optical distortions induced by the very wide field. Finally and fourth, the optimisation of the measurement strategy for the critical but time consuming determination of the best focus temperature, and characterisation of the image quality at the optimal focus temperature~\citep{borsa2022}.

\platosim{} is also used to prepare the operations at spacecraft level, for instance in simulations to test the compression algorithms to run on the Instrument Control Unit on-board the spacecraft, or in simulations of real-time operations of the Fine Guiding System, that is in the feedback loop between the fast cameras and the AOCS \citep{griessbach2021fine}. Figure~\ref{fig:alignment} displays an example of a hartmann pattern simulated for an ambient temperature test (left), as well as a simulation of the extended ghost image caused by an very bright star (right).

\subsection{Pipeline validation}\label{sec:pipeline_validation}

According to PLATO's main objectives the pipeline chain needs to be able to preserve signals in the light curve belonging to transiting planets, stellar activity, and stellar pulsations. These are time varying phenomena ranging from seconds to months. PLATO must deal with a huge photometric dynamical range of more than ten orders of magnitude. Furthermore, a strategy for merging the light curves across multiple cameras and mission quarters, in combination with the limited telemetry capabilities, require that most of the scientific analysis must be done autonomously on-board the spacecraft. Altogether this demands a sublime pipeline performance. The construction of a versatile pipeline can only be done prior to launch using a test harness of simulations for which \platosim{} plays a key role.

In comparison to the built in pre-processing steps of \platosim{}'s on-board photometry module (see Sect.~\ref{sec:preprocessing}), the removal of outliers and flux corrections from short term and long term pixel displacements are algorithms yet to be implemented. We highlight that the full PLATO pipeline chain consist of three main branches for the light curve generation: i) on-ground; ii) on-board; and iii) for saturated stars (\citealt{rauer2014plato}, Rauer et al. in prep.). To accommodate the need for generating fully calibrated data product of non-saturated stars, a computationally bridge between \platosim{} and the (preliminary) PLATO reduction and extraction pipeline has been established. 

For the validation of the photometric extraction of saturated stars \platosim{} is heavily used due to its ability to simulate non-linear effects. To validate the pipeline performance of non-saturated stars, \textit{stitching} and \textit{detrending} of light curves is required. Indeed, a wealth of events will leave large data gaps and introduce systematic errors that can be highly correlated and thus almost impossible to model and remove. Poor attempts can easily hinder the detection of the astrophysical signals. Hence light curve stitching and detrending are well studied topics in space  photometry \citep{garcia2011preparation, vanderburg2014technique, handberg2014automated, lund2015k2p2, aigrain2017robust, hippke2019wotan, lund2021tess}. They are also extremely instrument dependent and science driven and thus deserve special attention for PLATO. As such \platosim{} simulations are currently being used for this aspect of the pipeline, which is vital for PLATO's ability to characterise the exoplanet host stars, and thus ultimately the planets themselves.

\subsection{Performance studies}\label{sec:performance_studies}

The capacity to investigate how the photometric precision internally depends on instrumental systematics on both short and long time scales shorter than the cadence is one of \platosim{}'s advantages. As an example we show a performance study of PLATO's expected photometric precision at BOL. 

With a premise that a \SI{24}{\hour} duration dataset is representative for estimating the averaged NSR of each light curve, we simulated $10\,000$ F5-K7 dwarf and sub-dwarf stars from the LOP south covering a large photometric dynamical range. To populate the NSR--$V$ diagram approximately evenly a total number $2\,500$ stars was drawn from their camera observability $\tx{n}{CAM}\in\{6,12,18,24\}$, meaning that in total $150\,000$ light curves were generated. To simulate the effect of stellar crowding a choice was made to include photometric contaminant stars brighter than $\Delta V < 5$ and within a relative radial distance to their target of \SI{45}{arcsec}, that is maximally three pixels away from the target barycentre within each simulated imagette.

All simulations are configured with a realistic AOCS jitter time series from OHB/TAS sampled at \SI{8}{\hertz} (being the model shown in the bottom panel of Fig.~\ref{fig:jitter}) to realistically include its impact on the photometric precision `as expected'. All random and systematic noise sources are configured `as required' of the instrument design. For computational alleviation the analytic PSF model was used and charge diffusion was activated. Furthermore, it is assumed that the properties of each TOU, CCD, and FFE among all cameras are identical and their noise sources are uncorrelated. However, we randomly displace each camera pointing to imitate camera misalignments to the interface of the optical bench. For all simulations a constant intrinsic intensity was considered and the photometric extraction was performed by the build-in photometry module of \platosim{} (as described in Sect.~\ref{sec:photometry}) due to its superior computational speed compared to the preliminary on-board PLATO pipeline.

\begin{figure*}[t!]
\centering
\includegraphics[width=\columnwidth]{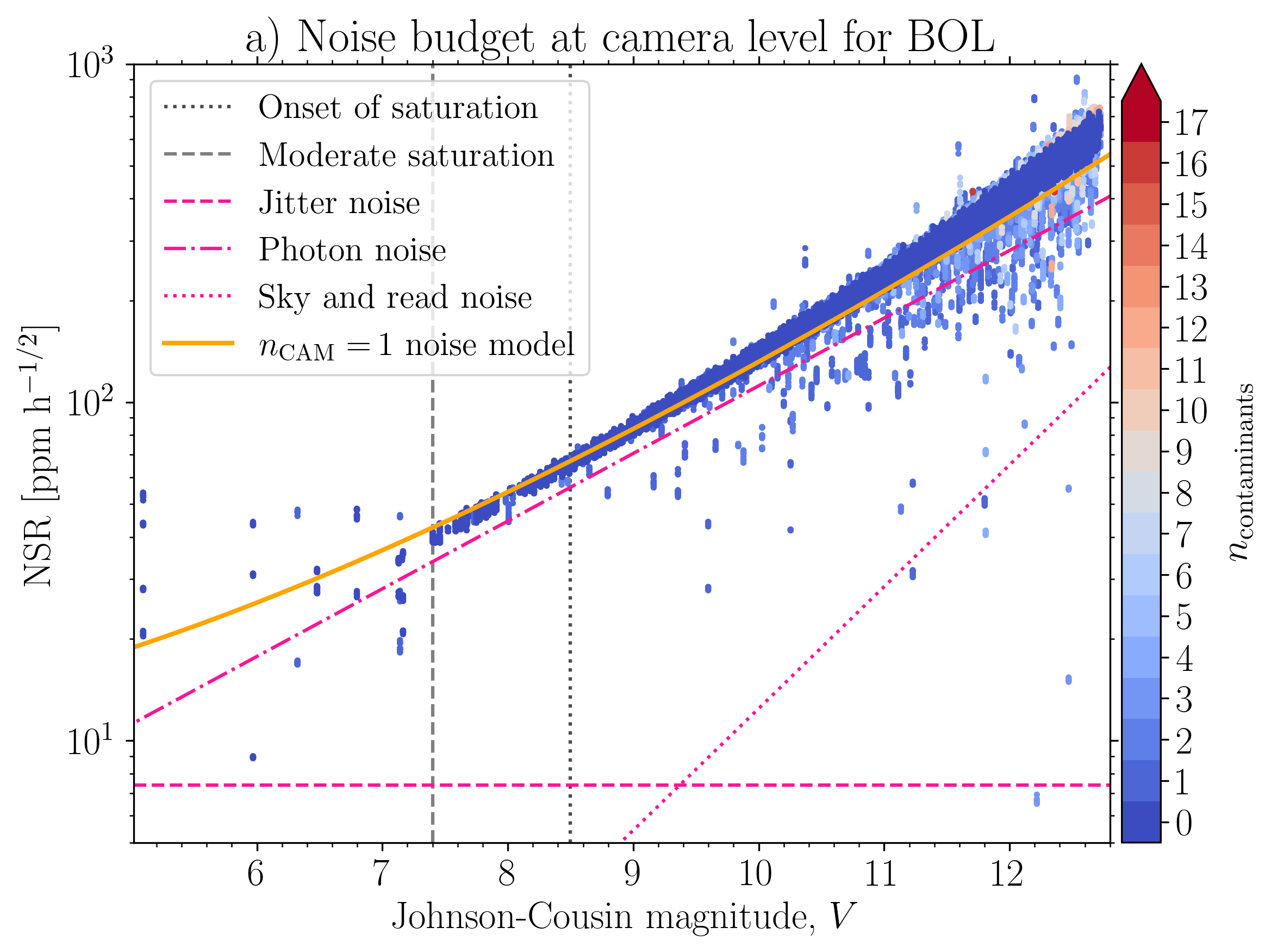}
\includegraphics[width=\columnwidth]{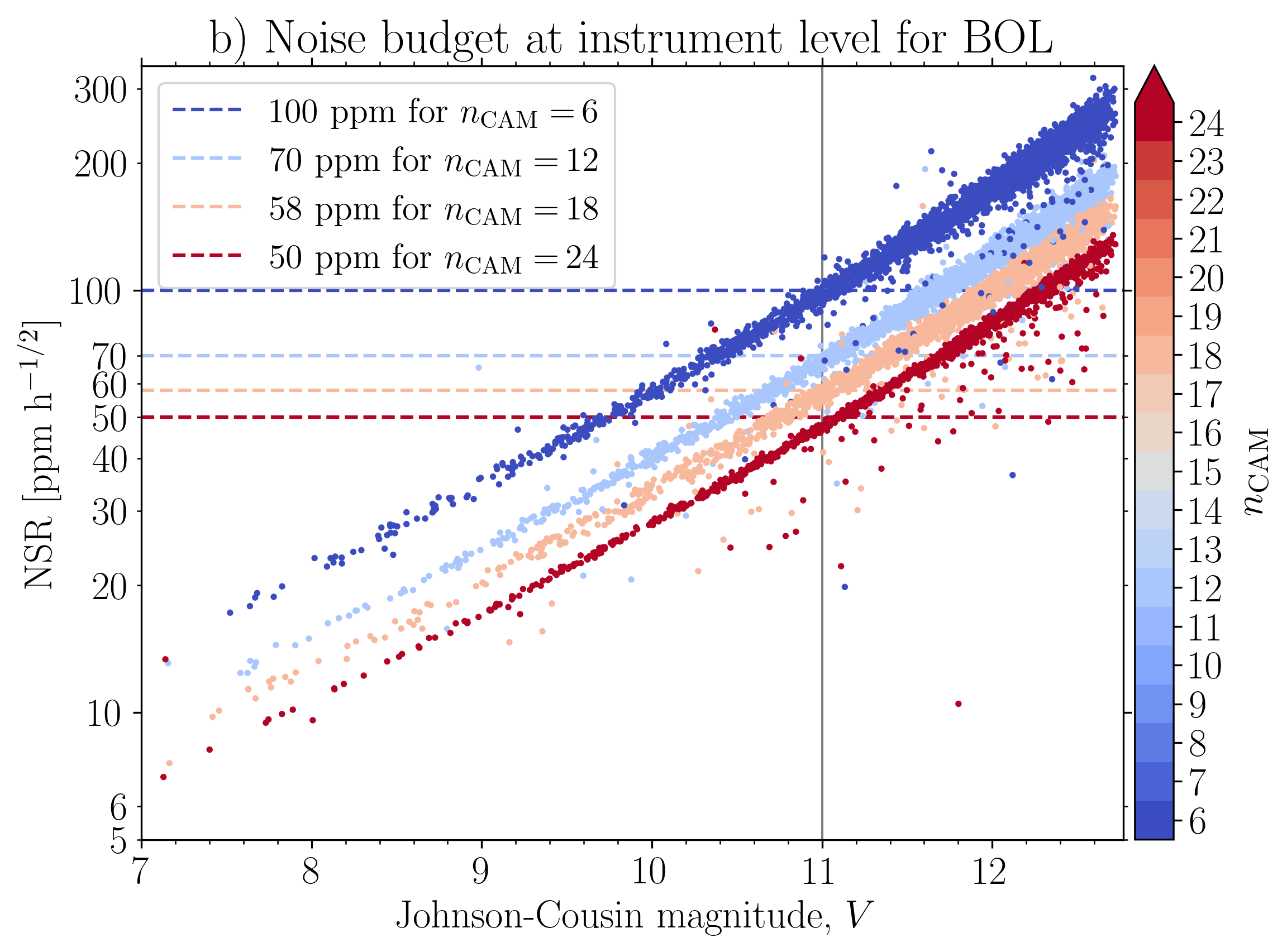}
\caption[]
{NSR$(V)$ simulation study at BOL as required by the mission. \textbf{a)} Noise budget at the camera level with each data point coloured after the number of stellar contaminants contained within \SI{3}{\pixel} and $\Delta V < 5$ of the target star. The model prediction of the noise (orange solid curve) consists of three photometric noise components: jitter noise (pink dashed line calculated using an averaged (yaw, pitch, and roll) jitter time series rms amplitude value of \SI{0.037}{\arcsec}) dominating in the bright regime, readout and sky background noise (pink dotted line using a sky background value of around \SI{60}{\electron\per\pixel\per\second}) dominating in the faint regime, and photon noise (pink dashed-dotted line) dominated between the two aforementioned regimes. Also the onset of saturation is indicated (grey dotted line) together with the onset of \textit{moderate saturation} (grey dotted line) here defined as where an extended mask expanding beyond the dimensions of an imagette is needed to capture total stellar flux due to blooming. \textbf{b)}~Noise~budget at the instrument level with each data point representing a multi-camera observation colour by the number of N-CAM observations used in the NSR calculation. The mission requirement of the photometry for $V<11$ is shown for an observability of $\tx{n}{CAM}\in\{6,12,18,24\}$ (horizontal dashed lines coloured after $\tx{n}{CAM}$).} 
\label{fig:performance}
\end{figure*}

For each star we compute the NSR by first combining the normalised light curves from each camera, averaging measurements of the same camera group (as they have identical time stamps). Next we resample the data into \SI{1}{\hour} bins, and compute the standard deviation. The resulting photometric noise is shown in Fig.~\ref{fig:performance}a for each individual camera observation (i.e. at camera level) and Fig.~\ref{fig:performance}b for the multi-camera observations (i.e. at instrument level). 

Taking a closer look at Fig.~\ref{fig:performance}a the colours represent the total number of stellar contaminants included within each simulation and the orange line is NSR model prediction given the mission requirements at BOL. In detail this model contain a component of noise from AOCS jitter (pink dashed line), photon noise (pink dashed-dotted line), and sky plus read noise (pink dotted line). As expected for dynamic range simulated, the photon noise dominated the noise budget, however, random noise from the sky background and readout noise will ultimately dominate beyond $V > 15$. The simulations shows a slight discrepancy with the model prediction in the bright end where noise from AOCS jitter starts to dominate. A thorough investigation shows a clear indication that the NSR at the onset of pixel saturation (grey dotted line) only starts to dependent on the CCD properties, such as the barycentric location of star, for relatively high rms amplitudes of the jitter Euler angles. The photometric prediction for saturated stars may thus behave very different to the in-flight measurements (especially considering asymmetric blooming not included here), nevertheless, it is clear that the photometric precision extracted from imagettes depends on the combination of saturation and AOCS jitter with an ultimate noise floor set for $V<7.4$. This relates to the onset of moderate saturation (grey dotted line), as defined here, to when blooming cause flux leakage out of an imagette and an extended mask is needed to conserve the flux measurement (for which an extended mask library are currently being designed using \platosim{}).

For the NSR at instrument level, Fig.~\ref{fig:performance}b displays a clear division between stars observed with $\tx{n}{CAM}\in\{6,12,18,24\}$. The applied camera misalignment model introduces significant barycentric displacements up to a few tens of pixels from camera to camera, which in turn may render some stars unobservable with one or more cameras (as expected in-flight). The figure illustrates that the mean noise level for stars with $V<11$ satisfy the requirements of $\text{NSR}\leq\{100, 70, 58, 50\}\,\si{\ppmh}$ for observations with $\tx{n}{CAM}\in\{6,12,18,24\}$, respectively. This agrees with a similar simulation study of Rauer et al. (in prep.) using the PINE simulator \citep{borner2022plato}. 

PINE is a theoretical noise estimator that uses the true (i.e. uncontaminated) flux of the target star to calculate the NSR, meanwhile such an approach is not readily possible at pixel level whilst extracting the photometry of stars from real stellar fields (such as that of the PIC). Since flux leakage from stellar contaminant(s) into the aperture mask is unavoidable (and can only partially be mitigated by the mask definition itself), the scattered data points below each of the corresponding NSR--$\tx{n}{CAM}$ curves of Fig.~\ref{fig:performance}b are PIC targets with one or more contaminants (as shown in Fig.~\ref{fig:performance}a). Naturally, accounting for stellar contamination is expected to place each of these measurements, above the corresponding NSR curve due to the addition of photon noise from the stellar contaminant(s).

\subsection{Hare and Hound exercises}\label{sec:hare_and_hound}

Hare and Hound exercises, also known as injection and retrieval exercises, have more pragmatically been performed with other PLATO simulators \citep[such as \texttt{PSLS};][]{samadi2019plato}. Although a full suite of such exercises using \platosim{} is beyond the scope of this paper, we make a realistic show case of PLATO's ability to detect an Earth-like planet orbiting a Sun-like host star. We simulate a $V=10$ star (i.e. $\Pb\simeq10.4$ cf. Eq.~\eqref{V-P}) observed with all 24 N-CAMs. We simulate 13 mission quarters ($\sim\SI{3.3}{\year}$) including four transits to test the retrieval efficiency of 2, 3, and 4 transits.

We employ the same computational setup as in Sect.~\ref{sec:performance_studies} with a change of the AOCS jitter to a red noise model sampled at \SI{0.1}{\hertz}, and an activation of the KDA model. The TED for each camera and quarter is included using a second order polynomial model whilst uniformly drawing the model coefficients under the restriction that the amplitude in yaw, pitch, and roll cannot exceed \SI{10}{\arcsec}. The latter is a conservative choice and was deliberately made to challenge the software that corrects for systematic long-term trends in the light curves. For a realistic representation of the noise budget for our P1 sample star, the preliminary PLATO pipeline (using PSF photometry) is employed. 

As the majority of targets from the PIC are expected to have convection-driven variability we model and inject stellar granulation and oscillations into the simulated target. A full asteroseismic Hare and Hound exercise is out of the scope of this paper, but a full model description of the solar-like oscillator is presented in Appendix \ref{sec:oscillations}. For the present showcase only the impact of the noise floor of stellar variability on the transit detection is of interest.

To generate the planet transits we use the open source software \texttt{batman}%
\footnote{\url{https://github.com/lkreidberg/batman}} %
\citep{kreidberg2015batman} tuning the radius, mass, and orbital period ($P=\SI{365.25}{\day}$) to that of the Sun-Earth system (assuming no third bodies) and choose the time of ephemeris to $t_0=\SI{10}{\day}$. For simplicity we consider circular orbits ($e=0$) and edge-on transits ($i=\SI{90}{\degree}$). We use a quadratic limb darkening (LD) law and calculate the LD coefficients using the software \texttt{PyLDTk}%
\footnote{\url{https://github.com/hpparvi/ldtk}} %
\citep{parviainen2015ldtk} to calculate the stellar LD profile in the PLATO passband. For our simulated Earth analogue orbiting a G2V host star this results in a transit depth of $\delta \approx\SI{103}{\ppm}$ and a total transit duration of $\tx{T}{tot} \approx \SI{13}{\hour}$. The transit depth overshoot of $\sim22\%$ compared to the planet-to-star radius ratio of \SI{84}{\ppm} has been explained by the effect of stellar limb darkening \citep{heller2019analytic}.

\begin{figure}[t!]
\centering
\includegraphics[width=\columnwidth]{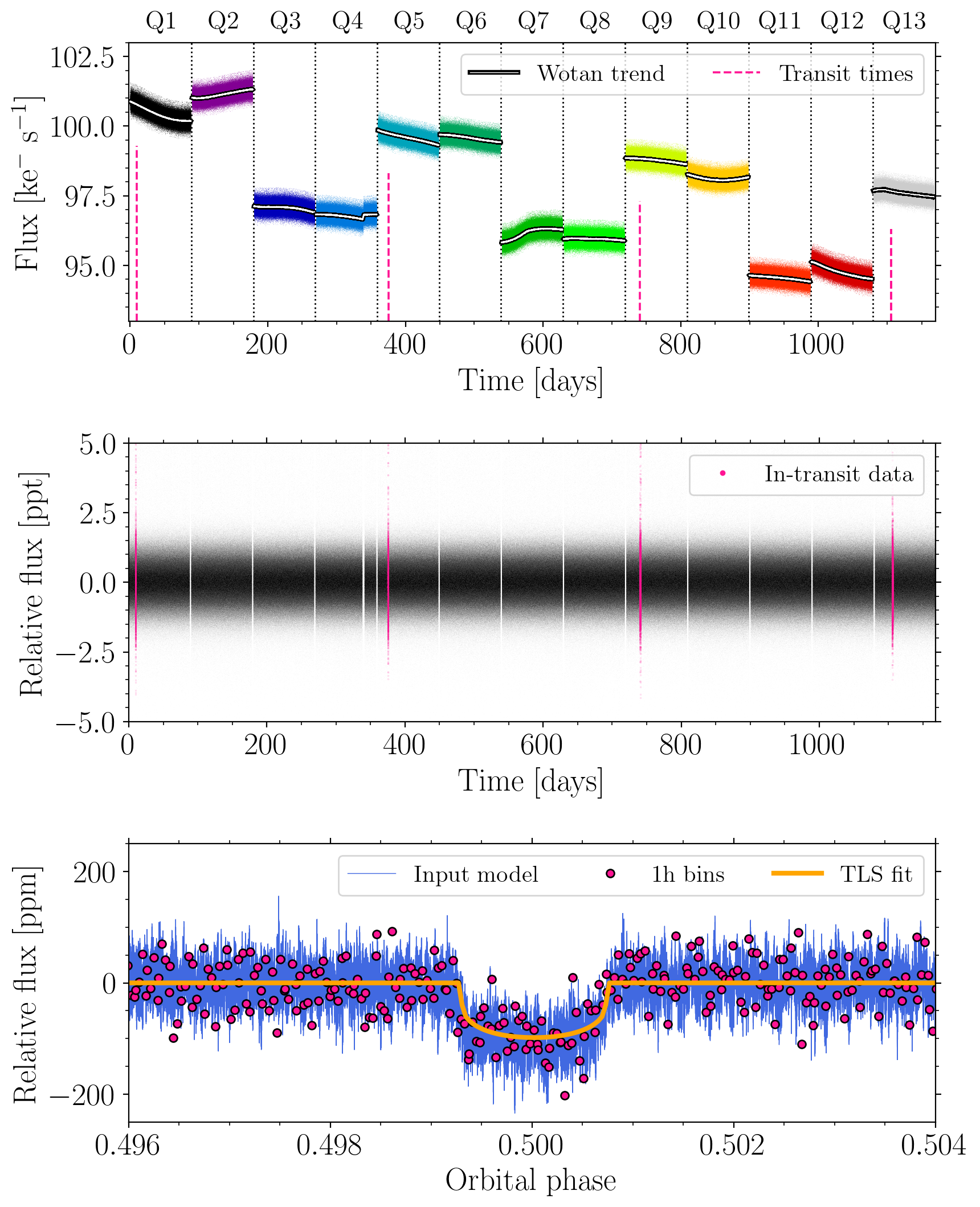}
\caption[]
{Results of the Hare and Hound (injection and retrieval) exercise of an Earth-sized planet transiting a $V=10$ Sun-like star. With an orbital period of \SI{365.25}{\day} the planet transits 4 times in the \SI{3.3}{yr} simulated light curve. \textbf{Top:} Light curves observed with a single camera (coloured per quarter) and the corresponding \texttt{Wōtan} trend (white/black lines). \textbf{Middle:} Detrended light of all 24 camera observations (black dots) with in-transit data highlighted (dots). \textbf{Bottom:} Phase folded light curve on the best fit period from \texttt{TLS}. For clarity a \SI{1}{\hour} binned representation of the light curve is shown (pink circles), together with the injected variability (blue line), and a best fit transit model (orange line).} 
\label{fig:hare_and_hound}
\end{figure} 

We follow the transit retrieval procedure of \cite{heller2022transit} and use the open source software \texttt{Wōtan}%
\footnote{\url{https://github.com/hippke/wotan}} \citep{hippke2019wotan} for detrending. \texttt{Wōtan} is optimised to preserve transiting signatures of exoplanets while effectively removing instrumental and stellar variability using a large library of available detrending filters. We here use Turkey's biweight method and a window size of $3\times\tx{T}{tot}$ shown by \cite{hippke2019wotan} to be an optimal choice for most transit searches. The detrending is performed on each individual camera and mission quarter segment. As a performance illustration of the detrending the top panel of Fig. \ref{fig:hare_and_hound} shows the simulated time series for a single camera across 13 mission quarters (coloured segments), the corresponding model trend (white/black line), and the mid transit times (dashed pink lines). Aside from the model trends caused primarily by the TED, Fig.~\ref{fig:hare_and_hound} shows two additional dominating features; i) large jumps in the mean flux level from one mission quarter to the next caused by a change in the optical throughput as the quarterly rotations relocate the stars to different distances from the optical axis in the FPA (whilst the corresponding change in PRNU is a second order effect here); and ii) an overall decrease in intensity due to the combined set of ageing effects.

The final detrended light curve merged across all cameras and quarters is shown in the middle panel of Fig.~\ref{fig:hare_and_hound} and is used for the transit vetting. The actual transit search is performed using the transit least-squares (TLS) method implemented in the open software \texttt{TLS}%
\footnote{\url{https://github.com/hippke/tls}} %
\citep{hippke2019optimized}. Compared to the traditional box least-squares (BLS) method, \texttt{TLS} outperforms the \texttt{BLS} algorithm \citep{kovacs2002box} especially in the domain of small planets \citep[e.g.][]{heller2019atransit, heller2019btransit} as it includes a precise modelling of the transiting signature (in particular the ingress, egress, and the LD profile). Upon execution \texttt{TLS} performs a grid search in the parameter space $\{P, \ t_0, \ \tx{T}{tot}\}$ to find the minimum $\chi^2$ value. While doing so, it calculates the Signal Detection Efficiency (SDE) being the key search metric of how significant the $\chi^2$ minimum is compared to its surrounding $\chi^2$ landscape as a function of orbital period. 

Before employing the actual vetting a $4\sigma$ clipping filter was used to remove large outliers and all measurements from the same camera group were averaged. Lastly, a \SI{1}{\hour} binned representation of the retrieved light curve was created (which here is justified for the planet detection however not for characterisation). The \texttt{TLS} vetting of 2, 3, and 4 transits resulted respectively in a SDE of around 24, 40, and 47. These are well beyond the previous reported detection threshold of $\text{SDE} \geq 9$ that result in a false-alarm probability of $\text{FPR} < 10^{-4}$ for Earth-sized planets \citep{hippke2019optimized}. As an illustration the bottom panel of Fig.~\ref{fig:hare_and_hound} shows the phase-folded light curve of the best \texttt{TLS} fit to the full dataset (orange line) together with the injected variability model (blue line) and a the \SI{1}{\hour} binned light curve (pink dots). With a signal-to-noise (S/N) ratio of $\text{S/N} > 7$ \citep[cf.][]{pont2006effect} in all above cases the presence of stellar granulation and pulsations does not avert the planet retrieval, which agrees well with the simulation study of \cite{morris2020stellar}.

\section{Discussion and conclusion}\label{sec:conclusion}


\platosim{} is an advanced end-to-end CCD and light-curve simulator dedicated to simulate PLATO measurements. Its algorithms and their underlying methodology aim to model realistic space-based photometric imaging. \platosim{} is available on GitHub%
\footnote{\url{https://github.com/IvS-KULeuven/PlatoSim3/tree/master}} %
together with a detailed documentation and tutorials.


The mission requirements of PLATO's innovative multi-camera configuration pushes the frontier of the instrumental design. Reliable simulations are therefore essential -- from the early design phases all the way to the in-flight operations -- to continuously assist with the design, assessment, and validation of the instrument. \platosim{} was developed to accommodate for this niche exactly. We have provided several examples of applications, ranging from dedicated studies to address the technical assessment and verification of the instrument (Sect.~\ref{sec:mission_preparation}), the development of data reduction and processing algorithms (Sect.~\ref{sec:pipeline_validation}), the performance assessment of the payload (Sect.~\ref{sec:performance_studies}), to the verification of a core science case relevant to the mission (Sect.~\ref{sec:hare_and_hound}).


To our knowledge, \platosim{} is one of the most feature-rich simulators currently available to simulate space-based photometry, including a wide range of instrumental noise sources both at platform level (e.g. AOCS jitter and TED), camera level (e.g. optical distortion and ghost images), and detector level (e.g. PRNU, CTI, and BFE), as well as astrophysical signals (granulation, stochastic oscillations, and exoplanet transits). A unique ability is the inclusion of a realistic PSF (as illustrated in Fig.~\ref{fig:platosim}b) either from Zemax simulations, or from a parametric description allowing a PSF that varies over the focal plane, while keeping the computational burden feasible. \platosim{} stands out due to its configurability of the N-CAM and F-CAM to simulate: i) a full suite of PLATO data products (i.e. images, meta data, housekeeping data, and light curves); ii) the different modes of operation (e.g. nominal observations and microscanning sessions); and iii) the different payload configurations. This together with its end-to-end functionality to generate light curves coming from 24 cameras makes it a key simulator for the PLATO mission consortium.  


\platosim{} chooses to perform its simulations in the time domain rather than in the Fourier domain. The advantage is the ability to easily include time-dependent instrumental variations that may interfere with the stellar signal. The corresponding challenge is the computational resources needed to generate long time series. \platosim{} largely manages to overcome the computational bottlenecks and drastically reduce the execution time by making extensive use of efficient parametric models while still preserving a realistic simulation of each physical phenomenon. An additional mitigation effort is the coherent use of wavelength-averaged quantities such as the PSF, the throughput, the detector efficiency, and the quantum efficiency, to avoid a time-consuming numerical integration over the wavelength in the PLATO passband. The monochromatic light approximation, however, makes it difficult to compare the increased level of noise expected for the extreme spectral types of PLATO’s targets (i.e. F5 and K7) had the wavelength dependence of the above-mentioned quantities been taken into account. Nevertheless, as \platosim{} is designed to simulate stars of spectral types similar to the Sun, quantitatively the impact on the photometry is expected to be small. Given PLATO’s spectral response, the baseline at camera level will be that the noise in the light curve of a F5 dwarf star will be slightly higher than for a K7 dwarf star.


Despite already being an advanced simulator, \platosim{} remains a simulation tool that is continuously being improved and extended upon. At the time of writing, the AIV of the PLATO cameras is well under way, and laboratory measurements of the instrument characteristics such as the PRNU and the morphology of ghost images are becoming available. These measured quantities will in the near future replace any simulated quantities in \platosim{} where possible. On the topic of the PRNU, if inter-pixel sensitivity measurements become available, they will allow \platosim{} to accurately account for a lower pixel sensitivity near the pixel borders opposed to the pixel centre. With \platosim{}'s ability to track time-dependent effects on time scales shorter than the exposure time, such an improved PRNU model will play a key role for future development concerning effects that are best described at the subpixel level. 

Charge injection will be implemented in order to investigate its potential as mitigation of the CTI in the final stage of the mission's lifespan when the CTI could be significantly worse depending on the solar activity. Additionally, a relevant noise source for photometric measurements is scattered light coming from the Earth and the Moon. Although measures will be taken to avoid this as much as possible, some residual light scatter may still occur. Since this scatter is time dependent and could interfere with the detected stellar signal, it is worth considering and implementing this in \platosim{}. Defective pixels are also on the list of future implementations. In particular, RTS pixels are important as their variable dark current behaviour for slow state switching (hours to days) will introduce flux jumps (similar to mask updates) and for fast state switching (from seconds to minutes) can introduce a significantly increased scatter in the extracted light curves. Finally, \platosim{} advantageously allows the user to simulate CCD subfields instead of full-frame images to reduce computational resources. As such, smearing and blooming trails from stars outside the subfield, plus light contribution from stars close to the pixel borders of the subfield, are not accounted for. Since the former is particularly important for designing the processing algorithms that account for the impact of smearing, it is a feature that belongs among the future enhancements of \platosim{}.


With the PLATO mission planned to see first light in less than three years the scientific justifications for the core and complimentary science programmes are constantly being verified with simulations. As demonstrated in this work, \platosim{} offers the opportunity to validate the performance of the PLATO payload and to undertake detailed simulation studies with a highly realistic approach that includes all foreseen random and systematic noise sources, an exact distribution of stars across the multi-camera arrangement in line with future planned pointing fields, and the inclusion of stellar variability for both target and contaminant stars. This makes \platosim{} a versatile software package under continuous development as a bedrock in the preparation for PLATO's future discoveries.

\begin{acknowledgements}

This work presents results from the European Space Agency (ESA) space mission
PLATO. The PLATO payload, the PLATO Ground Segment and PLATO data processing
are joint developments of ESA and the PLATO mission consortium (PMC). Funding for
the PMC is provided at national levels, in particular by countries participating in the PLATO Multilateral Agreement (Austria, Belgium, Czech Republic, Denmark, France, Germany, Italy, Netherlands, Portugal, Spain, Sweden, Switzerland, Norway, and United Kingdom) and institutions from Brazil. Members of the PLATO Consortium can be found at \url{https://platomission.com/}. The ESA PLATO mission website is
\url{https://www.cosmos.esa.int/plato}. We thank the teams working for PLATO for all their work.

The research behind these results has received funding from the BELgian federal Science Policy Office (BELSPO) through PRODEX grant PLATO: ZKE2050-01-D01.

RH acknowledges support from the German Aerospace Agency (Deutsches Zentrum f\"ur Luft- und Raumfahrt) under PLATO Data Center grant 50OO1501.

This project additionally made use of the following published Python packages: \texttt{NumPy} \citep{harris2020array}, \texttt{Numba} \citep{lam2015numba}, \texttt{Pandas} \citep{mckinney2011pandas, team2022pandas}, \texttt{SciPy} \citep{virtanen2020scipy}, \texttt{Matplotlib} \citep{hunter2007matplotlib}, and \texttt{Astropy} \citep{price2022astropy,price2018astropy,robitaille2013astropy}.

\end{acknowledgements}

\bibliographystyle{aa}
\bibliography{bibliography}

\begin{appendix}

\section{Reference frames}\label{app:reference_frames}

This Appendix gives a detailed description of the rotation matrices in Eq.~(\ref{rf_eq2ccd}) which are used to determine the focal plane
positions of the stars.

\subsection{The payload module reference frame}

The first rotation matrix $\mathbf{R}_\equa^\plm$ in Eq.~(\ref{rf_eq2ccd}) describes a passive rotation from the equatorial sky reference frame (EQ) to a reference frame associated with the payload module (PLM). Both reference frames have their origin in the geometric centre of the interface between the bottom of the optical bench and the service module. The former one is the base reference system of \platosim{} and refers to the standard right-handed celestial equatorial coordinate system $(X_\equa, Y_\equa, Z_\equa)$ where $X_\equa$ points towards the vernal equinox, and $Z_\equa$ to the equatorial north pole. We define a new right-handed Cartesian coordinate system $(X_\plm, Y_\plm, Z_\plm)$ which we call the payload module reference frame by performing the following rotations starting from the equatorial reference frame $(X_\equa, Y_\equa, Z_\equa)$. 

We first rotate over an angle $\alpha_\plm$ around the $Z_\equa$ axis, to obtain the $(X_\A, Y_\A, Z_\A)$ reference frame where $Z_\A = Z_\equa$. The corresponding passive rotation matrix is: 
\begin{equation}
              \mathbf{R}_\equa^\A = 
             \begin{pmatrix} 
                 \cos\alpha_\plm & \sin\alpha_\plm & 0 \\
                 -\sin\alpha_\plm &  \cos\alpha_\plm & 0 \\
                 0                  &  0                  & 1
             \end{pmatrix} \,.
            \label{eq:rotmatrix_EQ_A}
\end{equation}
Secondly, we then rotate over an angle $\pi/2-\delta_\plm$ around the $Y_\A$ axis, to obtain the $(X_\B, Y_\B, Z_\B)$ reference frame where $Y_\B = Y_\A$. The passive rotation matrix is:
\begin{equation}
              \mathbf{R}_\A^\B = 
             \begin{pmatrix} 
                 \sin\delta_\plm  & 0 & -\cos\delta_\plm \\
                 0                   & 1 & 0                  \\ 
                 \cos\delta_\plm  & 0 & \sin\delta_\plm \\
             \end{pmatrix} \,.
            \label{eq:rotmatrix_A_B}
\end{equation}
The reason for $\pi/2-\delta_\plm$ rather than $\delta_\plm$ is that $\delta_\plm$ is measured from the equator rather than the north pole, while the rotation angle should increase clockwise when looking in the positive direction of the rotation axis. Lastly, we rotate over a roll angle $\kappa_\plm$ around the $Z_\B$ axis, to obtain the $(X_\plm, Y_\plm, Z_\plm)$ reference frame, where $Z_\plm = Z_\B$. The passive rotation matrix is:
\begin{equation}
              \mathbf{R}_\B^\plm = 
             \begin{pmatrix} 
                  \cos\kappa_\plm & \sin\kappa_\plm & 0 \\
                 -\sin\kappa_\plm & \cos\kappa_\plm & 0 \\
                 0                  &  0                  & 1
             \end{pmatrix} \,.
             \label{eq:rotmatrix_B_SC}
\end{equation}

The first two rotations take care that the roll axis $Z_\plm$ of spacecraft is pointing in the direction of the mean LOS $(\alpha_\plm, \delta_\plm)$ in the sky given by the spacecraft operator. The third rotation rolls the spacecraft so that its yaw axis $X_\plm$, which points to the highest point of the sunshield, is orientated towards the Sun.

Given a target with celestial equatorial sky coordinates $(\alpha, \delta)$, the Cartesian coordinates $(v_x, v_y, v_z)_\plm$ of its unit direction vector in the payload module reference frame are computed using the following transformation
\begin{equation}
\label{EQ2SC}
\begin{pmatrix}
v_{x} \\
v_{y} \\
v_{z}
\end{pmatrix}_{\!\!\plm}
=
\mathbf{R}_\equa^{\plm}(\alpha_\plm, \delta_\plm, \kappa_\plm) \
\begin{pmatrix}
\cos\delta \cos\alpha \\
\cos\delta \sin\alpha \\
\sin\delta
\end{pmatrix}_{\!\!\equa} \,,
\end{equation}
where the rotation matrix $\mathbf{R}_\equa^\plm(\alpha_\plm, \delta_\plm, \kappa_\plm)$ is given by
\begin{equation}
    \mathbf{R}_\equa^\plm(\alpha_\plm, \delta_\plm, \kappa_\plm)
    = \mathbf{R}_\B^\plm \cdot  \mathbf{R}_\A^\B \cdot  \mathbf{R}_\equa^\A \,,
\end{equation}
and the rotation matrices $\mathbf{R}_\equa^\A$, $\mathbf{R}_\A^\B$, and $\mathbf{R}_\B^\plm$ are given by Eqs.~(\ref{eq:rotmatrix_EQ_A}), (\ref{eq:rotmatrix_A_B}), and (\ref{eq:rotmatrix_B_SC}).

\subsection{The camera reference frame}

The second rotation matrix $\mathbf{R}_\plm^\cam$ in Eq.~(\ref{rf_eq2ccd}) describes a passive rotation from the payload module reference frame 
to the right-handed Cartesian camera coordinate system $(X_\cam, Y_\cam, Z_\cam)$. It is associated with one of the cameras
and has its origin exactly in the middle of the 4 CCDs in the associated focal plane. The mounting of the normal cameras on the platform is such
that they are tilted and are therefore not pointing in the same direction of the pointing (roll) axis $Z_\plm$ of the payload module. Instead,
the optical axis $Z_\cam$ of a camera is pointing towards a field-of-view that is defined by an azimuth angle $\eta_\cam$ and a tilt angle 
$\rho_\cam$ with respect to the PLM pointing. These angles are listed for the different camera groups in Table \ref{CameraGroupInfo}. 
The tilt angle $\rho_\cam$ of each normal camera is fixed at 9.2 degrees, and the azimuth angle is $\eta_\cam = 45 + n\cdot 90$ degrees where $n$ 
is the normal camera group ID number.
\begin{table}
\caption[]{Overview of camera group angles.}
\begin{center}
\begin{tabular}{c c c c}
\Hline
    Camera Group\tablefootmark{a} & $\eta_\cam \ (^{\circ})$\tablefootmark{b}  & $\rho_\cam \ (^{\circ})$\tablefootmark{c} \\
\Hline
1    &  45 & 9.2  \\
2    & 135 & 9.2  \\
3    & 225 & 9.2  \\
4    & 315 & 9.2  \\
Fast &   0 & 0.0  \\
\hline
\end{tabular}
\label{CameraGroupInfo}
\end{center}
\tablefoot{Description of each column: \\
\tablefoottext{a}{The camera group ID number.} \\
\tablefoottext{b}{The azimuth rotation angle $\eta_\cam$ in degrees.} \\
\tablefoottext{c}{The tilt rotation angle $\rho_\cam$ in degrees.}
} 
\end{table}

To define the rotation matrix $\mathbf{R}_\plm^\cam$, we start from the payload module reference frame $(X\plm, Y_\plm, Z_\plm)$, and perform the following rotations we rotate over a position angle $\eta_\cam$ around the $Z_\plm$ axis, to obtain the $(X'_\plm, Y'_\plm Z'_\plm)$ reference frame where $Z'_\plm = Z_\plm$. The corresponding passive rotation matrix is
\begin{equation}
              \mathbf{R}_1 = 
             \begin{pmatrix} 
                 \cos\eta_\cam  &  \sin\eta_\cam & 0 \\
                 -\sin\eta_\cam &  \cos\eta_\cam & 0 \\
                 0             &  0            & 1
             \end{pmatrix} \,.
            \label{eq:rotmatrix_1}
\end{equation}
Secondly, we rotate over a tilt angle $\rho_\cam$ around the $Y'_\plm$ axis, to obtain the $(X''_\plm, Y''_\plm, Z''_\plm)$ reference frame where $Y''_\plm = Y'_\plm$. The corresponding passive rotation matrix is:
\begin{equation}
              \mathbf{R}_2 = 
             \begin{pmatrix} 
                 \cos\rho_\cam  & 0 & -\sin\rho_\cam \\
                 0              & 1 & 0             \\ 
                 \sin\rho_\cam & 0 & \cos\rho_\cam \\
             \end{pmatrix} \,;
            \label{eq:rotmatrix_2}
\end{equation}
We finally rotate over an angle $2\pi - \eta_\cam$ radians around the $Z''_\plm$ axis, to obtain the $(X_\cam, Y_\cam, Z_\cam)$ reference frame where $Z_\cam = Z''_\plm$, using the passive rotation matrix 
\begin{equation}
              \mathbf{R}_3 = 
             \begin{pmatrix} 
                 \cos\eta_\cam  &  -\sin\eta_\cam & 0 \\
                 \sin\eta_\cam  &   \cos\eta_\cam & 0 \\
                 0             &  0             & 1
             \end{pmatrix} = \mathbf{R}_1^t  \,.
            \label{eq:rotmatrix_3}
\end{equation}

\begin{figure}
\centering
\includegraphics[width=0.45\textwidth,keepaspectratio]{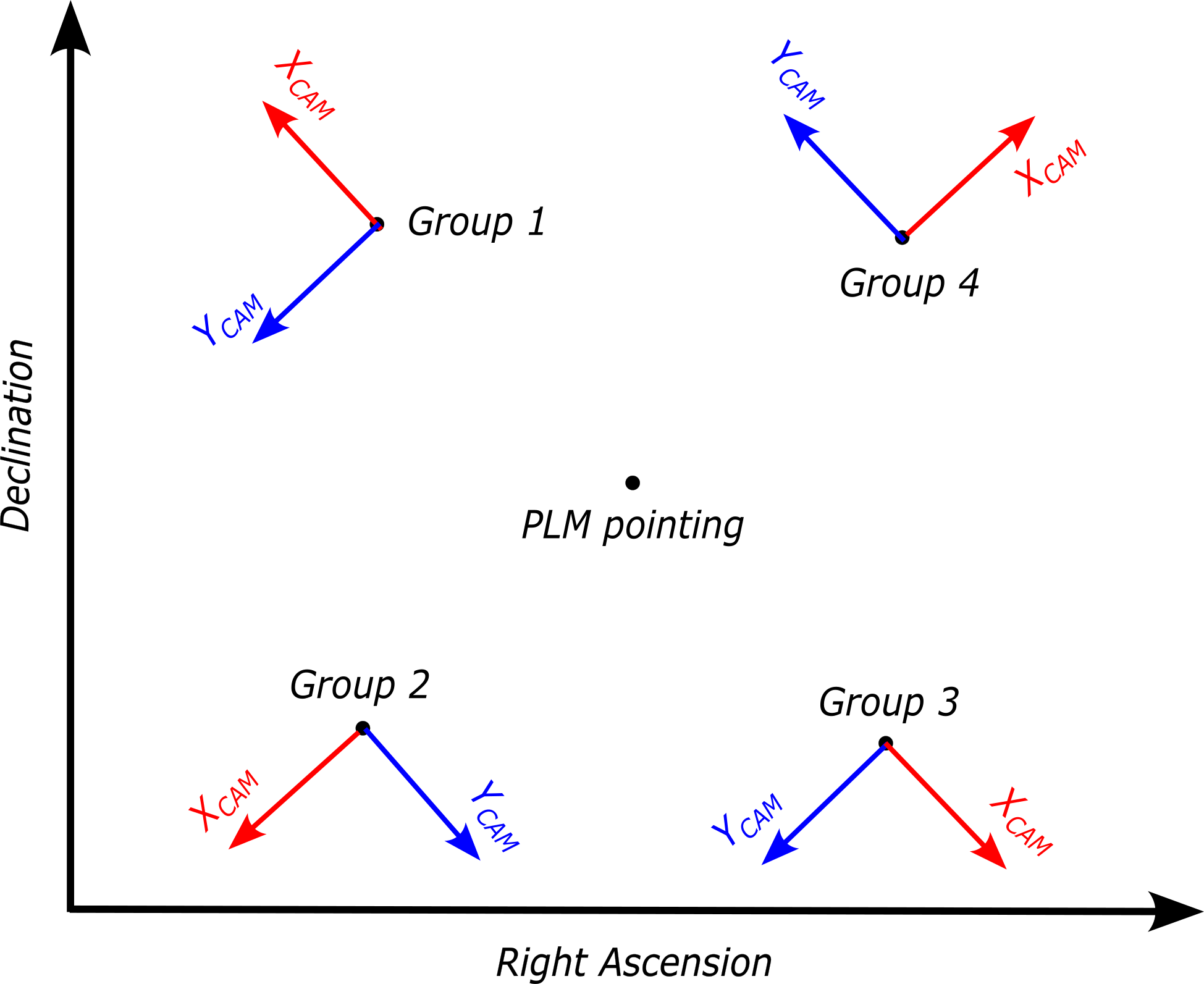}
\hspace{5mm}
\includegraphics[width=0.45\textwidth,keepaspectratio]{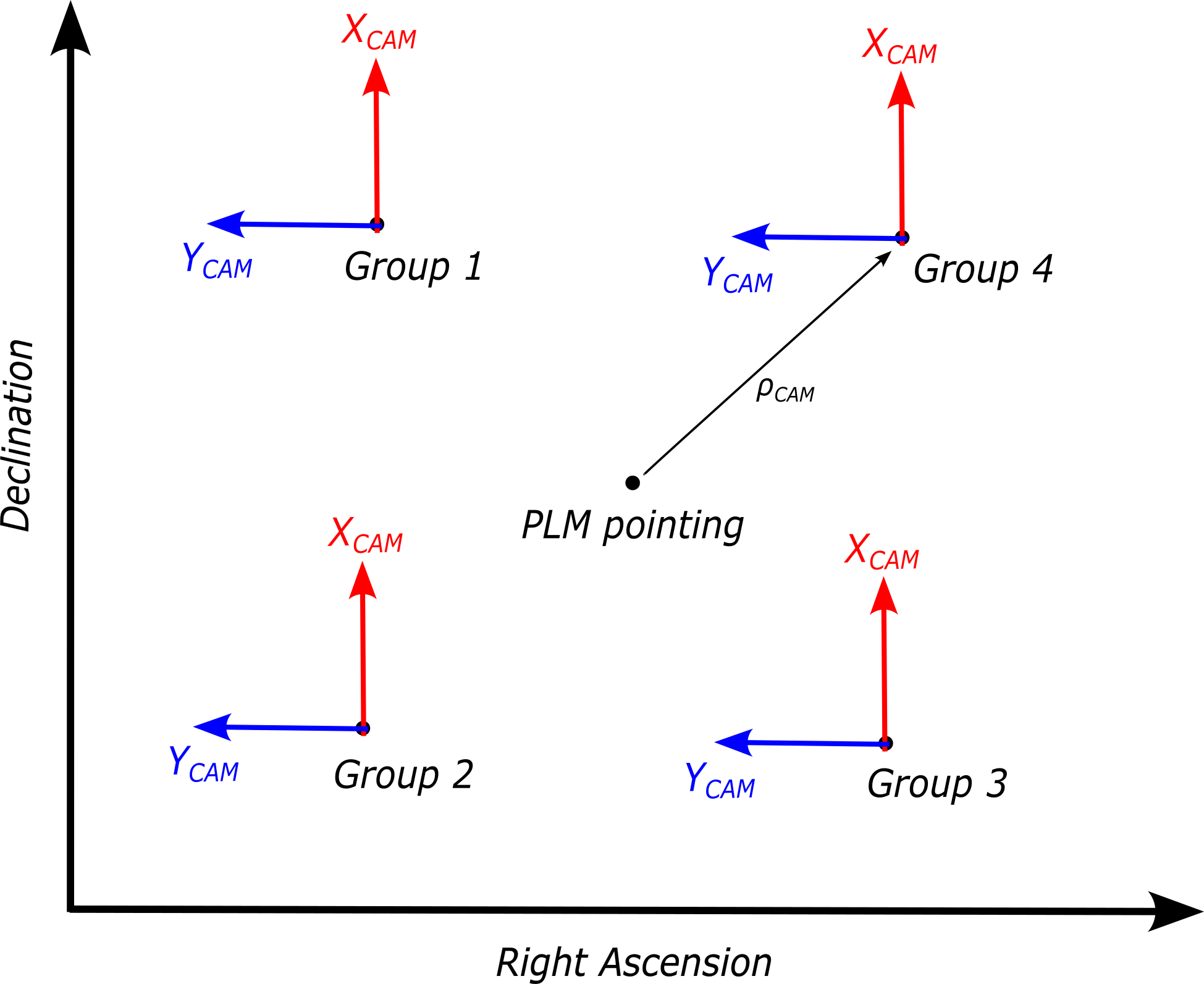}
\caption[]{Reference frames in equatorial coordinates. \textbf{Top:} the orientation of the camera reference frame for each of the camera groups when there would be no extra rotation over an angle $-\eta_\cam$. \textbf{Bottom:} the orientation of the same reference frames with the extra rotation (\ref{eq:rotmatrix_3}). $\rho_\cam$ is the camera tilt angle. The right-hand side orientation allows for nearly aligned camera axes $X_\cam$ and $Y_\cam$  of the different camera groups on the sky.}
\label{RefTLwithAndWithoutExtraAzimuth}
\end{figure}

The first two rotations serve to point the camera on the platform to the line-of-sight defined by the two angles $\eta_\cam$ and $\rho_\cam$. The last
rotation takes care that the $X_\cam$ and $Y_\cam$ axes of the different camera groups are aligned. For example, without the extra rotation, the
location of the spacecraft pointing axis $(x_\plm, y_\plm, z_\plm) = (0,0,1)$ in the camera reference frame would be $(x_\cam, y_\cam, z_\cam) =
(-\sin\varrho, 0, \cos\varrho)$ (cf.~Eq.~(\ref{SC2TL})). That is, in the camera xy-plane the payload module pointing axis would cross the negative
$x_\cam$-axis, independent of the camera group. This is illustrated in Fig.~\ref{RefTLwithAndWithoutExtraAzimuth}. The extra rotation allows to align
the focal planes of the different camera groups in the sky, and has the additional benefit that the payload module pointing coordinates always fall
on (rather than in between) a CCD for each of the 4 camera groups.

A vector with coordinates $(x_\plm, y_\plm, z_\plm)$ in the payload module reference frame, has coordinates $(x_\cam, y_\cam, z_\cam)$ 
in the camera reference frame that can be computed using the following transformation:
\begin{equation}
\label{SC2TL}
\begin{pmatrix}
x_\cam \\
y_\cam \\
z_\cam
\end{pmatrix} =
\mathbf{R}_\plm^\cam(\eta_\cam, \rho_\cam)
\begin{pmatrix}
x_\plm \\
y_\plm \\
z_\plm
\end{pmatrix} \,,
\end{equation}
where
\begin{equation}
    \label{R_SC2TL}
    \mathbf{R}_\plm^\cam(\eta_\cam, \rho_\cam) = \mathbf{R}_1^t \cdot \mathbf{R}_2 \cdot \mathbf{R}_1 \,,
\end{equation}
where we used the rotation matrices (\ref{eq:rotmatrix_1}), (\ref{eq:rotmatrix_2}) and (\ref{eq:rotmatrix_3}) enumerated above.

\subsection{The focal plane reference frame}\label{Sec:FocalPlane}

\begin{figure}
\centering
\includegraphics[width=0.5\textwidth,keepaspectratio]{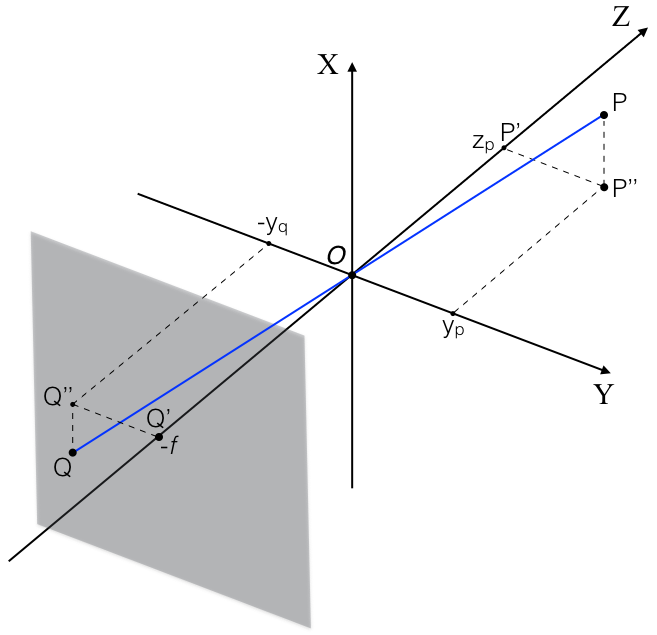}
\caption{\label{PinHoleCamera} Standard pinhole camera model. The $Z$ axis corresponds to the optical axis of the camera. 
The origin of the reference frame is the pinhole $O$, the focal plane is located behind the pinhole at a distance $f$ where $f$ is the 
focal length of the camera. Light rays of a star at point $P$, go through the pin hole $O$ and are projected to a point $Q$ on the focal plane.}
\end{figure}

The third rotation matrix $\mathbf{R}_\cam^\focal$ in Eq.~(\ref{rf_eq2ccd}) describes the projection of the sky on the focal plane of a camera. 
To do so, we use the standard pinhole camera model as is shown in Fig.~\ref{PinHoleCamera}.
The $Z$ axis is the optical axis of the camera, the $(X,Y)$ plane is parallel to the focal plane, but has its origin in the pinhole. 
The focal plane (with the CCDs) is at a distance $f$ behind the pinhole, where $f$ is the focal length of the camera. 
Light rays of a star at point $P$ go through the pin hole $O$ and are projected to a point $Q$ on the focal plane. In the $(Y,Z)$ plane, 
the triangles $\widehat{OP'P''}$ and $\widehat{OQ'Q''}$ are similar triangles, so that we have $|P'P''|/|OP'| = |Q'Q''|/|OQ'|$ or
\begin{equation}
\frac{-y_q}{f} = \frac{y_p}{z_p} \,,
\end{equation}
where the minus sign indicates that the pinhole reverses the image on the focal plane. 
A similar argument holds in the $(X,Z)$ plane, so that we can write
\begin{equation}
\label{pinholeProjection}
\begin{pmatrix}
x_q \\
y_q \\
-f
\end{pmatrix}
=
- \frac{f}{z_p}
\begin{pmatrix}
x_p \\
y_p \\
z_p
\end{pmatrix} \,.
\end{equation}

With normal coordinates, the non-linear division by $z_p$ prevents expressing the pinhole projection as a convenient matrix multiplication. We
therefore go to augmented space and use homogeneous coordinates $(x_q, y_q, 1)$ and $(x_p, y_p, z_p, 1)$
\begin{equation}
\begin{pmatrix}
x_q \\ y_q\\ z_q \\ 1
\end{pmatrix}
\sim
\mathbf{C} \cdot
\begin{pmatrix}
x_p \\ y_p \\ z_q \\ 1
\end{pmatrix}
\end{equation}
where the equality $=$ has been replaced by the equivalence relation $\sim$. After any transformation a homogenisation of the coordinates is
executed, which means dividing all coordinates by the fourth one. The camera matrix $\mathbf{C}$ used by \platosim{} is
\begin{equation}
\mathbf{C} =
\begin{pmatrix}
    -f & 0 & 0 & 0 \\
    0 & -f & 0 & 0 \\
    0 & 0 & -f & 0 \\
    0 & 0 & 1 & 0 \\
\end{pmatrix} \,.
\end{equation}
which mirrors the image after the pinhole projection.

We can now define the right-handed Cartesian coordinate system $(X_\focal, Y_\focal, Z_\focal)$ associated with the focal plane of one camera with
its origin exactly in the middle of the 4 CCDs (see Fig.~\ref{fig:ccdFocalPlane} for an illustration). The $Z_\focal$ axis coincides with the
$Z_\cam$ optical axis of the camera, which points towards the FOV. The focal plane $(X_\focal, Y_\focal)$ is obtained by rotating the
$(X_\cam, Y_\cam)$ plane over an angle $\gamma_\focal$ called the focal plane orientation around the $Z_\focal = Z_\cam$ axis. The corresponding
transformation (using homogeneous coordinates) is
\begin{equation}
\label{TL2FP}
\begin{pmatrix}
x_\focal \\
y_\focal \\
-f \\
1
\end{pmatrix}
=
\mathbf{R}_{\cam}^{\focal}(\gamma_\focal)\
\begin{pmatrix}
x_\cam \\
y_\cam \\
z_\cam \\
1
\end{pmatrix} \,,
\end{equation}
where the rotation matrix $\mathbf{R}_{\cam}^{\focal}(\gamma_\focal)$ is given by:
\begin{equation}
\mathbf{R}_{\cam}^{\focal}(\gamma_\focal) =
\begin{pmatrix}
    -f & 0 & 0 & 0 \\
    0 & -f & 0 & 0 \\
    0 & 0 & -f & 0 \\
    0 & 0 & 1 & 0
\end{pmatrix}
\cdot
\begin{pmatrix}
    \cos\gamma_\focal & \sin\gamma_\focal  & 0  & 0 \\
    -\sin\gamma_\focal & \cos\gamma_\focal & 0 & 0 \\
    0 & 0 & 1 & 0 \\
    0 & 0 & 0 & 1
\end{pmatrix} \,.
\end{equation}
Here $(x_\cam, y_\cam, z_\cam, 1)$ are the homogenised coordinates of a vector in the camera reference frame, and $(x_\focal, y_\focal, -f, 1)$ are the coordinates of the homogeneous projected vector in the focal plane reference frame. We note that $\mathbf{R}_\cam^\focal$ is not invertible: projecting loses the $z_\cam$ information. However, since \platosim{} only uses (sky) directions, the reverse transformation can be obtained by enforcing the length of the vector in the camera 
reference frame to be unity.

\subsection{The CCD reference frame}

The fourth transformation matrix $\mathbf{R}_\focal^\ccd$ in Eq.~(\ref{rf_eq2ccd}) describes a passive rotation from the focal plane reference frame
to the right-handed Cartesian CCD coordinate system $(X_\ccd, Y_\ccd)$, which is associated with one of the CCDs in a camera. Its origin is in one of
the corners of the CCD. The charge transfer direction during readout of the CCD happens in the negative $Y_\ccd$ direction towards $y_\ccd = 0$, and
the readout of the serial readout register happens in the negative $X_\ccd$ direction (towards the origin) for the left side of the CCD, and in the
opposite direction for the right side of the CCD. The orientation of the CCD with respect to the $(X_\focal, Y_\focal)$ coordinate system is defined
by the angle $\gamma_\ccd$, as shown in Fig.~\ref{fig:ccdFocalPlane}.
The transformation (in homogeneous coordinates, including a rotation as well as a translation) between the focal plane and the CCD reference 
frames is implemented by
\begin{equation}
\label{Rfp2ccd}
\begin{pmatrix}
x_\ccd \\
y_\ccd \\
1
\end{pmatrix}
=
\mathbf{R}_\focal^\ccd(\gamma_\ccd) \cdot
\begin{pmatrix}
x_\focal \\
y_\focal \\
1
\end{pmatrix} \,,
\end{equation}
where
\begin{equation}
\mathbf{R}_\focal^\ccd(\gamma_\ccd)
=
\begin{pmatrix}
    \phantom{+}\cos\gamma_\ccd  & \sin\gamma_\ccd  & \Delta x_\ccd \\
    -\sin\gamma_\ccd  & \cos\gamma_\ccd & \Delta y_\ccd \\
    0 & 0 & 1
\end{pmatrix} \,,
\end{equation}

and where $(\Delta x_\ccd, \Delta y_\ccd)$ are the coordinates of the CCD zeropoint in the $(X_\focal, Y_\focal)$ frame. Equation~(\ref{Rfp2ccd}) allows to
use the same $\Delta x_\ccd$ and $\Delta y_\ccd$ for each CCD, independent of the CCD rotation angle $\gamma_\ccd$. An overview of the zero point
coordinates and the orientation angle $\gamma_\ccd$ is given in Table \ref{tab:ccdInformation}. \platosim{} chooses to define the Focal Plane
reference frame such that its axes run parallel with the CCD axes of CCD $\#3$.

\begin{table}
\centering
\caption[CCD reference frame parameters]
{Overview of the CCD reference frame parameters.}
\begin{tabular}{ccccccc}
\Hline
CCD\tablefootmark{a} & $N_{\rm rows}$\tablefootmark{b} & $N_{\rm cols}$\tablefootmark{b} & $r_{\rm exp}$\tablefootmark{c} & $\Delta x_\ccd$\tablefootmark{d} & $\Delta y_\ccd$\tablefootmark{d} & $\gamma_\ccd$\tablefootmark{e} \\
ID & (pix) & (pix) & (pix) & (mm) & (mm) & (\si{degree}) \\
\Hline
1      & 4510 & 4510 &    0 &  -1.3 & 82.48  & 180  \\
2      & 4510 & 4510 &    0 &  -1.3 & 82.48  & 270  \\
3      & 4510 & 4510 &    0 &  -1.3 & 82.48  & 0    \\
4      & 4510 & 4510 &    0 &  -1.3 & 82.48  & 90   \\
1 (F)  & 4510 & 4510 & 2255 &  -1.3 & 82.48  & 180  \\
2 (F)  & 4510 & 4510 & 2255 &  -1.3 & 82.48  & 270  \\
3 (F)  & 4510 & 4510 & 2255 &  -1.3 & 82.48  & 0    \\
4 (F)  & 4510 & 4510 & 2255 &  -1.3 & 82.48  & 90   \\
\hline
\end{tabular}
\tablefoot{Description of each column: \\
\tablefoottext{a}{The CCD codes 1, 2, 3, and 4 for a N-CAM, 1 (F), 2 (F), 3 (F), and 4 (F) for a F-CAM.} \\
\tablefoottext{b}{The number of rows and columns for for each CCD.} \\
\tablefoottext{c}{First row that is exposed; for the fast camera only rows 2255 until 4510 are exposed.} \\
\tablefoottext{d}{The CCD origin coordinates in the $(X_\focal, Y_\focal)$ frame.} \\
\tablefoottext{e}{The CCD rotation angle.}
} 
\label{tab:ccdInformation}
\end{table}

\section{The PLATO PSF}\label{app:plato_psf}

\begin{figure*}[t!]
\centering
\includegraphics[width=\columnwidth]{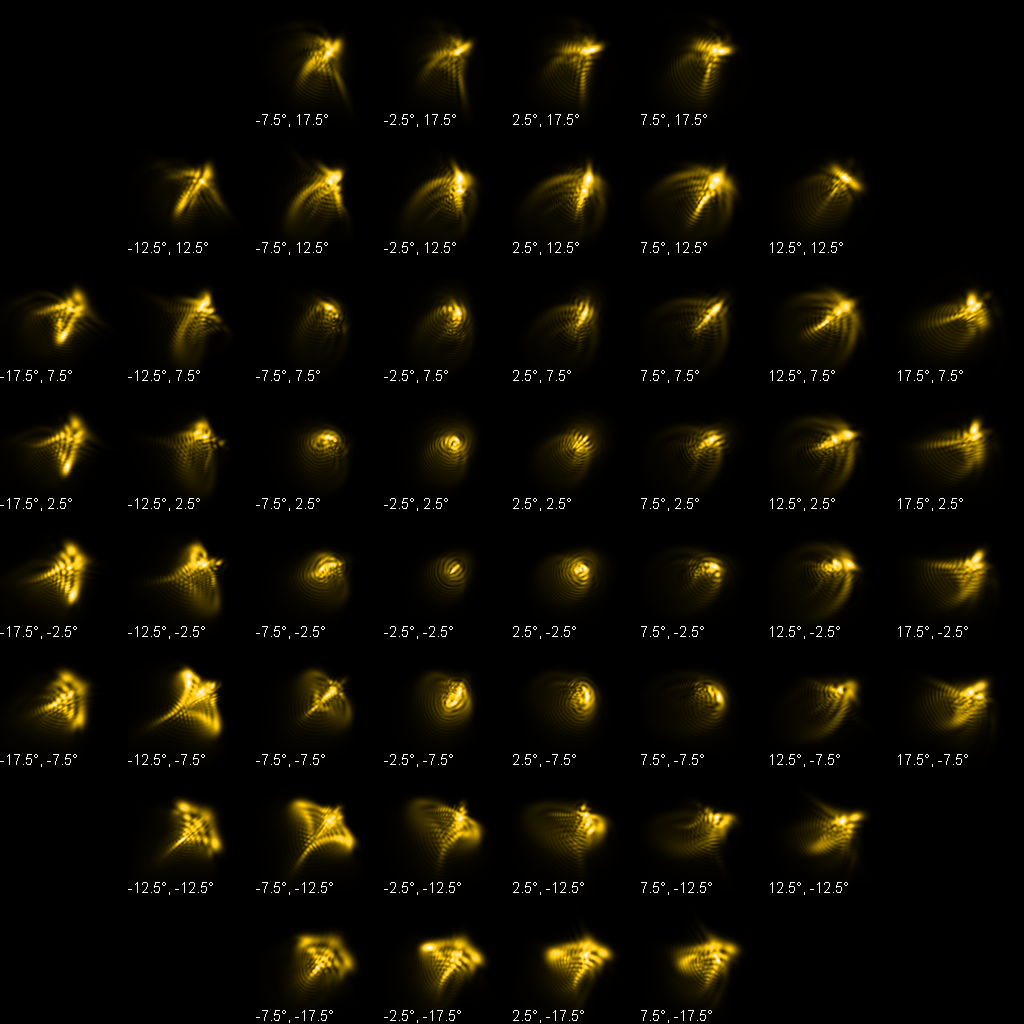}
\includegraphics[width=\columnwidth]{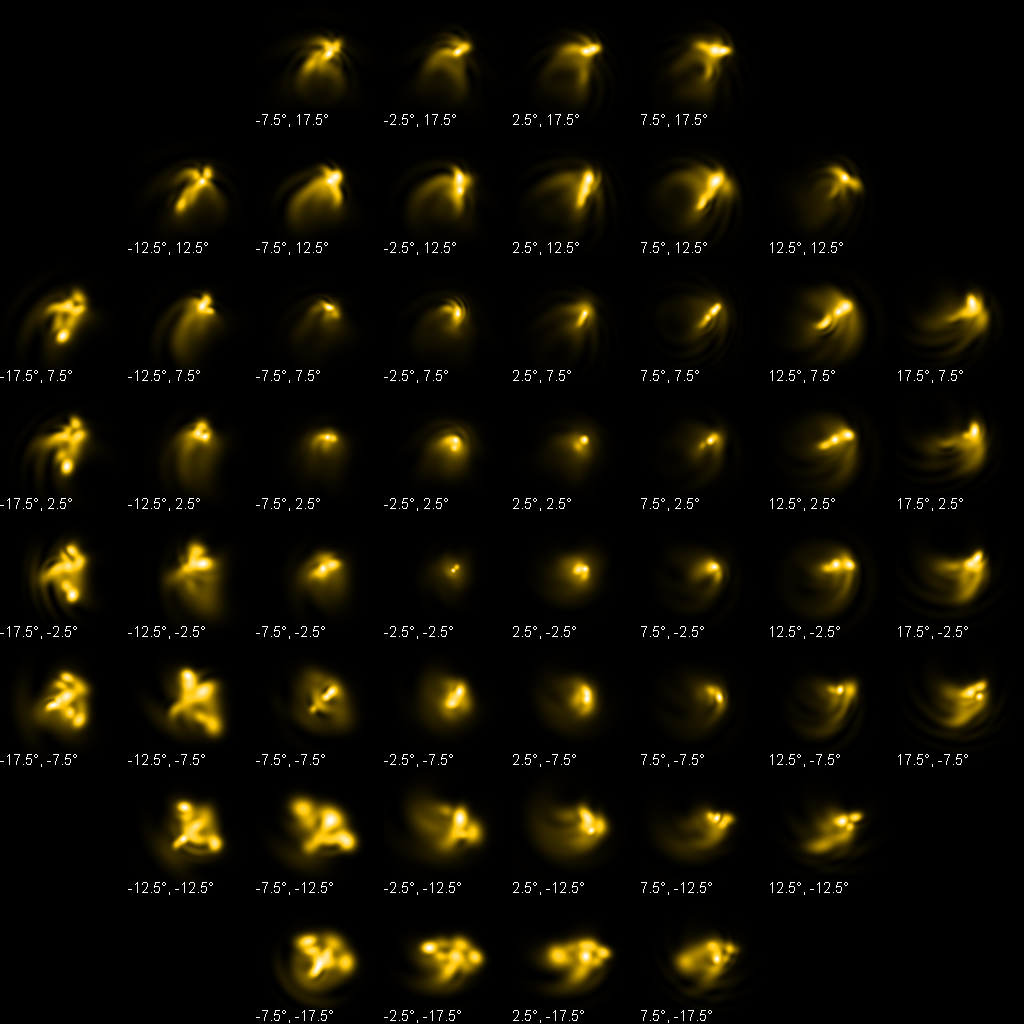}
\begin{picture}(1,1)
\put( 50, 0){\comment{a)}}
\put(100, 0){\comment{b)}}
\end{picture}
\caption[]
{Normalised non-diffused PLATO PSFs. \textbf{Left:} PSFs generated with Zemax OpticStudio. \textbf{Right:} PSFs generated with the analytic model. Both plots display the PSF across the full FPA and the focal plane coordinate is shown below the corresponding PSF.} 
\label{fig:PSF_Carsten}
\end{figure*}
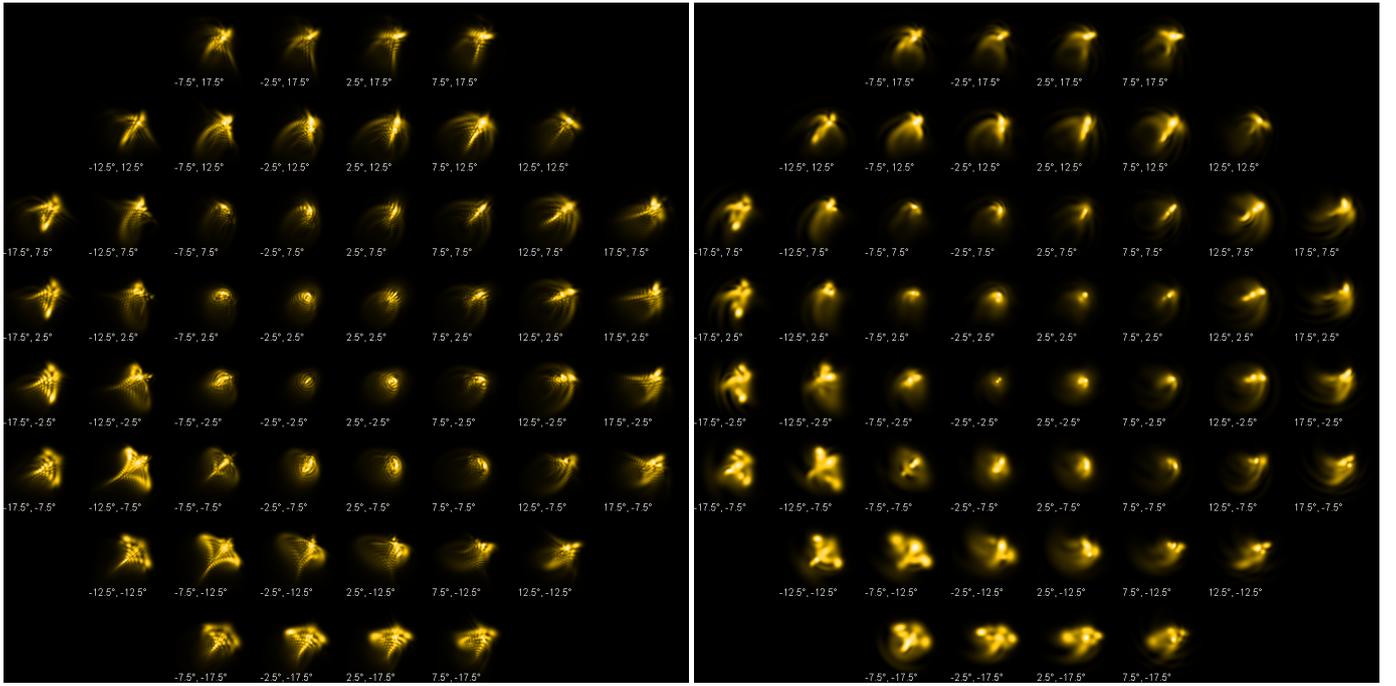

As for most cameras, the PSF depends on the angular radial distance between the barycentre of the PSF in the focal plane $(x, y)$ to the optical axis. The latter is assumed to go through the origin $(0,0)$ of the focal plane. This angular distance $\vartheta$ can be computed in the focal plane reference
frame with
\begin{equation}
\cos\vartheta = \left[ 1 + \left(\frac{x_\focal}{f}\right)^2 + \left(\frac{y_\focal}{f}\right)^2 \right]^{-1/2} \,,
\end{equation}
where $(x_\focal, y_\focal)$ are the pinhole-projected focal plane coordinates of the centre of the PSF (cf Sect.~\ref{Sec:FocalPlane}). 
Moreover, the exact morphology of the PSF is not only dependent on $\vartheta$ but also on the orientation angle $\tx{\theta}{OA}$ of the PSF. 
A PSF therefore has to be properly rotated at its location in the focal plane. 

\platosim{} has two options for incorporating PSFs: synthetic and analytical ones. As part of the mission preparation, synthetic PSFs of the PLATO
cameras have been extensively generated with the ray-tracing software package Zemax for many different positions in the focal plane, for both
in-focus and out-off-focus PSFs, and for different stellar effective temperatures. For a given subfield with point sources in the focal plane, the
optimal PSF corresponding to the subfield centre is selected from the library, and applied to the entire subfield using a convolution with a Fast
Fourier Transform. As a result, with this option the PSF is the same across the complete subfield. Being a compromise of computational cost, 
the assumption of a constant PSF becomes less valid with increasing subfield size.

In the case of analytic PSF models a parametric model is used, which is defined as
\begin{equation}\label{analyticNonGaussian}
\tx{G}{N}(x,y) = \frac{1}{Z} \left(\sum\limits_{n=1}^{N_G} f_{G,n}(x,y) +  \sum\limits_{l=1}^{N_S} f_{S,l}(x,y) +  \sum\limits_{k=1}^{N_C} 
f_{C,k}(x,y)\right) \,,
\end{equation}
where the three base functions are given by
\begin{align}\label{analyticNonGaussianComponents}
f_G(x,y) & = A \ e^{-\beta_0} \,, \\
f_S(x,y) & = A \ e^{-\beta_0} \sin^2(\delta \ \beta_s) \,, \nonumber \\
f_C(x,y) & = A \ e^{-\beta_0} \cos^2(\delta \ \beta_s) \,, \nonumber
\end{align}
where $\sigma$ is the generalised standard deviation of the PSF, $\delta$ describes the morphology of the PSF side lopes, and ($\beta_0$,
$\beta_s$) are abbreviations for the following expressions:
\begin{align*}
\beta_0 &\equiv \frac{(x-x_0)^2+(y-y_0)^2}{2\sigma^2} \,, \\
\beta_s  &\equiv \frac{(x+x_s-x_0)^2+(y+y_s-y_0)^2}{4\sigma^2} \,.
\end{align*}
The normalisation constant $Z$ in Eq. \eqref{analyticNonGaussian} is given by
\begin{equation}\label{analyticNonGaussianZ}
Z = \sum\limits_{n=1}^{N_G} Z_{G,n} +  \sum\limits_{l=1}^{N_S} Z_{S,l} +  \sum\limits_{k=1}^{N_C} Z_{C,k} \,,
\end{equation}
where
\begin{align}\label{analyticNonGaussianNorm}
Z_G & = 2 \, A \, \pi \, \sigma^2 \,, \\
Z_S & = A \, \pi \, \sigma^2 \, \paren{1 - \kappa_0} \nonumber \,, \\
Z_C & = A \, \pi \, \sigma^2 \, \paren{1 + \kappa_0} \nonumber \,,
\end{align}
with $\kappa_0$ and $\kappa_s$ defined as
\begin{align*}\label{analyticNonGaussianAlpha}
\kappa_0 &\equiv \frac{e^{-\delta \, \kappa_s} \, (\cos\kappa_s - \delta\sin\kappa_s)}{1+\delta^2} \,, \\
\kappa_s &\equiv \frac{\delta \, (x_s^2+y_s^2)}{2\sigma^2 (1+\delta^2)} \,.
\end{align*}
We note that $\lim\limits_{\delta\rightarrow 0} f_C = f_G$ and that if two base function $f_S$ and $f_C$ have the same parameters $(x_0, y_0, A, \sigma, 
\delta, x_s, y_s)$ their sum $f_S + f_C = f_G$. 

The advantage of this parametric model is that it allows for a wide variety of PSF morphologies, and that it can be efficiently integrated over 
a pixel. For example, in the case of the base function $f_G(x,y)$ we have  
\begin{equation}\label{analyticIntegral}
\int\limits_{x_i}^{x_{i+1}}\int\limits_{y_i}^{y_{i+1}} \frac{1}{Z_G}\,f_G(x,y)\,dx dy = \frac{1}{4} \ \mathcal{E}_x \ \mathcal{E}_y \,,
\end{equation}
where $\mathcal{E}_x$ and $\mathcal{E}_y$ are given by
\begin{align}
\mathcal{E}_x &= {\rm erf}\left(\frac{x_{i+1}-x_0}{\sqrt{2}\sigma}\right) - {\rm erf}\left(\frac{x_{i}-x_0}{\sqrt{2}\sigma}\right) \,, \\ 
\mathcal{E}_y &= {\rm erf}\left(\frac{y_{i+1}-y_0}{\sqrt{2}\sigma}\right) - {\rm erf}\left(\frac{y_{i}-y_0}{\sqrt{2}\sigma}\right) \,. \nonumber
\end{align}
Here ${\rm erf}(x)$ is the standard error function given by
\begin{equation}
{\rm erf}(x) \equiv \frac{2}{\sqrt{\pi}}\,\int_0^x e^{-t^2} dt \,,
\end{equation}
which can be very efficiently computed using the \texttt{Faddeeva}%
\footnote{\url{http://ab-initio.mit.edu/wiki/index.php/Faddeeva_Package}} %
package. 

In practice, the model parameters $(A, \sigma, \delta, x_s, y_s)$ need to be fitted, which was done using the Zemax PSFs as a reference model. The
metric used was the sum of the absolute values of all subpixel value differences between the Zemax PSF and the analytic PSF. This led to a typical
residual deviation between the two of $\sim12\%$ for the PSFs without charge diffusion, and $\sim6\%$ for PSFs with
charge diffusion.

The analytic PSFs are not convolved, but simply summed over the subfield. This has the advantage that the PSF variation over the focal plane can be
taken into account, which is particular relevant for a large subfield.

For a comparison, Fig. \ref{fig:PSF_Carsten} shows the PSF across the focal plane modelled with Zemax (left) and the analytic (right) one. The
relatively large PLATO plate scale of \SI{15}{\arcsec\per\pixel} means that on average across the FOV 99\% of the PSF flux is enclosed in a subfield of
$5\times5$ pixels. The sparse sampling of the PSF makes the flux distribution of a star very sensitive to its barycentre location within a
pixel (see Sec. \ref{sec:photometry}). Furthermore, asymmetric PSFs that contains large spikes (more typical for larger angular distance from the
optical axis) further enhance effects of pixel-to-pixel and intra-pixel sensitivities as one or more pointing errors moves the PSF relative to the
pixel array.

\section{Stochastic oscillations}\label{sec:oscillations}

Since both the P1, P2, and P5 samples of the asPIC comprise stars of spectral type F5-K7 \citep{montalto2021all}, we expect all stars to show
photometric variability due to granulation and stochastic oscillations. To efficiently model and include these signals in simulations with
\platosim{}, BiSON observations%
\footnote{\texttt{Main\_Fits\_BiSON\_8640d\_lbest\_UseInSolarCycle.dat}}%
of 96 distinct pulsation mode frequencies of the Sun were used as a benchmark. We model the frequency of maximum power ($\tx{\nu}{max}$) and the
primary frequency spacing ($\Delta\nu$) from the asteroseismic scaling relation of \cite{kjeldsen1994amplitudes}. The granulation signal component
is modelled using 2 super-Lorentzian functions as per \cite{kallinger2014connection} and the pulsation amplitudes are modelled using the methodology
of \cite{corsaro2013bayesian} (which use \textit{Kepler} red giant stars in the open clusters NGC 6791 and NGC 6819, together with a sample of main
sequence and sub-giant \textit{Kepler} field stars). The granulation and pulsation amplitude spectrum is rescaled to the PLATO passband
outlined in \cite{sarkar2018stellar} (initially developed for the ARIEL mission), with the general methodology structured around determining the
bolometric coefficient from an interpolation of synthetic spectra -- here we use the PHOENIX library%
\footnote{\url{https://phoenix.astro.physik.uni-goettingen.de/}} %
\citep{husser2013new}. 
Compared to \cite{sarkar2018stellar}'s use of the PHOENIX NextGen library of Atmospheric models, we here use the PHOENIX high resolution spectral
models for a more complete space of parameters \{$\tx{M}{s}$, $\tx{R}{s}$, $\tx{T}{eff}$, $\log\, g$, $Z$, $\alpha$\}.  
\begin{figure}
\centering
\includegraphics[width=\columnwidth]{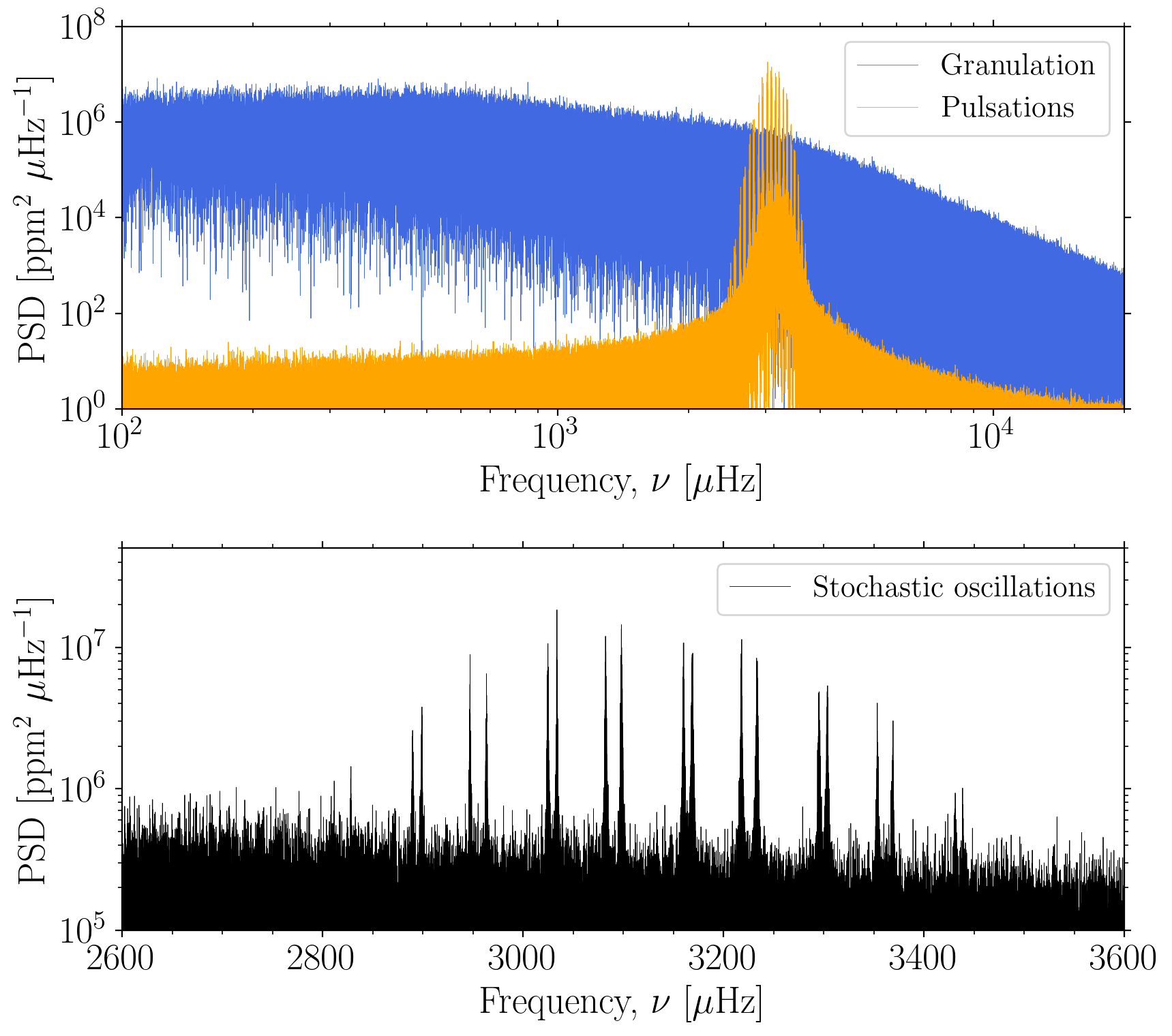}
\caption[]
{Power spectral density (PSD) plots of a \SI{2}{\year} simulated light curve due to convection-driven variability. \textbf{Top:} Components of the
simulated granulation (blue) and stochastic oscillations (orange). \textbf{Bottom:} Zoom-in on the granulation and pulsation model clearly showing
the recognisable envelope of frequency peaks (used to estimate $\tx{\nu}{max}$) and the small ($\delta\nu$) and large ($\Delta\nu$) frequency
separations.}
\label{fig:oscillations}
\end{figure}  

We simulate time series of stochastically damped modes using the formalism described in \cite{de2006modelling}. Being equivalent to the traditional
formalism of modelling solar-like oscillations in the Fourier domain \citep{anderson1990modeling} the code produces time series directly from the
time domain. We emphasise that the applied formalism (and the one formulated by \citealt{anderson1990modeling}) are simplistic models which do not
take into account a full physical description of turbulent convection nor rotation. Nevertheless, comparing our results using a solar-like
analogue, we note that our methodology results in a granulation amplitude of $\sim\SI{48.9}{\ppm}$ which is higher than the value reported by
\cite{kallinger2016precise} of $\SI{41\pm2}{\ppm}$. However, it resembles more closely the value of $\SI{46}{\ppm}$ from a recent study by
\cite{diaz2022scaling} which uses 3D stellar atmosphere models. Regarding the pulsation amplitude we find that our value of $\SI{10.6}{\ppm}$ is in
good agreement with the result from \cite{kallinger2016precise} of $\SI{11\pm6}{\ppm}$.

\end{appendix}

\end{document}